\documentclass[pdftex,twocolumn,epjc3]{svjour3}  

\usepackage[dvipsnames]{xcolor}
\usepackage{amsmath}
\usepackage{amssymb}
\usepackage[T1]{fontenc}
\usepackage{graphicx}
\usepackage{amsmath}
\usepackage{xspace}
\usepackage[numbers,sort&compress]{natbib}
\usepackage{subfig}
\usepackage[shortlabels]{enumitem}
\usepackage{textpos}
\usepackage[export]{adjustbox}

\usepackage{pstricks}
\usepackage{graphicx}
\usepackage{xspace}
\usepackage[pdfborder={0 0 0}]{hyperref}

\journalname{Eur. Phys. J. C}

\newcommand{\GF}{{G_\mu}}

\newcommand{\mix}{\text{mix}\xspace}
\newcommand{\LO}{\text{LO}\xspace}
\newcommand{\NLO}{\text{NLO}\xspace}

\newcommand{\QCD}{\text{QCD}\xspace}

\newcommand{\EW}{\text{EW}\xspace}

\newcommand{\Collier}{{\rmfamily\scshape Collier}\xspace}
\newcommand{\Munich}{{\rmfamily \scshape Munich}\xspace}
\newcommand{\Sherpa}{{\rmfamily\scshape Sherpa}\xspace}
\newcommand{\SherpaOpenLoops}{{\rmfamily\scshape Sherpa+OpenLoops}\xspace}
\newcommand{\OpenLoops}{{\rmfamily\scshape OpenLoops}\xspace}
\newcommand{\MunichOpenLoops}{{\rmfamily \scshape Munich+OpenLoops}\xspace}

\newcommand{\PV}{\ensuremath{V}\xspace}

\newcommand{\jet}{\ensuremath{j}\xspace}
\newcommand{\jets}{\mathrm{jets}}
\newcommand{\taucut}{{\tau}_{\mathrm{cut}}}

\newcommand{\ri}{\mathrm{i}}
\newcommand{\rF}{\mathrm{F}}
\newcommand{\rR}{\mathrm{R}}

\newcommand{\rT}{\mathrm{T}}
\newcommand{\rd}{\mathrm{d}}

\newcommand{\rS}{\mathrm{S}}

\newcommand{\mur}{\mu_{\rR}}
\newcommand{\muf}{\mu_{\rF}}

\newcommand{\GeV}{\text{GeV}\xspace}
\newcommand{\TeV}{\text{TeV}\xspace}

\newcommand{\alphaS}{\alpha_{\rS}}
\newcommand{\ord}{\mathcal{O}}

\newcommand{\HTprimehat}{\hat{H}_{\mathrm{T}}'}

\newcommand{\pT}{\ensuremath{p_{\mathrm{T}}}\xspace}

\def\parx{\frac{\rm d}{{\rm d}x}}
\def\pary{\frac{\rm d}{{\rm d}\vec y}}

\def\eps{\varepsilon}
\def\MC{\mathrm{MC}}
\def\TH{\mathrm{TH}}
\def\QCD{\mathrm{QCD}}
\def\mix{\mathrm{mix}}
\def\EW{\mathrm{EW}}
\def\PDF{\mathrm{PDF}}
\def\siv{\sigma^{(V)}}

\def\nnlo{\mathrm{NNLO}}
\def\nLO{\mathrm{nLO}}
\def\nNLO{\mathrm{nNLO}}

\def\NNLO{\mathrm{NNLO}}

\def\rshape{\mathrm{shape}}
\def\jet{\mathrm{jet}}

\def\Sud{\mathrm{Sud}}
\def\hard{\mathrm{hard}}
\def\max{\mathrm{max}}
\def\min{\mathrm{min}}
\def\NkLO{\mathrm{N}^k\mathrm{LO}}
\def\NkmLO{\mathrm{N}^{k-1}\mathrm{LO}}

\newcommand{\beqar}{\begin{eqnarray}}
\newcommand{\eeqar}{\end{eqnarray}}
\newcommand{\beq}{\begin{equation}}
\newcommand{\eeq}{\end{equation}}
\newcommand{\bit}{\begin{itemize}}
\newcommand{\eit}{\end{itemize}}

\def\refpar#1{\mbox{(\ref{#1})}}
\def\refeq#1{\mbox{Eq.~(\ref{#1})}}
\def\refeqs#1#2{\mbox{Eqs.~(\ref{#1})--(\ref{#2})}}
\def\reffi#1{\mbox{Figure~\ref{#1}}}
\def\reffis#1#2{\mbox{Figures~\ref{#1}--\ref{#2}}}
\def\refta#1{\mbox{Table~\ref{#1}}}

\def\refse#1{\mbox{Section~\ref{#1}}}
\def\refses#1#2{\mbox{Sections~\ref{#1}--\ref{#2}}}
\def\refapp#1{\mbox{\ref{#1}}}
\def\citere#1{\mbox{Ref.~\cite{#1}}}
\def\citeres#1{\mbox{Refs.~\cite{#1}}}

\def\ie{i.e.\ }

\def\max{\mathrm{max}}
\def\min{\mathrm{min}}

\def\thetaw{\theta_{\mathrm{w}}}

 \newcommand{\lsim}
 {\;\raisebox{-.3em}{$\stackrel{\displaystyle <}{\sim}$}\;}
 \newcommand{\gsim}
 {\;\raisebox{-.3em}{$\stackrel{\displaystyle >}{\sim}$}\;}

\def\ratiotextwidthapp{0.39}
\def\ratiotextwidth{0.39}
\def\ratiotextwidthbig{0.45}

\usepackage[left,mathlines]{lineno} % line numbering
\usepackage{soul} %text highlighting

\title{Precise predictions for $V$+jets dark matter backgrounds}
\author{J.~M.~Lindert\thanksref{addr1}
        \and
        S.~Pozzorini\thanksref{addr2}
        \and
        R.~Boughezal\thanksref{addr3}
        \and
        J.~M.~Campbell\thanksref{addr4}
        \and
        A.~Denner\thanksref{addr5}
        \and
        S.~Dittmaier\thanksref{addr6}
        \and
        A.~Gehrmann-De~Ridder\thanksref{addr2,addr7}
        \and
        T.~Gehrmann\thanksref{addr2}
        \and
        N.~Glover\thanksref{addr1}
        \and
        A.~Huss\thanksref{addr7}
        \and
        S.~Kallweit\thanksref{addr8}
        \and
        P.~Maierh\"ofer\thanksref{addr6}
        \and
        M.~L.~Mangano\thanksref{addr8}
        \and
        T.A.~Morgan\thanksref{addr1}
        \and
        A.~M\"uck\thanksref{addr9}
        \and
        F.~Petriello\thanksref{addr3,addr10}
        \and
        G.~P.~Salam\thanksref{e1,addr8}
        \and
        M.~Sch\"onherr\thanksref{addr2}
        \and
        C.~Williams\thanksref{addr11}
}

\thankstext{e1}{on leave from CNRS, UMR 7589, LPTHE, F-75005, Paris, France}

\institute{Institute for Particle Physics Phenomenology, Department of Physics, University of Durham, Durham,~DH1~3LE, UK\label{addr1}
\and
Physik-Institut, Universit\"at Z\"urich, Winterthurerstrasse 190, CH-8057 Z\"urich, Switzerland\label{addr2}
\and
High Energy Physics Division, Argonne National Laboratory, Argonne, IL 60439, USA\label{addr3}
\and
Fermilab, P.O.Box 500, Batavia, IL 60510, USA\label{addr4}
\and
Universit\"at W\"urzburg, Institut f\"ur Theoretische Physik und Astrophysik, 97074 W\"urzburg, Germany\label{addr5}
\and
Albert-Ludwigs-Universit\"at Freiburg, Physikalisches Institut, 79104 Freiburg, Germany\label{addr6}
\and
Institute for Theoretical Physics, ETH, CH-8093 Z\"urich, Switzerland\label{addr7}
\and
Theoretical Physics Department, CERN, CH-1211 Geneva 23, Switzerland\label{addr8}
\and
Institut f\"ur Theoretische Teilchenphysik und Kosmologie, RWTH Aachen University,
D-52056~Aachen,~Germany\label{addr9}
\and
Department of Physics \& Astronomy, Northwestern University, Evanston, IL 60208, USA\label{addr10}
\and
Department of Physics, University at Buffalo, The State University of New York, Buffalo 14260 USA\label{addr11}
\\
}

\hypersetup{
	pdfauthor={R. Boughezal, 
		J. M.Campbell,
		A. Denner,
		S. Dittmaier,
		A. Gehrmann-De Ridder
		T. Gehrmann,
	         N. Glover,
	         A. Huss,
		S. Kallweit,
		J. M. Lindert,
		P. Maierh\"ofer,
		M. L. Mangano,
		T.A. Morgan, 
		A. M\"uck,
		F. Petriello
		F. Petriello,
		S. Pozzorini,
		G. Salam
		M. Sch\"onherr,
		C. Williams
		},
		pdftitle={QCD and EW corrections for V+jets DM backgrounds},
		pdfkeywords={...}
}

\date{Received: date / Accepted: date}

\begin{document}
%%%%%%%%%%%%%%%%%%%%%%%%%%%%%%%%%%%%

\maketitle

%preprint numbers
\begin{textblock*}{10cm}(12.5cm,-10.1cm)
   {\footnotesize \flushleft 
CERN-TH-2017-102, 
CERN-LPCC-2017-02, \\
FERMILAB-PUB-17-152-T,
IPPP/17/38, \\
ZU--TH 12/17
}
\end{textblock*}

\begin{abstract} 

High-energy jets recoiling against missing transverse energy (MET) are powerful
probes of dark matter at the LHC. Searches based on large MET signatures require
a precise control of the $Z(\nu\bar\nu)+$\,jet background in the signal region.
This can be achieved by taking accurate data in control regions dominated by
$Z(\ell^+\ell^-)+$\,jet, $W(\ell\nu)+$\,jet and $\gamma+$\,jet production, and
extrapolating to the $Z(\nu\bar\nu)+$\,jet background by means of precise
theoretical predictions. In this context, recent advances in perturbative
calculations open the door to significant sensitivity improvements in dark
matter searches. In this spirit, we present a combination of state-of-the-art
calculations for all relevant $V+$\,jets processes, including throughout NNLO
QCD corrections and NLO electroweak corrections supplemented by Sudakov
logarithms at two loops. Predictions at parton level are provided together with
detailed recommendations for their usage in experimental analyses based on the
reweighting of Monte Carlo samples. Particular attention is devoted to the
estimate of theoretical uncertainties in the framework of dark matter searches,
where subtle aspects such as correlations across different $V+$\,jet processes
play a key role. The anticipated theoretical uncertainty in the
$Z(\nu\bar\nu)+$\,jet background is at the few percent level up to the TeV
range.
\end{abstract}

%\end{titlepage}

%\tableofcontents

\section{Introduction}

The signature of missing transverse energy (MET) is one of the most powerful
tools in the interpretation of data from hadron colliders.  In the Standard
Model (SM), MET arises from the neutrinos from the decay of $W$ and $Z$
bosons, and it can be used in their identification and study, as well as in
the identification and study of Higgs bosons, top quarks and other
SM particles whose decay products include $W$ or $Z$ bosons.
But MET is also an almost omnipresent feature of
theories beyond the SM (BSM), where it can be associated to the decay of new
particles to $W$ and $Z$ bosons, or directly to the production of new
stable, neutral and weakly interacting particles.  Typical examples are
theories with dark matter (DM) candidates, or Kaluza-Klein theories with
large extra dimensions.  Depending on the details, MET is accompanied by
other model-discriminating features, such as the presence of a small or
large multiplicity of hard jets, or of specific SM particles.  The
experimental search for these extensions of the SM relies on a proper
modeling of the SM backgrounds to the MET signature.  The determination of
these backgrounds is ideally done by using data control samples, but
theoretical input is often helpful, or even necessary, to extend the
experimental information from the control to the signal regions, or to
extend the application range of the background predictions and to improve
their precision~\cite{Bern:2011pa,Ask:2011xf,Malik:2013kba}.

In this paper we focus on the theoretical modeling of the SM $V+$\,jet
backgrounds to inclusive production of large MET recoiling against one or
more hadronic jets.  These final states address a broad set of BSM models,
where the production of an otherwise invisible final state is revealed by
the emission of one or more high-$p_\rT$ jets from initial state
radiation, where $p_\rT$ is the momentum in the transverse plane.\footnote{For a recent comprehensive review of DM models leading
to this class of signatures, see e.g.~\cite{Abercrombie:2015wmb}.} Recent
publications by ATLAS~\cite{Aaboud:2016tnv} and CMS~\cite{Khachatryan:2016mdm,Sirunyan:2017hci},
relative to LHC data collected at $\sqrt{s}=13$\,TeV, document in detail the
current experimental approaches to the background evaluation.  The leading
background is $Z(\nu\bar\nu)+$\,jet production, followed by
$W(\ell\nu)+$\,jet (in particular for $\ell=\tau$ or when the lepton is
outside of the detector).\footnote{Other backgrounds (such as QCD multijets,
$t\bar{t}$ or pairs of gauge bosons) are suppressed, and their contribution
to the overall uncertainty is well below the percent level.} The
experimental constraints on $Z(\nu\bar\nu)+$\,jet production at large
MET can be obtained from accurate measurements of $V+$\,jet production
processes with visible vector-boson signatures.  It is quite obvious, for example, that
the measurement of $Z(\ell^+\ell^-)+$\,jets with $\ell=e,\mu$ is the most
direct and reliable proxy for $Z(\nu\bar\nu)+$\,jets.  This control sample,
however, is statistics limited, due to the smaller branching ratio of $Z$
bosons to charged leptons relative to neutrinos.  To extrapolate the shape
of the $Z$ spectrum to the largest $p_\rT$ values, therefore, requires a
theoretical prediction.  The larger statistics of $W(\ell\nu)+$\,jets and
$\gamma+$\,jets events makes it possible to directly access the relevant
$p_\rT$ range, but the relation between their spectra and the $Z$ spectrum
needs, once again, theoretical guidance.

\begin{figure}[t]   
\centering
  \includegraphics[width=.48\textwidth]{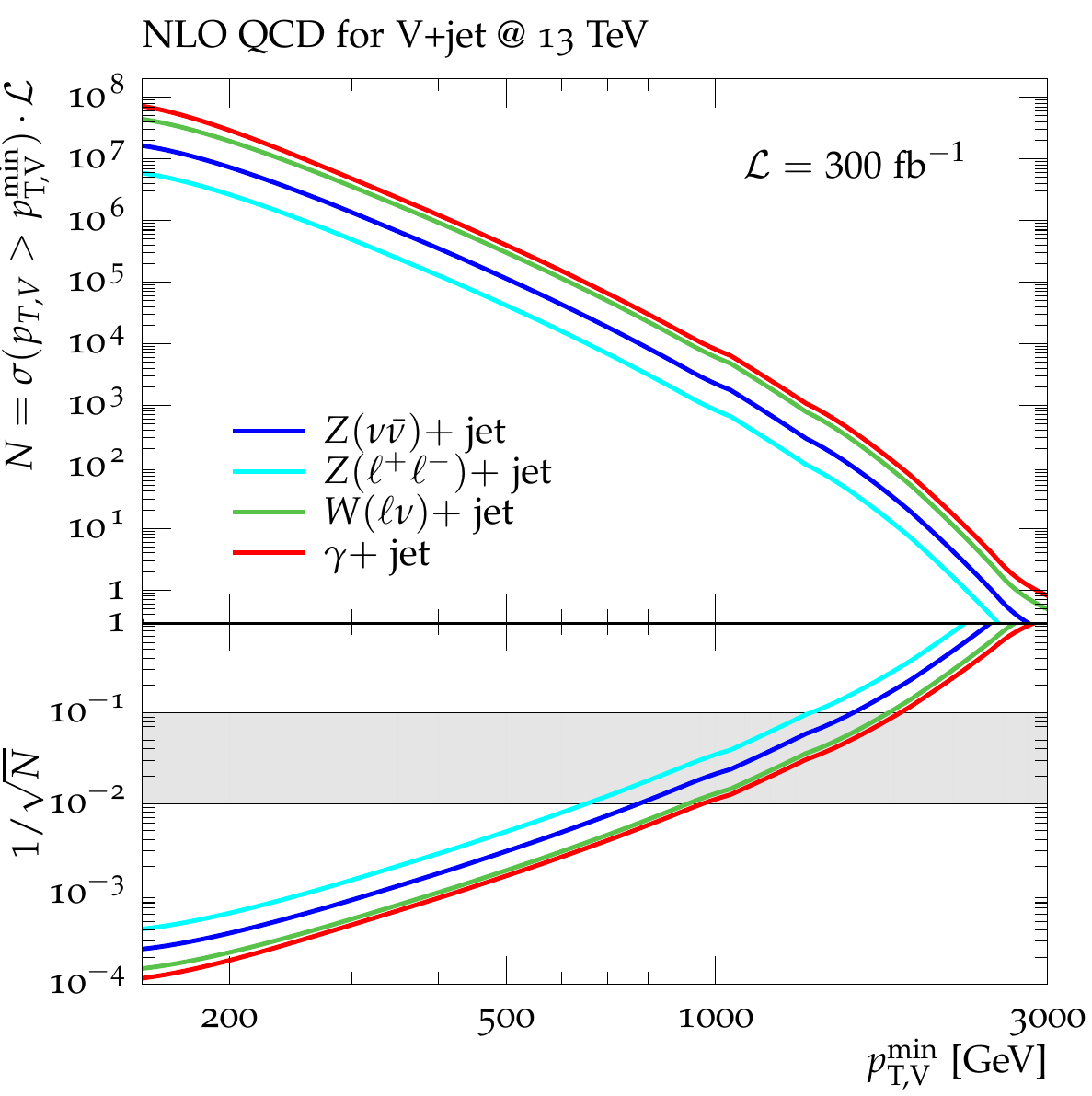}
\caption{Production rates for $V+$\,jet(s), for various decay channels, as a
function of the minimum $p_\rT$ of the vector boson. Decays into 
$\ell^\pm=e^\pm,\mu^\pm$ and $\nu_e,\nu_\nu,\nu_\tau$ are included.
The number of events,
$N$, is normalized to 300\,fb$^{-1}$ of LHC data at $\sqrt{s}=13$\,TeV, and
includes the basic selection cuts listed in the main body of the paper.  The log 
lower panel shows the statistical uncertainties, calculated as $1/\sqrt{N}$. 
The gray band in the lower panel indicates the regime of 1--10\% statistical
uncertainty.
}
\label{fig:stats}
\end{figure}

To put things into a concrete perspective,  \reffi{fig:stats} shows the
expected
event rates, and the relative statistical uncertainty, for 300\,fb$^{-1}$ of
integrated luminosity at 13\,TeV. The extrapolation to the $\ord$(100\,fb$^{-1}$)
and $\ord$(3000\,fb$^{-1}$) expected from the full run 2 and at the end of the full
LHC programme, respectively, is straightforward.  The
$Z(\ell^+\ell^-)+$jets data allow 
for
a direct estimate of the
$Z(\nu\bar\nu)+$jets rate with a statistical precision below 1\% for $p_\rT$
up to 
about
$600$\,GeV.  
Using the $W(\ell\nu)+$jets or $\gamma+$jets data
could in principle extend this range up to about $900$\,GeV.  Beyond this
value, the statistical precision of the $W(\ell\nu)+$jets and $\gamma+$jets
events remains a factor of two better than that of the $Z(\nu\bar\nu)+$jets
signal.  In order to ensure that the theoretical systematics in the
extrapolation from the $W+$jets and $\gamma+$jets rates to the $Z+$jets
rates remains negligible with respect to the statistical uncertainty, the
former should be kept at the level of a few percent up to $p_\rT\sim 2$\,TeV,
and around 10\% up to $p_\rT\sim 2.5-3$\,TeV, which is the ultimate
kinematic reach for the $Z(\nu\bar\nu)+$jets signal at the end of LHC data
taking.

%%%%%%%%%%%%%%%%%%%%
%%%%%%%%%%%%%%%%%%%%

The main result of this work is to prove that, thanks to the recent
theoretical advances, these goals can be met.  This proof requires the
analysis of a series of possible effects.  On the one hand, the theoretical
extrapolation to larger $p_\rT$ of the very precise $Z(\ell^+\ell^-)$+jets
data requires firm control over the shape of the distribution.  Several
effects, from the choice of 
parton distribution functions
(PDFs) to the choices made
for the renormalisation and factorisation scales used in the calculations,
can influence the extrapolation.  On the other hand, the level of
correlation between the $W$, $\gamma$ and $Z$ spectra must be kept under
control.  At large $p_\rT$, in particular, large and process-dependent
corrections arise due to the growth of the electroweak 
(EW)
corrections, and
these may spoil the correlation induced by pure QCD effects.  For our
analysis we shall use the most up-to-date theoretical predictions available
today for the description of vector boson production at large $p_\rT$.  On the
QCD side, we rely on the next-to-next-to-leading order (NNLO) calculations,
which appeared recently for
$Z+$jet~\cite{Ridder:2015dxa,Ridder:2016nkl,Gehrmann-DeRidder:2016jns,Boughezal:2015ded,Boughezal:2016isb},
$W+$jet~\cite{Boughezal:2015dva,Boughezal:2016dtm} and $\gamma+$jet~\cite{Campbell:2016lzl,Campbell:2017dqk}
production.  On the EW side, we apply full
NLO calculations for $Z+$jet~\cite{Denner:2011vu,Denner:2012ts,Kallweit:2015dum},
$W+$jet~\cite{Denner:2009gj,Kallweit:2015dum}
and $\gamma+$jet~\cite{Badger:2016bpw}
production with off-shell decays of the $Z$ and $W$ bosons.
Given the strong enhancement of EW Sudakov effects in the TeV region,
we also include 2-loop logarithmic terms at next-to-leading logarithmic (NLL)
accuracy for all $V+$\,jet processes~\cite{Kuhn:2005gv,Kuhn:2005az,Kuhn:2007qc,Kuhn:2007cv}. 
An extensive assessment and discussion of the estimates of missing
higher-order terms, and of the relative systematics, is given in the main
body of this paper. 
In particular, in order to address non-trivial issues that arise in the context of 
dark matter searches, we introduce a global framework for the estimate of
theoretical uncertainties in all $V+$\,jet processes, taking into 
account correlation effects across different processes and $p_\rT$ regions.  Also the
uncertainties associated with the combination of QCD and EW corrections 
are discussed in detail.

From the experimental perspective, the determination of the background
composition in signal and control regions, and the modeling of other key
aspects of experimental analyses (e.g.~lepton identification and
reconstruction, missing energy, etc.) require a theoretical description of
the various $V+$\,jets processes at the particle level.  Typically, this is
provided by Monte Carlo (MC) samples based on multi-jet merging at LO or NLO
QCD, 
%
%and the reweighting of MC events is a standard approach that allows one
%to implement various possible improvements, such as higher-order QCD or EW
%corrections.  
%
and improvements based on higher-order theoretical calculations
can be implemented through reweighting of MC events.
For the fit of MC predictions to data, ATLAS and CMS analyses
rely on the profile likelihood approach, where experimental and theoretical
uncertainties are described in terms of nuisance parameters with Gaussian
distributions.  In this context, the correlations of theoretical
uncertainties across $p_\rT$ bins (shape uncertainties) and across different
$V+$\,jets processes play a key role for searches at large MET.  

For the implementation of higher-order QCD and EW
corrections and for the estimate of theoretical uncertainties
in the experimental analysis framework,
we propose a procedure based on a one-dimensional
reweighting of MC samples.
The proposed
framework should enable the experiments to carry out their profile
likelihood approach, quantifying the impact of the theoretical systematics
in their analyses, and validating directly with data the reliability and
robustness of the theoretical inputs.  In this respect, we would like to
stress that, independently of the application to BSM searches, the results
in this paper provide a framework for incisive validations of the
theoretical calculations.  
Furthermore, these results might allow for further
constraints on PDFs~\cite{Malik:2013kba,Boughezal:2017nla}.

If the experimental analyses of the MET+jets channel should confirm the
usefulness of the approach we propose, the same framework could be adapted
to more complex or exclusive final states, in which for example MET is
accompanied by a large number of (hard) jets or by specific objects
(photons, heavy quarks, Higgs, etc).  These extensions are left for future
studies.

The structure of this paper is as follows: In \refse{se:reweighting}
we introduce the reweighting technique, to incorporate in a MC analysis the
effect of higher-order corrections and of their systematic uncertainties
including correlations.  
\refse{se:setup}
describes details of the setup for our numerical calculations, the employed tools and methods,
as well as the detailed 
definition of physics objects and observables to be used 
in the context of MC reweighting.
In \refse{se:ho} we discuss higher-order QCD
and EW corrections, including the contribution of photon-\linebreak initiated processes
and real vector boson emission.  We present here our approach to the
estimate of the various systematics, covering QCD scale, shape and
process-dependent uncertainties, as well as uncertainties arising from
higher-order EW and mixed QCD--EW corrections.  
\refse{se:conclusions} contains our summary and conclusions.
As detailed in \refapp{se:numpredictions},
results for all $V$+jets processes
are available in form of one-dimensional histograms in 
the vector-boson 
$p_\rT$ covering
central predictions and all mentioned uncertainties. 
Technical plots on the individual sources of QCD and EW uncertainties 
are documented in \refapp{app:unc}.

\section{Reweighting of Monte Carlo samples}
\label{se:reweighting}

The reweighting of MC samples is an approximate, but straightforward  
and easy to implement method of combining (N)LO MC simulations with 
(N)NLO QCD+NLO EW perturbative calculations and to account for the respective uncertainties in a systematic way.
The following formula describes the one-dimensional
reweighting of MC samples for $V+$\,jet production
($V=\gamma,Z,W^\pm$) in a generic variable $x$,
\begin{align}
\label{eq:rew}
\parx\pary\,&\siv(\vec\eps_\MC,\vec\eps_\TH)= \\ & \parx\pary \siv_{\MC}(\vec\eps_\MC)  \nonumber
\left[\frac{\parx\siv_{\TH}(\vec\eps_\TH)}{\parx\siv_{\MC}(\vec\eps_\MC)}\right].
\end{align}
In the case at hand, i.e.~$V+$\,jet production, the one-dimensional parameter $x$ should be understood as the vector-boson transverse momentum, $x=p^{(V)}_\rT$, 
while $\vec y$ generically denotes the remaining variables of the fully differential kinematic dependence of the accompanying QCD and QED activity, including both extra jet and photon radiation, as well as leptons and neutrinos from hadron decays. It is implicitly understood that $\parx\pary \sigma$ depends on $x$ and $\vec y$, while in
$\parx \sigma$ the variables $\vec y$ are integrated out.

The labels MC and TH in \refeq{eq:rew}
refer to Monte Carlo and higher-order theoretical predictions, respectively,
and the related uncertainties are parametrised through nuisance parameters $\vec\eps_\TH, \vec\eps_\MC$.
Our recommendations for theory uncertainties in \refse{se:ho}
are formulated in terms of intervals 
for the related nuisance parameters, 
\begin{equation}
-1<\eps_{\TH,k}<1,
\end{equation}
which pragmatically should be understood as the $1\sigma$ range of Gaussian uncertainties.

Monte Carlo uncertainties, described by $\vec\eps_\MC$, must be correlated in the numerator and denominator on the r.h.s.\ of \refeq{eq:rew}, while they  
can be kept uncorrelated across different processes (apart from $Z(\nu\bar\nu)+\,\jet$ and $Z(\ell^+\ell^-)+\,\jet$).

We note that, as opposed to an approach based only on ratios of $p_\rT$ distributions, where theory is used for extrapolations across different processes
at fixed $p_\rT$, 
MC reweighting is more powerful as it supports all possible
extrapolations across different processes and $p_\rT$ regions. In
particular, it makes it possible to exploit $V+$\,jet precision measurements at moderate $p_\rT$ in order to constrain $Z(\nu\bar\nu)+\jet$ production in the TeV region. 

A further advantage of the reweighting approach~\refpar{eq:rew} lies in the fact that the 
three terms on the r.h.s.\ of \refeq{eq:rew}
do not need to be computed with the same numerical  setup (parameters, cuts, observables, etc.).
More precisely, only the definition of the variable $x$ and the binning of its distribution need to be the same in all three terms. Scale choices, 
QCD and EW input parameters and PDFs should be the same only in the numerator and denominator of
\beq 
\label{eq:rmc}
R_\MC(x,\vec y)=\frac{\parx\pary\siv_{\MC}}{\parx\siv_{\MC}},
\eeq  
but can be chosen in a different way in $\parx\siv_\TH$, provided that QCD and EW corrections themselves are computed using the same settings.
Vice versa,  
possible cuts must be identical only in the numerator and denominator of 
\beq 
\label{eq:rth}
R_{\TH/\MC}(x)=\frac{\parx\siv_{\TH}}{\parx\siv_{\MC}},
\eeq  
while particle-level MC predictions, $\parx\pary\siv_\MC$, can be subject to more exclusive or inclusive cuts in the experimental analysis.

For an optimal combination of higher-order calculations and MC predictions, two conditions should be fulfilled. On the one hand,
theory calculations should describe the distribution in the reweighting variable with higher (or at least equal) precision as compared to the MC sample,
\beq
\label{eq:rewc1}
\Delta\left[\parx\siv_{\TH}
\right]\le \Delta\left[\parx\siv_{\MC}\right].
\eeq
On the other hand, the MC sample should be more
accurate than TH calculations in describing the correlation between $x$ and all other variables $\vec y$,  
\beq
\label{eq:rewc2}
\Delta\left[\frac{\parx\pary\siv_{\MC}}{\parx\siv_{\MC}}\right]
\le
\Delta\left[\frac{\parx\pary\siv_{\TH}}{\parx\siv_{\TH}}\right].
\eeq
More precisely, condition \refpar{eq:rewc2} needs to be fulfilled only for those aspects of $V+\,$jet events that are relevant for the actual experimental analysis.

As concerns the first condition, 
we note that, depending on the choice of 
the observable $x$,
using state-of-the-art theory calculations that involve higher-order QCD and EW corrections may not guarantee 
that \refeq{eq:rewc1} is fulfilled. 
In fact, there are a 
number of aspects, i.e.\ resolved multi-jet emissions, the resummation of soft logarithms in the region of small vector-boson $p_\rT$, soft QCD radiation of non-perturbative origin, multiple photon radiation, 
or neutrinos and charged leptons resulting from hadron decays,
for which fixed-order perturbative calculations of $pp\to V+$\,jet are less accurate than MC simulations.

Thus, the reweighting variable $x$ should be defined such as to have minimal sensitivity to the above-\linebreak mentioned aspects. In this respect, due to its reduced sensitivity to multiple jet emissions, the vector-boson $p_\rT$ is a natural choice. 
However, in order to fulfil 
\refeq{eq:rewc1}, 
the  region $p_\rT^{(V)}\ll M_V$
should be excluded from the reweighting procedure, 
unless QCD Sudakov logarithms are resummed to all orders
in the theoretical calculations.
Moreover, in order to simultaneously fulfil conditions~\refpar{eq:rewc1} and~\refpar{eq:rewc2}, any aspect of 
the reconstructed vector-boson $p_\rT$ that is better described at MC level
should be excluded from the definition of $x$ and included in $\vec y$. This applies, as discussed in \refse{se:setup}, to multiple photon emissions off leptons, and to possible isolation prescriptions for the soft QCD radiation that surrounds leptons or photons.
In general, purely non-perturbative aspects of MC simulations, i.e.~MPI, UE, hadronisation and hadron decays, should be systematically excluded from the definition of the reweighting variable  $x$. Thus, impact and uncertainties related to this 
non-perturbative modelling 
will remain as in the original MC samples.

It should be stressed that the above considerations are meant for dark-matter searches based on the {\it inclusive} MET distribution, while more exclusive searches that exploit additional  information on hard jets
may involve additional subtleties.
In particular,
for analyses that are sensitive to multi-jet emissions, using the
inclusive vector-boson $p_\rT$ as the reweighting variable would still fulfil 
\refeq{eq:rewc1}, but the lack of QCD and EW corrections to $V+2$\,jets production in MC simulations could lead to a violation of 
\refeq{eq:rewc2}.
In analyses that are sensitive to the tails of inclusive jet-$p_\rT$ and $H_\rT$ distributions this issue is very serious,
and QCD+EW corrections should be directly implemented at MC level using multi-jet merging~\cite{Kallweit:2015dum}.

In general, as a sanity check of the reweighting procedure, we
recommend verifying that, for reasonable choices of input parameters and QCD scales,
(N)NLO QCD calculations and (N)LO merged MC predictions for vector-boson
$p_\rT$ distributions are in reasonably good agreement within the respective
uncertainties. Otherwise, in case of significant MC mismodelling of the 
$\parx\siv$ distribution, one should check the reliability of the 
MC in extrapolating TH predictions from the reweighting distribution 
to other relevant observables.

In general, one could check whether the one-\linebreak dimensional reweighting via the variable~$x$
in \refeq{eq:rew} can in fact reproduce the dependence of the corrections
in other kinematic variables that are relevant for the experimental analysis.
To this end, distributions of $\siv$ w.r.t.\ another kinematic 
variable~$x'$ should be calculated upon integrating \refeq{eq:rew}. Switching on and off
the corrections on the r.h.s.\ of \refeq{eq:rew} in $\siv_{\TH}$ and taking the ratio of the
obtained differential cross sections $\siv$,
produces the relative correction to the $x'$~distribution that could be 
compared to the corresponding result directly calculated from $\siv_{\TH}$.\footnote{This 
procedure should be restricted to variables $x'$ that can be described with good accuracy 
both in perturbative calculations and in the MC simulations.}

Finally, it is crucial to check that state-of-the-art predictions for
absolute $\rd\sigma/\rd p_{\rT}$ distributions agree with data for the
various visible final states.

\section{Setup for numerical predictions}
\label{se:setup}
In this section we specify the physics objects (\refse{se:objects}), acceptance cuts and observables (\refse{se:cutsnadobs}), input parameters (\refse{se:inputs}) and tools (\refse{se:tools}) used in the theoretical calculations for $pp\to W^\pm/Z/\gamma+$\,jet.

The definitions of physics objects, cuts and \linebreak observables---which specify the setup
for the reweighting procedure discussed in
\refse{se:reweighting}---should be adopted both for
theoretical calculations and for their Monte Carlo counterpart
in the reweighting factor~\refpar{eq:rmc}.
The details of the reweighting setup are designed 
such as to avoid any possible deficit in the
perturbative predictions (e.g.~due to lack of resummation at
small $p_\rT$) and any bias due to non-perturbative aspects of Monte Carlo simulations
(e.g.~leptons and missing energy from hadron decays).
Let us also recall 
that this setup is completely independent of the physics objects, cuts and observables 
employed in the experimental analyses.

As concerns input parameters and PDFs, the recommendation of 
\refse{se:inputs} should be applied to all QCD and EW higher-order calculations.
In particular, it is mandatory to compute (N)NLO QCD and EW corrections in the
same EW input scheme, otherwise NLO EW accuracy would be spoiled.
Instead, Monte Carlo simulations and the corresponding $\parx\siv_{\MC}$ contributions 
to the reweighting factor~\refpar{eq:rmc} do not need to 
be based on the same input parameters and PDFs used for theory predictions.

We recommend handling $W/Z+\jet$ production and decay on the Monte Carlo side as the full 
processes $pp\to \ell\ell/\ell\nu/\nu\nu+\jet$,
i.e.\ with a consistent treatment of off-shell effects, as is done on the theory side.

\subsection{Definition of physics objects}
\label{se:objects}
In the following we define the various physics objects relevant for 
higher-order perturbative calculations and for the reweighting in the Monte Carlo 
counterparts in \refeq{eq:rmc}.

\subsubsection*{Neutrinos}
\label{se:neutrinos}
In parton-level calculations of $pp\to \ell\ell/\ell\nu/\nu\nu+\jet$,
neutrinos originate only from vector-boson decays, while
in Monte Carlo samples they can arise also from 
hadron decays.  In order to avoid any bias in the reweighting procedure, 
only neutrinos arising
from $Z$ and $W$ decays at Monte Carlo truth level should be considered.

\subsubsection*{Charged leptons}
\label{se:dressing}
Distributions in the lepton $p_\rT$ and other leptonic observables are known
to be highly sensitive to QED radiative corrections, and the differences in
the treatment of QED radiation on Monte Carlo and theory side can lead to a bias in
the reweighting procedure.  This should be avoided by using dressed
leptons, i.e.~recombining all leptons with nearly collinear photons that lie
within a cone
\def\rrec{R_{\mathrm{rec}}}
\beq 
\Delta R_{\ell\gamma}=\sqrt{\Delta \phi^2_{\ell\gamma}
+\Delta \eta^2_{\ell\gamma}}<\rrec.
\eeq 
For the radius of the recombination cone we employ the standard value $\rrec=0.1$,
which allows one to capture the bulk of the collinear final-state radiation, 
while keeping contamination from large-angle photon radiation at a negligible level.
All lepton observables as well as the kinematics of reconstructed $W$ and $Z$ bosons are defined in terms of dressed leptons,
and, in accordance with standard experimental practice, both muons and electrons should be
 dressed.  In this way differences between electrons and muons, $\ell=e,\mu$,
become negligible, and the reweighting function needs to be computed only
once for a generic lepton flavour $\ell$.

Similarly as for neutrinos, only charged leptons that arise from $Z$ and $W$
decays at Monte Carlo truth level should be considered. 
Concerning QCD radiation in the vicinity of leptons, 
no lepton isolation requirement should be imposed
in the context of the reweighting procedure. Instead,
in the experimental analysis
lepton isolation cuts can be applied in the usual manner.

\subsubsection*{$Z$ and $W$ bosons}
\label{se:ptVdef}
The off-shell four-momenta of $W$ and $Z$ bosons are defined as
\beqar
p^\mu_{W^+}&=&p^\mu_{\ell^+}+p^\mu_{\nu_\ell},\qquad
p^\mu_{W^-}=p^\mu_{\ell^-}+p^\mu_{\bar\nu_\ell},\\
p^\mu_{Z}&=&p^\mu_{\ell^+}+p^\mu_{\ell^-},\qquad
%\quad\mbox{or}\quad
p^\mu_{Z}=p^\mu_{\nu_\ell}+p^\mu_{\bar\nu_\ell},
\eeqar
where the leptons and neutrinos that result from $Z$ and $W$ decays are defined as discussed 
above.
%in \refses{se:neutrinos}{se:dressing}.

\subsubsection*{Photons}
\label{se:photons}
\def\rdyn{R_{\mathrm{dyn}}}
\def\dyn{\mathrm{dyn}}
\def\fix{\mathrm{fix}}

At higher orders in QCD, photon production involves final-state 
$q\to q\gamma$ splittings that lead to collinear singularities when 
QCD radiation is emitted in the direction of the photon momentum.
Since such singularities are of QED type, they are not cancelled
by corresponding virtual QCD singularities.
Thus, in order to obtain finite predictions in perturbation theory,
the definition of the $pp\to \gamma+$\,jet cross section requires a
photon-isolation prescription that vetoes collinear $q\to q\gamma$ 
radiation while preserving the cancellation of QCD infrared singularities.

To this end, in this study we adopt Frixione's isolation 
prescription~\cite{Frixione:1998jh}, which 
limits the hadronic transverse energy within a smooth cone around the photon
by requiring 
\begin{align}
\label{eq:standard_isolation}
\sum\limits_{i={\rm partons/hadrons}} & p_{\rm T, i}\,\Theta(R-\Delta  R_{i\gamma})\\  &\leq 
\epsilon_0\, p_{\rm T,\gamma} \left(\frac{1-\cos R}{1-\cos R_0}\right)^n \nonumber
\quad\forall\; R\le R_0,
\end{align}
where the sum runs over all quarks/gluons and hadrons at parton level and Monte Carlo level, respectively,
while $p_{\rm T,i}$ and $p_{\rm T,\gamma}$ denote the transverse momenta of partons/hadrons and photons.
The $p_\rT$-fraction $\eps_0$, the cone size $R_0$, and the exponent $n$ are free parameters that 
allow one to control the amount of allowed QCD radiation in the vicinity of the photon.

The photon-isolation prescription is applicable to QCD as well as to EW higher-order corrections.
At NLO EW, $\gamma+$\,jet production involves bremsstrahlung contributions with 
two final-state photons.
In this case, at least one isolated photon is required. The other photon might become soft, guaranteeing  
the  cancellation of related soft and collinear singularities in the virtual EW corrections.
In case of two isolated photons in the final state, the hardest photon is considered.
In particular, an explicit photon isolation prescription is mandatory at NLO EW in order to prevent uncancelled singularities 
from $q\to q\gamma$ splittings in the $\ord(\alpha^2\alphaS)$ mixed EW--QCD contributions from 
$qq\to qq\gamma$ and crossing-related channels.

As a consequence of $q\to q\gamma$ collinear singularities and the need to
apply a photon isolation prescription, QCD corrections to
$pp\to\gamma+$\,jet behave differently as compared to $Z/W+$\,jet
production. Such differences can be important even 
at the TeV scale,  where one might naively expect that 
massive and massless vector bosons behave in a universal way 
from the viewpoint of QCD dynamics. Instead,
the presence of collinear $q\to q V$ singularities
at (N)NLO QCD implies a logarithmic sensitivity to the vector-boson masses,
which results, respectively, in $\ln(R_0)$ and $\ln(p_{\rT,V}/M_V)$ terms for
the case of massless and massive vector bosons
at $p_{\rT,V}\gg M_{W,Z}$.

A quantitative understanding of these differences and their 
implications on the correlation of QCD uncertainties between $\gamma+$\,jet
and $Z+$\,jet production is crucial for the extrapolation of $\gamma+$\,jet
measurements to $Z+$\,jet dark-matter backgrounds.  
To this end, as discussed in \refse{se:ho}, 
%
%in order to quantify the correlation of QCD uncertainties
%across different $V+$\,jet processes, 
%
we propose a systematic approach based on the idea that,
at large $p_{\rT,V}$, the
$pp\to \gamma+$\,jet process can be split into a dominant part
with universal QCD dynamics (in the 
sense that QCD effects in $\gamma+$\,jet and $Z/W+$\,jet production are strongly correlated) 
and a remnant contribution that 
has to be handled as uncorrelated in the treatment of QCD
uncertainties.
To achieve this, we introduce 
a modified
photon isolation prescription, which is designed such as to render the QCD
dynamics of $\gamma+$\,jet and $Z/W$+\,jet production as similar as possible
at high $p_\rT$.
To this end we define a dynamic cone radius
\beqar
\label{eq:dynrad}
\rdyn(p_{\rT,\gamma},\eps_0)=\frac{M_Z}{p_{\rT,\gamma}\sqrt{\eps_0}},
\eeqar
which is chosen in such a way that the invariant mass of a photon-jet pair with 
$R_{\gamma j}=\rdyn$ and $p_{\rT,j}=\eps_0\, p_{\rT,\gamma}$ corresponds to the $Z$-boson mass, \ie
\beqar
\label{eq:dynrad2}
M^2_{\gamma j}\simeq p_{\rT,\gamma}\,p_{\rT,j} R^2_{\gamma j}=\eps_0\, p^2_{\rT,\gamma} \rdyn^2 = M_Z^2,
\eeqar
where the first identity is valid in the small-$R$ approximation.
In this way, using a smooth isolation with $R_0=\rdyn(p_{\rT,\gamma},\eps_0)$
mimics the role of the $Z$- and $W$-boson masses as regulators of collinear singularities in 
$Z/W$+jet production at high $p_\rT$, while using a fixed cone radius $R_0$ would 
correspond to an effective $M_{\gamma j}$ cut well beyond $M_{Z,W}$, resulting 
in a more pronounced suppression of QCD radiation in 
$\gamma+$\,jet production as compared to $Z/W+$\,jet.

Specifically, as default photon selection for the theoretical 
predictions\footnote{The same isolation prescription used for theory predictions should be applied also
to their MC counterparts $\rd\sigma_{\rm MC}/$d$ x$ in the context of the reweighting procedure.}
in this study
we use the dynamic cone isolation defined 
through \refeq{eq:standard_isolation} and \refeq{eq:dynrad},
with parameters
\beqar
\label{eq:dynisolation}
\eps_{0,\dyn}&=&0.1,\qquad \nonumber
n_\dyn=1, \\
R_{0,\dyn}&=& \min\left\{1.0, \rdyn(p_{\rT,\gamma},\eps_{\dyn,0})\right\}.
\eeqar
Note that, in order to prevent that the veto against collinear 
QCD radiation is applied to an excessively large region of phase space,
the dynamic cone radius in \refeq{eq:dynisolation} 
is limited to $\rdyn\le 1.0$. 
As a result of this upper bound,
for $p_{\rT,\gamma}< M_Z\eps^{-1/2}_{0,\dyn}\simeq 290$\,GeV
the cone radius is kept fixed, and the impact of collinear QCD radiation 
starts to be significantly enhanced as compared 
to the case of $Z/W+$\,jet production. Vice versa, for  $p_{\rT,\gamma}> M_Z\eps^{-1/2}_{0,\dyn}$,  
thanks to the dynamic isolation cone~\refpar{eq:dynisolation},
QCD effects in $\gamma+$\,jet and $Z/W+$\,jet production become 
closely related, and the degree of correlation between QCD uncertainties 
across all $V+$\,jet processes can be described 
with the prescription of~\refeqs{eq:proccorr1}{eq:dKqcd3}.

For a realistic assessment of theoretical uncertainties, one should also 
consider the fact that photon isolation prescriptions used in experimental analyses 
differ in a significant way from the dynamic prescription of~\refeq{eq:dynisolation}.
To this end, we recommend to repeat the reweighting procedure using 
theory predictions for $\gamma+$\,jet based on a standard Frixione 
isolation~\refpar{eq:standard_isolation} 
with fixed cone radius and parameters that 
mimic typical experimental selections at particle level~\cite{Khachatryan:2015ira},
\beqar
\label{eq:standard_isolation_parameters}
\eps_{0,\fix}=0.025,\qquad 
n_\fix=2,\qquad 
R_{0,\fix}= 0.4.
\eeqar
The difference between $\gamma+$\,jet MC samples reweighted in the dynamic- and fixed-cone 
setup should be taken as an additional uncertainty for $pp\to\gamma+$\,jet.
As ingredients for this uncertainty estimate
 we provide higher-order QCD predictions (without uncertainties) with fixed-cone
isolation~\refpar{eq:standard_isolation_parameters} 
besides the full set of $pp\to
\gamma+$\,jet predictions and uncertainties with dynamic photon isolation
(see~\refapp{se:numpredictions}).
In the EW corrections, differences between the two photon isolation prescriptions
 are well below the percent level.
Thus predictions for $\gamma+$\,jet at (n)NLO EW are provided only with 
the dynamic cone prescription of~\refeq{eq:dynisolation}.

%%%%%%%%%%%%%%%%%%%%
\begin{figure*}[t]   
\centering
\includegraphics[width=\ratiotextwidth\textwidth]{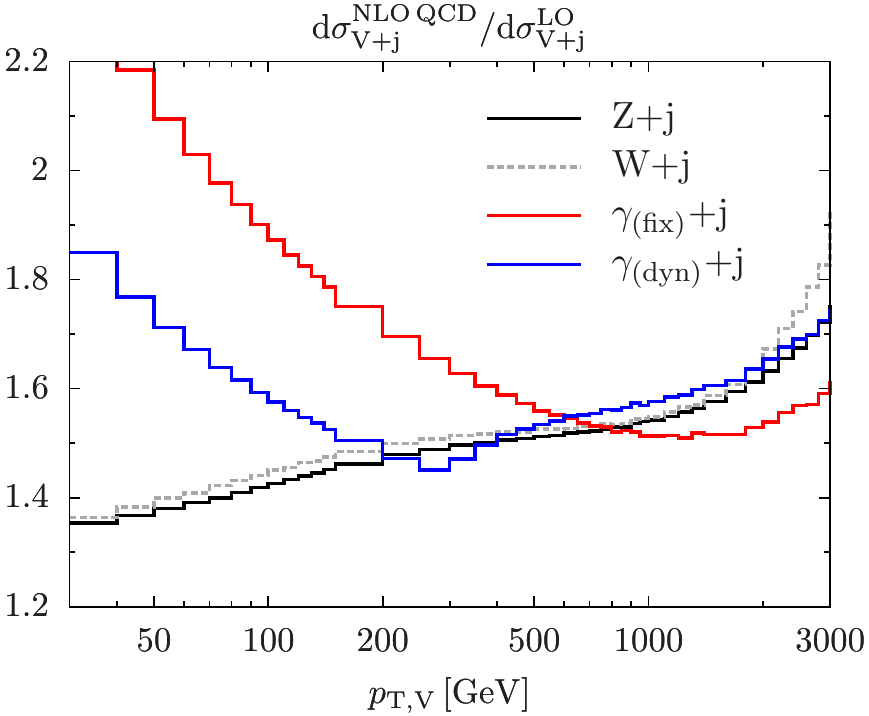}
\includegraphics[width=\ratiotextwidth\textwidth]{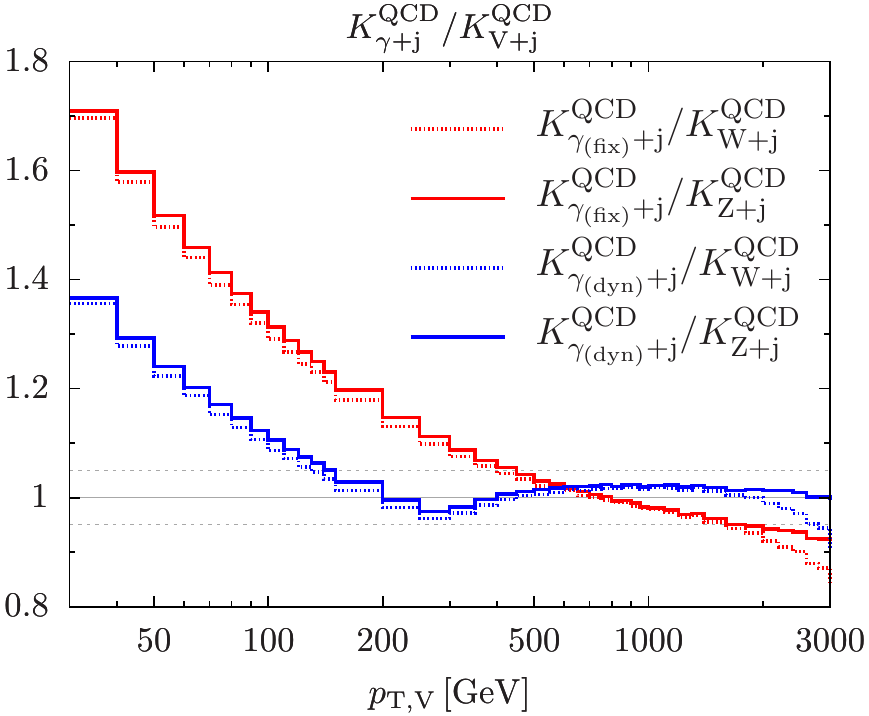}
\caption{Comparison of NLO QCD $K$-factors (left) for $W+$\,jet,  $Z+$\,jet,  and 
$\gamma+$\,jet production with dynamic photon isolation~\refpar{eq:dynisolation}
and standard fixed-cone isolation~\refpar{eq:standard_isolation_parameters}. On the right corresponding ratios of $K$-factors are shown, and the dotted lines indicate $K$-factor variations of $\pm 0.05$.
}
\label{fig:dynvsfix_isolations}
\end{figure*}
%%%%%%%%%%%%%%%%%%%%

%%%%%%%%%%%%%%%%%%%%
\begin{figure}[t]   
\centering
  \includegraphics[width=\ratiotextwidthbig\textwidth]{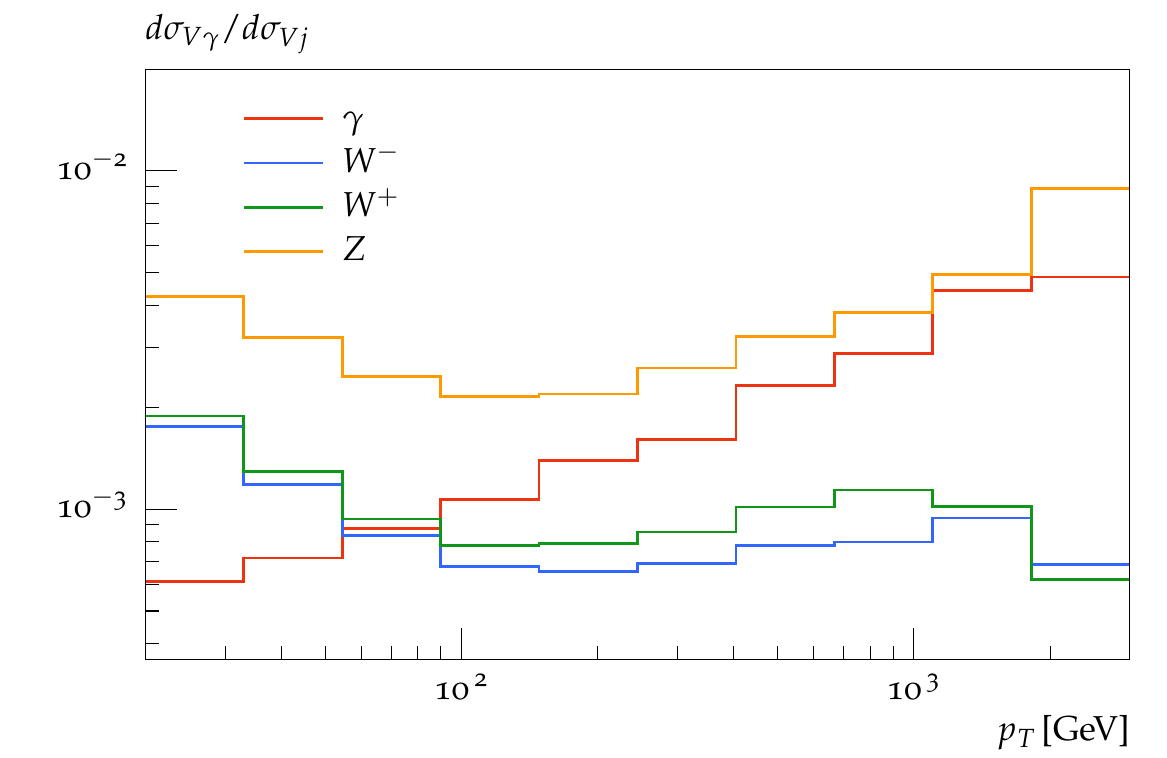}
\caption{
Ratios of distributions in the vector-boson transverse momenta
for $pp\to V\gamma$ versus $pp\to Vj$ at LO with $\mu_{\rR,\rF}=H_\rT/2$. 
The vector bosons $V=W^\pm,Z,\gamma$ are on shell and $\sqrt{s}=13\,\TeV$.
}
\label{fig:ratios_Vgamma}
\end{figure}
%%%%%%%%%%%%%%%%%%%%

In~\reffi{fig:dynvsfix_isolations} we present a comparison of the NLO QCD
$K$-factors for $W/Z+$\,jet and $\gamma+$\,jet production with dynamic and
fixed cone isolation. 
For $p_{\rT,\gamma}<290$\,GeV, where both isolation prescriptions
correspond to a fixed cone radius, the QCD corrections to $pp\to \gamma+$\,jet 
grow rapidly with decreasing $\pT$. At low $\pT$, due to the smaller cone size, 
fixed isolation ($R_0=0.4$) leads to more pronounced corrections as compared to 
dynamic isolation ($R_0=1.0$), but the slopes of the corresponding $\gamma+$\,jet $K$-factors 
are quite similar to each other and very different as compared to the ones for $pp\to W/Z+$\,jet.
In the case of fixed isolation, this difference persists also 
in the high-$\pT$ regime (apart form the accidental agreement of
$K$-factors at $p_{\rT,V}\approx 800$\,GeV). Instead,
in the case of dynamic photon isolation, 
at large $\pT$  the QCD corrections to 
$\gamma+$\,jet and 
$W/Z+$\,jet 
production turn out to be remarkably similar, both in shape and size.
As expected, the onset of this universal behaviour is located close to
$p_{\rT,\gamma}=290$\,GeV, where the isolation radius $R_{0,\dyn}$
starts varying with $\pT$ in a way that rejects QCD radiation with $M_{\gamma j}\lsim M_{W,Z}$.
The differences between $\gamma+$\,jet and 
$W/Z+$\,jet $K$-factors 
remain as small as a few percent
up to the TeV scale.

\subsubsection*{QCD partons and photons inside jets}
\label{se:photonsinjets}

In order to avoid any bias due to the different modelling of jets in MC
simulations and perturbative calculations, theory calculations and
reweighting should be performed at the level of inclusive vector-boson
$p_\rT$ distributions, without imposing any requirement on the recoiling
jet(s).  Predictions presented in this study are 
thus independent of specific jet definitions or jet cuts.

Concerning the composition of the recoil, we observe that, at NLO EW, $q\to
q\gamma$ splittings can transfer an arbitrary fraction of the recoiling momentum
from QCD partons to photons.
In particular, in $pp\to V\gamma j$ contributions of $\ord(\alpha^2\alphaS)$,
the photon can carry up to 100\% of the recoil momentum.
Such contributions involve soft QCD singularities that are 
cancelled by including  also virtual QCD corrections to $pp\to V\gamma$.
In order to minimise double counting
with diboson production,\footnote{Diboson backgrounds, including $pp\to V\gamma$,
can be included through separate Monte Carlo 
samples in the experimental analyses.}
 $V\gamma$ production at LO is not included in the EW corrections to 
$pp\to Vj$.
In practice, as demonstrated in \reffi{fig:ratios_Vgamma}, 
the relative weight of $pp\to V\gamma$ at $\ord(\alpha^2)$ versus $pp\to Vj$ at $\ord(\alpha\alphaS)$  
is well below the percent level. 
Thus the impact of 
$\ord(\alpha^2\alphaS)$ contributions from hard $V\gamma$ production, which are included in this 
study, 
should be completely negligible.

\subsection{Cuts and observables}
\label{se:cutsnadobs}

 \begin{table*}[t]
\begin{center}
\begin{tabular}{c|c|c|c}
process & extra cuts  & observable  & comments
\\\hline
$pp\to \ell^+\nu_\ell+$\,jet & none & $p_{\rT,\ell^+\nu_\ell}$ & $\ell= e$ or $\mu$
\\[2mm]\hline
$pp\to \ell^-\bar\nu_\ell+$\,jet & none & $p_{\rT,\ell^-\bar\nu_\ell}$ &   $\ell= e$ or $\mu$
\\[2mm]\hline
$pp\to \nu_\ell\bar\nu_\ell+$\,jet & none & $p_{\rT,\nu_\ell\bar\nu_\ell}$ &  $\ell=e+\mu+\tau$
\\[2mm]\hline
$pp\to \ell^+\ell^-+$\,jet &  $m_{\ell\ell} > 30\,\GeV$ & $p_{\rT,\ell^+\ell^-}$ &   $\ell= e$ or $\mu$
\\[2mm]\hline
$pp\to \gamma+$\,jet & dynamic isolation %\refpar{eq:dynrad}--\refpar{eq:dynisolation}
 & $p_{\rT,\gamma}$ &
\\
& \refpar{eq:dynrad}--\refpar{eq:dynisolation} & & 
\\[2mm]\hline
\end{tabular}
\caption{Extra selection cuts, in addition 
to \refeq{eq:ptcut},
and observables for the various $V+$\,jet processes.
Alternative predictions for $\gamma+$jet production are provided also 
for the case of  a standard Frixione isolation with parameters~\refpar{eq:standard_isolation_parameters}.}
\label{Tab:observables}
\end{center}
\end{table*}

Theoretical calculations and the reweighting of MC samples should be performed 
in a fully inclusive $V+$\,jet setup, 
imposing a single cut
\beqar 
\label{eq:ptcut}
p_{\rT,V}>30\,\GeV\quad\mbox{for}\quad V=W^\pm,Z,\gamma,
\eeqar 
with $p_{\rT,W^\pm}$ and $p_{\rT,Z}$ defined as in \refse{se:ptVdef}. The cut \refpar{eq:ptcut}
is crucial in order to avoid the region where perturbative predictions suffer form the lack of QCD resummation.\footnote{See e.g.~the comparison of NNLOPS against fixed-order predictions in Figure~3 of~\citere{Karlberg:2014qua}.} 

For leptons and MET we do not apply any $p_\rT$ or rapidity cuts. Moreover, we do not impose any restrictions on QCD radiation in the vicinity of leptons and MET. 
Also QCD radiation is handled in a fully inclusive way, i.e.~the presence of a recoiling jet is not explicitly required, and, 
as discussed in~\refse{se:photonsinjets},
at NLO EW the recoil can be entirely carried by a photon. 
Here we want to stress again that of course the particle-level analysis of the reweighted
Monte Carlo samples can (and will) involve a more exclusive event selection than used for the reweighting itself.

\begin{figure*}[t]   
\centering
  \includegraphics[width=\ratiotextwidth\textwidth]{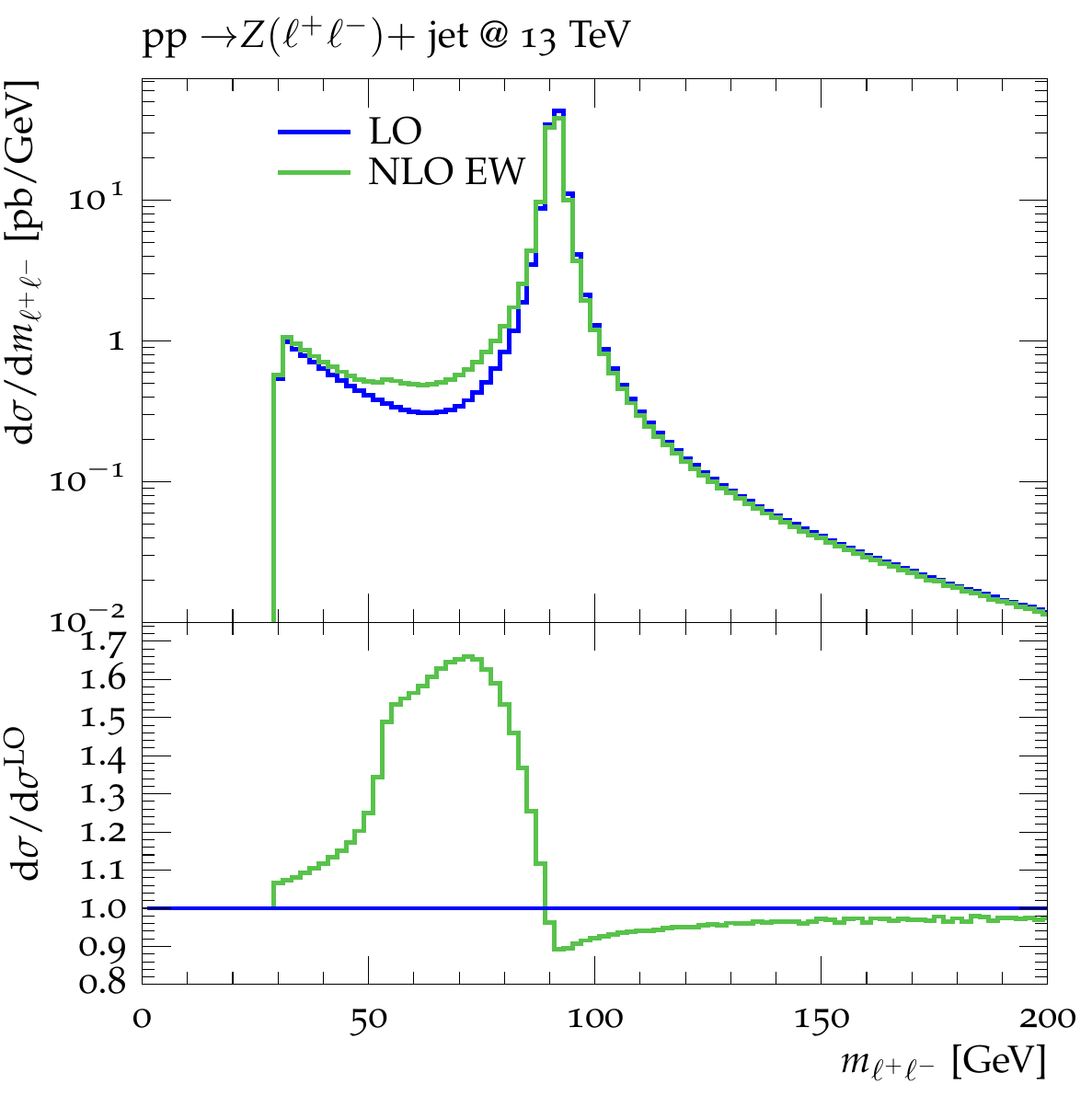}
\caption{
Dilepton invariant-mass distribution in $pp\to\ell^+\ell^-+$jet for $m_{\ell\ell}\in [30,200]\,\GeV$ comparing LO and NLO EW. Collinear lepton--photon pairs 
with $R_{\gamma \ell} < 0.1$ are 
recombined. 
}
\label{fig:mll}
\end{figure*}

The differential distributions to be 
used for the \linebreak reweighting 
of the various $pp\to V+$\,jet processes and process-specific selection cuts to be applied in addition to~\refeq{eq:ptcut} are summarised in \refta{Tab:observables}. 
In the case of $pp\to \nu\bar\nu+$\,jet
all three neutrino species are added, while for all other $Z$ and $W$ decays only a single lepton generation is considered.  
For 
$pp\to\ell^+\ell^-+$\,jet an extra invariant-mass cut is applied in order to 
avoid far off-shell contributions, especially from  $\gamma^*\to \ell^+\ell^-$ at low invariant mass.
The relatively low value of the lower cut, $m_{\ell\ell}> 30\,\GeV$,
is intended to minimise
cross section loss due to photon radiation that shifts events from the 
$Z$-peak region down to lower invariant mass (see~\reffi{fig:mll}). This choice guarantees 
a reduced sensitivity with respect to the modelling of QED radiation.

%%%%%%%%%%%%%%%%%%%%
%%%%%%%%%%%%%%%%%%%%

The following binning is adopted for distributions in the reconstructed vector-boson transverse momenta,
\beqar
\label{eq:binning}\nonumber
\frac{p_{\rT}}{\GeV} & \in  [30, 40, \dots, 140, 150, 200, 250 \dots, 950, 1000, \\ &1100, 1200, 1300, 1400, 1600\dots, 2800, 3000, 
6500]\,.\nonumber\\
\eeqar

\subsection{Input parameters, PDFs and QCD scales}
\label{se:inputs}

Input parameters and PDFs employed for theoretical predictions in this study are specified in the following. 
Let us recall that, as discussed in \refse{se:reweighting},
Monte Carlo samples used in the experimental analyses 
do not need to be generated with the same input parameters and PDFs used 
for higher-order theoretical predictions. 

In the calculation of  $pp\to\ell\ell/\ell\nu/\nu\nu/\gamma\,+$\,jet we use the 
gauge-boson masses~\cite{Agashe:2014kda}
\beqar\label{eq:massesew}
M_Z&=&91.1876~\GeV,\quad   
M_W=80.385~\GeV,
\eeqar
and the corresponding widths,
\beqar
\Gamma_Z&=&2.4955~\GeV,\quad
\Gamma_W=2.0897~\GeV.
\eeqar
The latter are obtained from state-of-the-art theoretical calculations.
For the top-quark~\cite{Agashe:2014kda} and Higgs-boson~\cite{Heinemeyer:2013tqa} masses and widths we use
\beqar\label{eq:massestop}
M_t&=&173.2~\GeV,\qquad
M_H=125~\GeV,
\eeqar
and\footnote{
Besides loop diagrams with top quarks and Higgs bosons, the NLO EW corrections to $pp\to W^\pm+$\,jet receive $\ord(\alpha^2\alphaS)$
bremsstrahlung contributions from
$qb\to q' W^\pm b$ channels that involve $s$-channel top-quark propagators and thus require a finite top-quark width, for which we use the 
NLO QCD value $\Gamma_t=1.339~\GeV$.  
However, at the perturbative order considered in this study,
such topologies arise only in QCD--EW interference terms that do not give rise to 
Breit--Wigner resonances.  The dependence of our results on $\Gamma_t$ is thus completely negligible.}
\beqar
\Gamma_t&=&1.339~\GeV,\qquad  
\Gamma_H=0~\GeV.
\eeqar

All unstable particles are treated in the complex-mass scheme~\cite{Denner:2005fg},
where width effects are absorbed into the complex-valued renormalised masses
\beqar\label{eq:complexmasses}
\mu^2_i=M_i^2-\ri\Gamma_iM_i \qquad\mbox{for}\;i=W,Z,t.
\eeqar

For $W$+jet and $Z$+jet production processes the 
EW
 couplings are derived from the gauge-boson masses and the Fermi constant,
$\GF=1.16637\times10^{-5}~\GeV^{-2}$, using 
\beq\label{eq:defalpha}
\alpha=\left|\frac{\sqrt{2}\sin^2\thetaw\,\mu^2_W G_\mu}{\pi}\right|,
\eeq
while for $\gamma$+jet production the 
EW
coupling is chosen to be~\cite{Agashe:2014kda}
\beq
\alpha=\alpha(0)=1/137.035999074\,.
\eeq
In both schemes the weak mixing angle $\thetaw$ is determined by 
\beq\label{eq:defsintheta}
\sin^2\thetaw=1-\cos^2\thetaw=1-\frac{\mu_W^2}{\mu_Z^2},
\eeq
and becomes complex-valued.
The $G_\mu$-scheme guarantees an  optimal description of pure SU(2) interactions
at the 
EW
scale. It is the scheme of choice for
$W+$\,jet production, and it provides a very good description of
$Z\,+$\,jet production as well. 
The $\alpha(0)$ scheme to be used for $\gamma$+jet, on the other hand, expresses the fact that on-shell photons effectively couple at a scale $Q^2$=0 .
The CKM matrix is assumed to be diagonal and we checked at LO and NLO QCD that for $W$+jet production the difference with respect to a non-diagonal CKM matrix  is always well
below 1\%.
For the choice of renormalisation and factorisation scales and variations thereof we refer to \refse{se:qcd}.

For the calculation of hadron-level cross sections at (N)NLO QCD\,+\,(n)NLO
EW we employ the
{\tt LUXqed$\_\linebreak$plus$\_$PDF4LHC15$\_$nnlo$\_$100}
PDF set, which is based on    
PDF4LHC NNLO PDFs~\cite{Butterworth:2015oua,Ball:2014uwa,Dulat:2015mca,Harland-Lang:2014zoa,Gao:2013bia,Carrazza:2015aoa} supplemented with QED effects~\cite{Manohar:2016nzj}.
The same PDF set, and the related $\alphaS$ value,  is used throughout, i.e.~also in the relevant LO and NLO 
ingredients used in the estimate of theoretical uncertainties.
At the level of precision discussed in this study also the uncertainty on the value of $\alphaS$ becomes relevant.
Given 1\% uncertainty on the measured value of $\alphaS$ this results in an overall 1--2\% normalisation uncertainty on the differential $\pT$ distributions. However, one should keep in mind that in the process ratios this uncertainty cancels completely and thus
 it is irrelevant for background estimates in DM searches at high-MET.
Consistently with the five-flavour number scheme employed in the PDFs,
$b$-quarks are treated as massless partons, and channels with initial-state $b$-quarks are taken into account.
All light quarks, including
bottom quarks, are treated as massless particles, and top-quark loops are included up to NLO 
throughout. 
Matrix elements at (N)NLO are evaluated using the five-flavour running of the strong coupling 
supported by the PDFs and, for consistency, top-quark loops are renormalised in the decoupling scheme.
For the NNLO QCD coefficient no top-quark loops are considered.  

For the assessment of PDF uncertainties the \linebreak PDF4LHC prescription~\cite{Butterworth:2015oua} is adopted.
In addition to standard PDF variations, also additional \texttt{LUXqed} variations 
for the photon PDF are applied. 
For more details see more details in~\refses{se:gammaind}{se:pdfuncert}.

%However, given the smallness  of photon-induced effects
%and related uncertainties in \texttt{LUXqed} PDFs (see~\refse{se:gammaind}) such variations can be safely neglected.
%For convenience, PDF uncertainties can safely be determined based on NLO instead of NNLO
%matrix elements. 

%%%%%%%%%%%%%%%%%%%%%%%%%%%%%%%%%%%%%%%%%%%%
\subsection{Computational frameworks}
\label{se:tools}

The theoretical predictions presented in~\refse{se:ho} 
include corrections up to NNLO\,QCD and NLO\,EW, as well 
as Sudakov EW effects at $\ord(\alpha^2)$. They have been obtained
by means of a variety of methods and tools, as detailed in the following.

The NLO QCD and NLO EW calculations for all $pp\to V+$\,jet processes have been performed with 
\MunichOpenLoops and/or \SherpaOpenLoops.
In these automated frameworks~\cite{Kallweit:2014xda,Kallweit:2015dum,Kallweit:2017khh}
virtual amplitudes are provided by the \OpenLoops
program~\cite{hepforge,Cascioli:2011va}, combined with the \Collier tensor reduction library
\cite{Denner:2016kdg} or with {\sc CutTools}~\cite{Ossola:2007ax}.
The remaining tasks are supported by the two independent and fully
automated Monte Carlo generators \Munich~\cite{munich} and \linebreak
\Sherpa~\cite{Gleisberg:2008ta,Krauss:2001iv,Gleisberg:2007md,sherpaqedbrems}.  
Additionally, we carefully validated the NLO
EW predictions against the results
of \linebreak\citeres{Denner:2009gj,Denner:2011vu,Denner:2012ts}.
The NLO EW calculations for $pp\to V+2$\,jets performed to test the factorisation of
QCD and EW corrections have been checked against the one
of \citere{Denner:2014ina} for $pp\to Z+2$\,jets in~\citere{Badger:2016bpw}.
The NLO EW amplitudes for all $V$+jet processes in \OpenLoops have been
supplemented with the one- and two-loop analytical Sudakov logarithms of
\citeres{Kuhn:2004em,Kuhn:2005gv,Kuhn:2005az,Kuhn:2007qc,Kuhn:2007cv}.

The NNLO QCD predictions for $Z$+jet production have been obtained with the
parton-level event generator {\rmfamily\scshape NNLOjet}, which 
provides the necessary infrastructure to perform fully differential
calculations at NNLO using the antenna subtraction
formalism~\cite{GehrmannDeRidder:2005cm,GehrmannDeRidder:2005aw,GehrmannDeRidder:2005hi,Daleo:2006xa,Daleo:2009yj,Gehrmann:2011wi,
Boughezal:2010mc,GehrmannDeRidder:2012ja,Currie:2013vh}.
The computation of $pp\to W$+jet through NNLO is based on the $N$-jettiness
subtraction scheme for NNLO calculations~\cite{Boughezal:2015dva}.  The above-cut contribution within the
$N$-jettiness subtraction was obtained using \linebreak \MunichOpenLoops.
The NNLO QCD prediction for the $pp\to \gamma$+jet process is based on the
calculations of~\citeres{Campbell:2016lzl,Campbell:2017dqk} and has been
obtained using MCFM~\cite{Boughezal:2016wmq}.
In order to ensure the correctness of the numerical implementation of cuts
and other parameters in the NNLO codes, a detailed comparison has been
performed at the level of the NLO QCD results as described above.

\section{Higher-order QCD and EW predictions}
\label{se:ho}
  
Precise theory predictions for $V+$\,jet production require 
QCD and EW higher-order corrections,
mixed QCD--EW contributions, as well as photon-induced contributions, 
\beq
\label{eq:th1}
\parx\siv_{\TH}=
\parx\siv_{\QCD}+ \parx\Delta\siv_{\EW} + \parx\Delta\siv_{\mix} + \parx\siv_{\gamma-{\rm ind.}}.
\eeq
In this section we present theoretical predictions that include 
corrections up to NNLO\,QCD and NLO\,EW supplemented by EW Sudakov logarithms at two loops.
Moreover, we introduce a coherent theoretical framework for the combination of
EW and QCD calculations for the various $V+$\,jet production processes
and for the assessment of the corresponding remaining sources of 
theoretical uncertainty.
State-of-the art QCD and EW predictions and the related theoretical uncertainties are discussed in
\refse{se:qcd} and \ref{se:ew} respectively. \refse{se:gammaind} is devoted to 
photon-induced channels and \refse{se:pdfuncert} to PDF uncertainties, while in 
\refse{se:dibosons} we discuss the real emission of vector bosons, and
mixed corrections of $\ord(\alpha\alphaS)$ are addressed in \refse{se:ewqcd} by means of 
a factorised combination of QCD and EW corrections.

To illustrate the effect of higher-order corrections and uncertainties we
present a series of numerical results for $pp\to V+$\,jet at a
centre-of-mass energy of 13\,TeV in the setup specified in \refse{se:setup}. 
In particular, $pp\to \gamma+$\,jet predictions are based 
on the dynamic photon isolation~\refpar{eq:dynisolation}.  As anticipated
in~\refse{se:objects}, 
this prescription provides a very convenient basis for the systematic
modelling of the correlation of QCD uncertainties between the various
$V+$\,jet production processes (see~\refse{se:qcd}).

Vector-boson $p_\rT$ spectra are plotted starting at \linebreak 80\,GeV, 
but for the sake of a complete documentation 
data sets are provided  above 30 GeV (see \refapp{se:numpredictions}).
However, 
we note that in the region of $p_\rT \lsim 100$~GeV there are potential sources of
systematics that we are not controlling or even discussing, as they would require a separate
study.  These arise from the resummation of QCD Sudakov logarithms or from
non-perturbative effects (e.g. an order $\Lambda_{\mathrm{QCD}}$ average shift of the vector boson $p_\rT$ associated with the asymmetry of colour flow in the final state). Furthermore, as shown later, a reliable correlation between
the $Z/W$ spectra and the photon spectrum requires $p_\rT$ to be large enough
so that fragmentation contributions in $\gamma +$jet production become small.
We also expect that in
the $p_\rT$ regions up to a few hundred GeV the statistics are sufficient to
guarantee that experimental analyses of missing-$E_\rT$ backgrounds can
entirely rely on the direct measurement of the $Z$ spectrum measured via
$Z\to\ell^+\ell^-$.  As a result, we believe that our conclusions on the
systematic uncertainties are most reliable and useful for experimental
applications in the region of $p_\rT$ larger than 100--200\,GeV.

\subsection{Higher-order QCD predictions}
\label{se:qcd}

For perturbative QCD predictions at LO, NLO and NNLO we use the generic notation
\beq
\label{eq:qcd0}
\parx\siv_{\QCD} = 
\parx\siv_{\NkLO\,\QCD},
\eeq
with $k=0,1$ or $2$. Wherever possible, nominal predictions are provided at NNLO QCD, 
i.e.~including terms up to\footnote{Here and in the following we adopt a power counting 
that does not include the extra factor $\alpha$ associated
with vector-boson decays.} $\ord(\alpha\alphaS^3)$.  
However, as ingredients for the 
assessment of some theory uncertainties, also LO and NLO QCD contributions will be used.

For convenience, results at $\NkLO$ QCD are systematically expressed in terms of 
LO predictions and relative correction factors defined through
\beqar\label{eq:kfactors}
\parx\siv_{\NkLO\,\QCD}(\vec\mu)&=& 
K^{(V)}_{\NkLO}(x,\vec\mu)
\parx\siv_{\LO\,\QCD}(\vec\mu_0).
\eeqar
We calculate all $\NkLO$ and LO cross sections with one and the same set of NNLO
PDFs as discussed in\linebreak \refse{se:inputs}.
The dependence on the renormalisation and factorisation scales,
$\vec\mu=(\mu_{R},\mu_{F})$, is absorbed into the $K$-factors,
while LO predictions on the r.h.s.~of \refeq{eq:kfactors} are taken at the central scale, 
$\vec\mu_0=(\mu_{R,0},\mu_{F,0})$. 
For the central scale we adopt 
the commonly used choice
\beq
\label{eq:htscale}
\mu_{R,0}=\mu_{F,0}=\mu_{0}=\HTprimehat/2  ,
\eeq
where the total transverse energy, $\HTprimehat$, 
is defined as the scalar sum of the transverse energy of all parton-level
final-state objects,
\beq\label{eq:BHscale}
\HTprimehat = E_{\rT,\PV}\,+\sum_{i\in \{\mathrm{q,g,\gamma}\}} |p_{\rT,i}|. 
\eeq
Also quarks (q), gluons (g) and photons that are radiated 
in the (N)NLO QCD or EW corrections
are included in $\HTprimehat$,
and the vector-boson transverse energy, $E_{\rT,\PV}$,  is computed using the
total (off-shell) four-momentum of the corresponding decay products, \ie
\beqar\label{ETboson}
E^2_{\rT,Z}&=&p^2_{\rT,\ell^+\ell^-}+m_{\ell^+\ell^-}^2,\nonumber\\
E^2_{\rT,W}&=&p^2_{\rT,\ell\nu}+m_{\ell\nu}^2,\nonumber\\
E^2_{\rT,\gamma}&=&p^2_{\rT,\gamma}.
\eeqar
In order to guarantee infrared safety at NLO EW, the scale
\refpar{eq:BHscale} must be insensitive to collinear photon emissions off
charged fermions.  To this end, the vector-boson transverse energies defined in 
\refeq{ETboson} should be computed in terms of dressed leptons as specified in\linebreak \refse{se:dressing}, while 
$|p_{\rT,\gamma}|$
contributions to \refeq{eq:BHscale} should involve only photons that have not been recombined
with charged leptons. It is worth to note that  
$\mu_{0} \approx  p_{\rT,V}$ at large $p_{\rT,V}$.

\subsubsection*{Pure QCD uncertainties}

\def\shape{\omega_\rshape}

The uncertainty associated with the truncation of the perturbative expansion in $\alphaS$
is estimated by means of factorisation and renormalisation scale variations. 
We consider standard seven-point variations applying, respectively, 
 factor-two rescalings, \ie
\beqar
\frac{\vec \mu_i}{\mu_0}&=& (1,1), (2,2), (0.5,0.5), (2,1), (1,2), (1,0.5), (0.5,1), \nonumber\\
\eeqar 
where $i=0,\ldots 6$.
Nominal predictions and related uncertainties are defined as
the centre and the half-width of the band resulting from the above variations. 
In terms of $K$-factors this corresponds to 
\beqar\label{eq:meanKfact}
K^{(V)}_{\NkLO}(x) &=&
\frac{1}{2}\left[K^{(V,\max)}_{\NkLO}(x)+K^{(V,\min)}_{\NkLO}(x)\right],
\\
\delta^{(1)} K^{(V)}_{\NkLO}(x)&=&
\frac{1}{2}\left[K^{(V,\max)}_{\NkLO}(x)-K^{(V,\min)}_{\NkLO}(x)\right],
\label{eq:var1Kfact}
\eeqar
with
\beqar
K_{\NkLO}^{(V,\max)}(x)&=&\max \left\{K^{(V)}_{\NkLO}(x,\vec\mu_i)\left| 0\le i\le 6\right.\right\},
\nonumber\\
K_{\NkLO}^{(V,\min)}(x)&=&\min \left\{ K^{(V)}_{\NkLO}(x,\vec\mu_i)\left| 0\le i\le 6\right.\right\}.
\eeqar
Since the shift resulting form the symmetrisation of scale variations in
\refeq{eq:meanKfact} is encoded in the $K$-factors, also the 
LO $K$-factor differs from one.

\begin{figure*}[t]   
\centering
\includegraphics[width=0.50\textwidth]{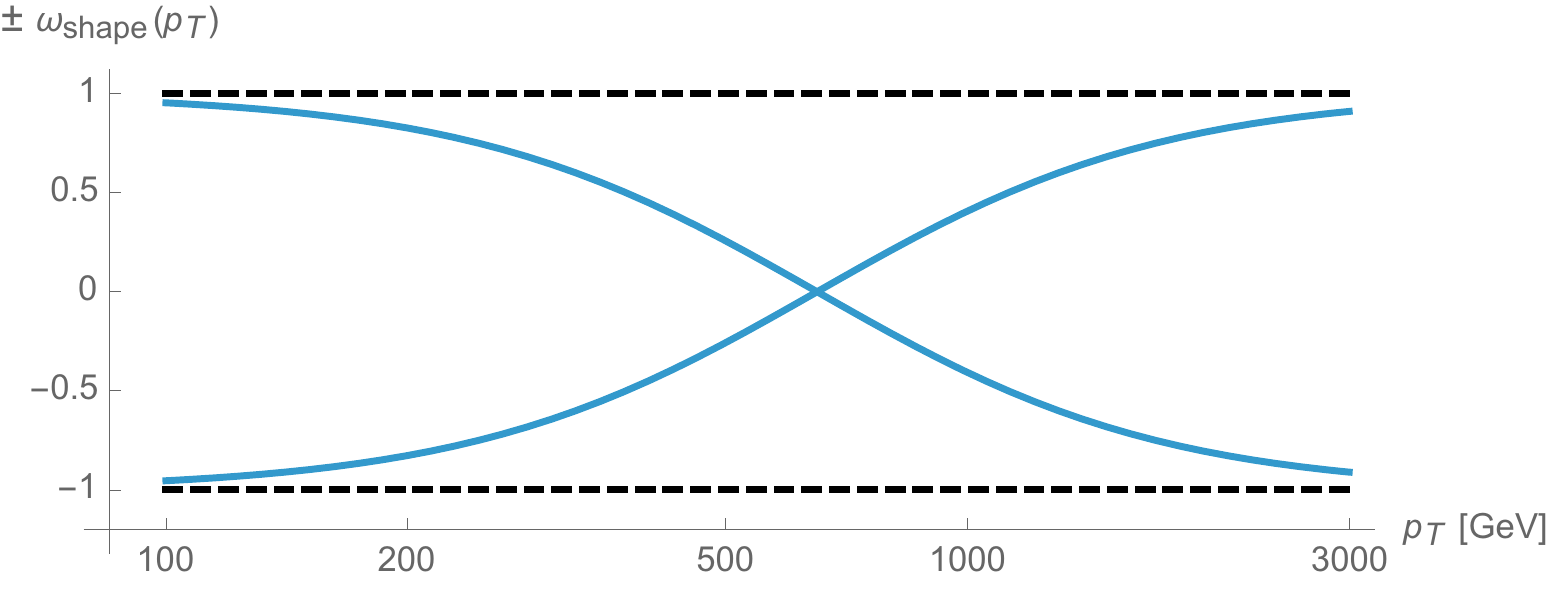}
\caption{Shape variation function $\shape(p_\rT)$ defined in \refeq{eq:shapedist}.
}
\label{fig:shapedist}
\end{figure*}

Constant scale variations mainly affect the overall normalisation
of $p_\rT$-dis\-tri\-bu\-tions and tend to underestimate shape uncertainties,
which play an important role in the extrapolation of low-$p_\rT$ measurements
to high $p_\rT$.
Thus, for a reasonably conservative estimate of shape uncertainties, 
we introduce an additional variation,
\beqar
\label{eq:dKqcd2}
\delta^{(2)} K^{(V)}_{\NkLO}(x)=
\shape(x)\,
\delta^{(1)} K^{(V)}_{\NkLO}(x),
\label{eq:var2Kfact}
\eeqar
where the standard scale uncertainty \refpar{eq:var1Kfact}
is supplemented by a shape distortion $\shape(x)$, with \linebreak
$|\shape(x)|\le 1$ and  $\shape(x)\to \pm 1$ at high and small transverse momentum, respectively.
The function $\shape$ is defined as
\beqar
\label{eq:shapedist}
\shape(p_\rT)= \tanh\left[\ln\left(\frac{p_\rT}{p_{\rT,0}}\right)\right]
=\frac{p^2_\rT-p^2_{\rT,0}}{p^2_\rT+p^2_{\rT,0}},
\eeqar
and as reference transverse momentum we choose the 
value $p_{\rT,0}=650$\,GeV, which corresponds (in logarithmic scale) to the 
middle of the range of interest, 0.2--2\,TeV.
As illustrated in \reffi{fig:shapedist}, 
the function $\shape(x)$ induces asymmetric variations that cover $\pm 75\%$ of the 
standard scale variation band for $p_\rT\in [250,1750]\,\GeV$. 
Note that, in the combination of the uncertainties \refpar{eq:var1Kfact} and \refpar{eq:dKqcd2},
our choice to have an additional shape variation augments 
the standard scale uncertainty by a factor
$1 \le \sqrt{1+\shape^2(p_\rT)}\le \sqrt{2}$.

%%%%%%%%%%%%%%%%%%%%
%%%%%%%%%%%%%%%%%%%%

Besides shape uncertainties, also the correlation of QCD uncertainties
across $V+$\,jet processes plays a key role in fits of the
$Z(\nu\bar\nu)+$\,jet dark matter background, and the quantitative
understanding of such process correlations belongs to the most important
theoretical aspects in dark matter searches.
From the viewpoint of QCD interactions, the processes $pp\to W+$\,jet and 
$pp\to Z+$\,jet are quite similar to each other at $p_{\rT,V}\gg M_{W,Z}$.
Thus, the respective QCD uncertainties are expected to be strongly correlated.
However, due to the presence of $q\to q\gamma$ collinear singularities and the need 
to suppress them with an appropriate photon-isolation prescription,
higher-order QCD contributions to $\gamma+$\,jet production
can behave in a significantly different way as compared to 
the case of $pp\to W/Z+$\,jet. 
In order to reduce such differences, 
we adopt the dynamic photon isolation approach defined in~\refeq{eq:dynisolation}.
As discussed in~\refse{se:objects}, this prescription
renders the QCD dynamics of $pp\to \gamma+$\,jet and $pp\to Z/W+$\,jet
processes almost universal. As a result,
QCD $K$-factors 
$K^{(V)}_{\NkLO}(x)$ and their uncertainties $\delta^{(i)}K^{(V)}_{\NkLO}(x)$ 
depend only very weakly\footnote{For what concerns
process correlations, it is crucial that (apart from the $M_V$ dependence) all $V+$\,jet 
processes are evaluated using equivalent dynamical scales.}
on $V$ at high $p_\rT$,
and in this situation the small process-dependent part of QCD $K$-factors can be used as 
an estimator of the degree of correlation across processes.
To this end we consider
the highest available term in the perturbative expansion,
\beqar
\label{eq:proccorr1}
\Delta K^{(V)}_{\NkLO}(x)=K^{(V)}_{\NkLO}(x)/K^{(V)}_{\NkmLO}(x)-1,
\eeqar
and as estimate of unknown process correlation effects 
we take the difference of the known QCD $K$-factors with respect to $Z+$\,jet production,
\beqar
\label{eq:dKqcd3}
\delta^{(3)} K^{(V)}_{\NkLO}(x)
&=&
%\Delta \bar K^{(V)}_{\NkLO}(x).
\Delta K^{(V)}_{\NkLO}(x)-\Delta  K^{(Z)}_{\NkLO}(x).
\eeqar
This process correlation uncertainty can be assessed using the central scale 
\refpar{eq:htscale} throughout. 
Applying it to nominal predictions, \ie replacing 
$K^{(V)}_{\NkLO} \to K^{(V)}_{\NkLO}\pm \delta^{(3)} K^{(V)}_{\NkLO}$, amounts to 
doubling or removing $K$-factor differences between processes.
The choice of $Z+$\,jet production as
reference process in \refeq{eq:dKqcd3} is arbitrary, but 
changing the reference process has very little impact on process correlations
since the resulting 
overall shift in $\delta^{(3)} K^{(V)}_{\NkLO}(x)$ cancels to a large extent 
in ratios of $V+$\,jet cross sections.

The above prescription should be regarded as conservative, since 
parts of the available $K$-factors are downgraded 
from the status of known higher-order corrections to uncertainties.
However, thanks to the 
fact that the $V+$\,jet $K$-factors of the same order $k$ are strongly
correlated, $\delta^{(3)} K^{(V)}_{\NkLO}(x) \ll \Delta K^{(V)}_{\NkLO}$,
the resulting losses of accuracy in the nominal $\NkLO$ predictions 
for individual processes are rather small.

For the application to experimental analyses, it is important to keep in mind that 
the above modelling of process correlations assumes a close similarity of QCD
effects between all $pp\to V+$\,jet processes, which is achieved, in the present study,
by means of the dynamic photon isolation~\refpar{eq:dynisolation}.
Thus, as discussed in \refse{se:photons}, 
experimental analyses that employ a different photon isolation
approach require 
an additional $\gamma+$\,jet specific uncertainty.

The above uncertainties can be 
parametrised through a set of independent nuisance parameters,
$\vec\eps_\QCD$,
and combined using
\beqar\label{ew:QCDcomb}
&\parx \siv_{\NkLO\,\QCD}(\vec\eps_{\QCD})=
\left[K^{(V)}_{\NkLO}(x)\right.\\
&\left.+\sum_{i=1}^3\eps_{\QCD,i}\,\delta^{(i)} K^{(V)}_{\NkLO}(x) \right]
\nonumber
\times \,\parx\siv_{\LO\,\QCD}(\vec\mu_0).
\eeqar
The nuisance parameters $\eps_{\QCD,1},\eps_{\QCD,2}$ and $\eps_{\QCD,3}$ \linebreak
should be 
Gaussian distributed 
with one standard deviation
corresponding to the range $\eps_{\QCD,i}\in[-1,+1]$.
These parameters should be kept uncorrelated, 
but each $\eps_{\QCD,i}$-variation should be applied in a correlated way across $p_\rT$ bins 
and processes, since correlation effects are consistently implemented in the 
$\delta^{(i)} K^{(V)}_{\NkLO}(x)$ terms.

\subsubsection*{Numerical results}

%%%%%%%%%%%%%%%%%%%%
\begin{figure*}[t]   
\centering
  \includegraphics[width=\ratiotextwidthbig\textwidth]{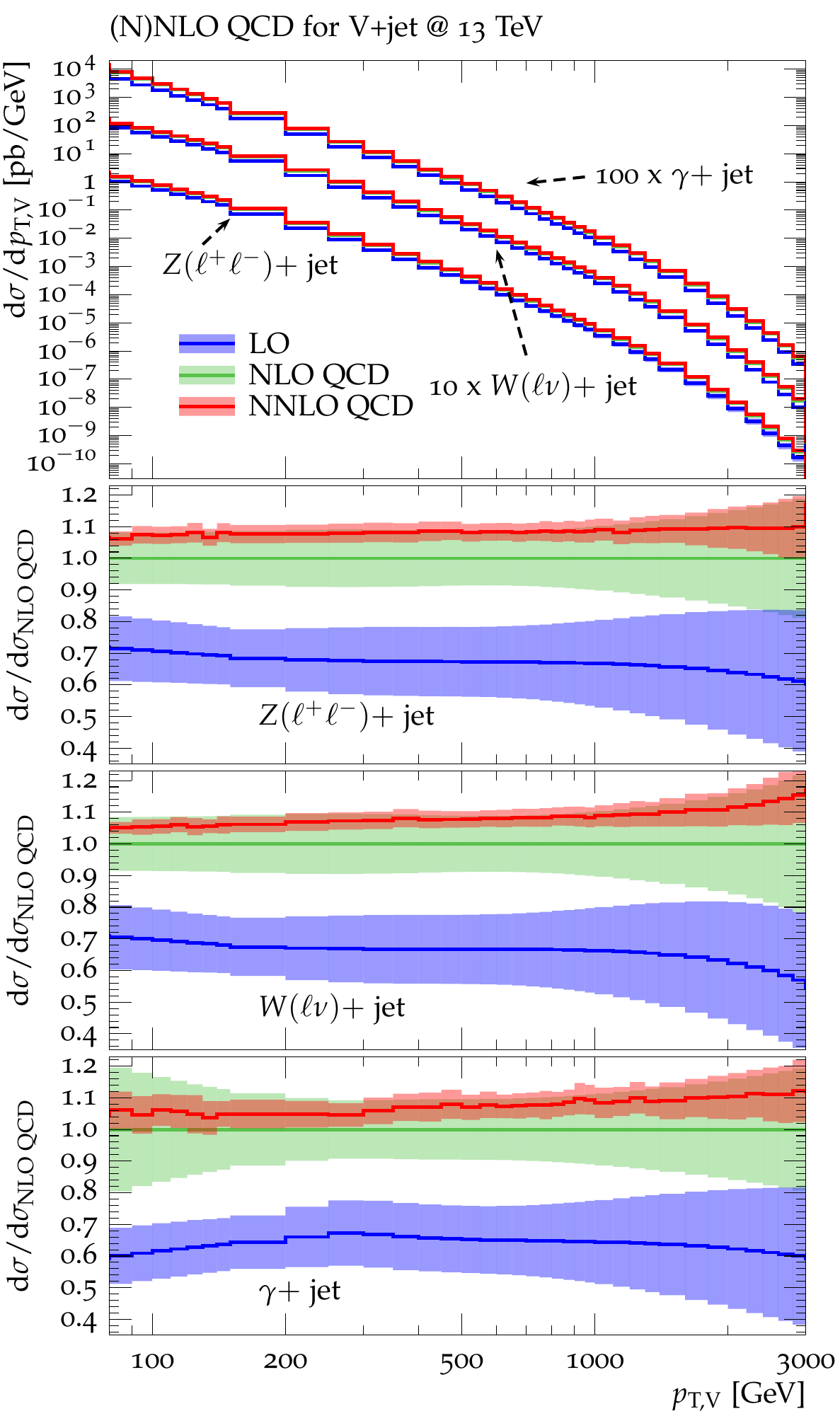}
\caption{Higher-order QCD predictions and uncertainties for $Z(\ell^+\ell^-)$+jet, 
$W^\pm(\ell\nu)$+jet, and 
$\gamma$+jet production at 13\,TeV.
Absolute predictions at LO, NLO and NNLO QCD are displayed in the main frame.
The ratio plots show results for individual processes normalised to NLO QCD.
The bands
 correspond to the combination (in quadrature) of the three types of QCD uncertainties, $\delta^{(i)}K_{\NkLO}$, i.e.~scale uncertainties according to \refeq{eq:var1Kfact},
shape uncertainties according to \refeq{eq:dKqcd2}, and process-correlation uncertainties according to \refeq{eq:dKqcd3}. 
}
\label{fig:QCD_error}
\end{figure*}
%%%%%%%%%%%%%%%%%%%%

\begin{figure*}[tp]   
\centering
  \includegraphics[width=\ratiotextwidth\textwidth]{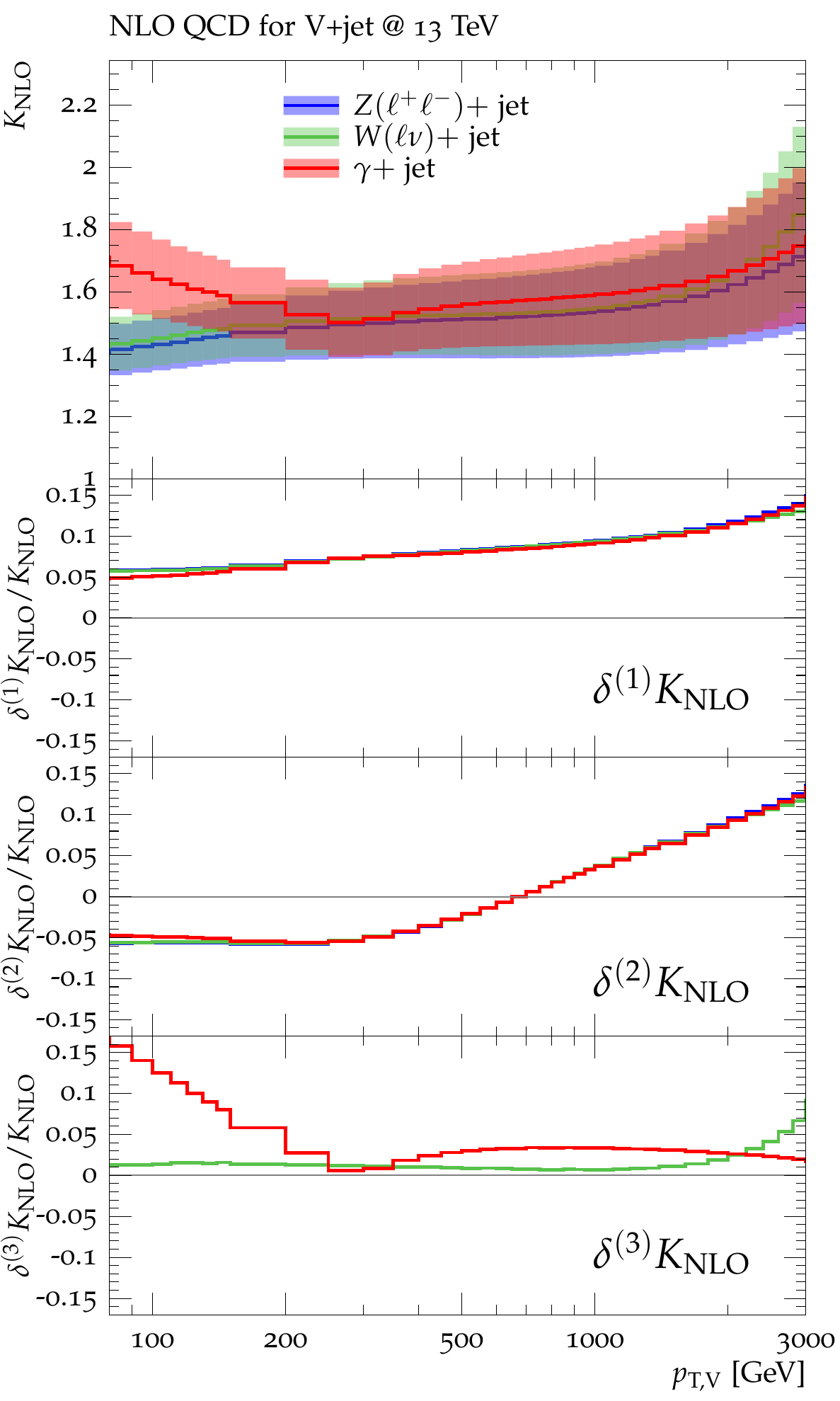}
  \includegraphics[width=\ratiotextwidth\textwidth]{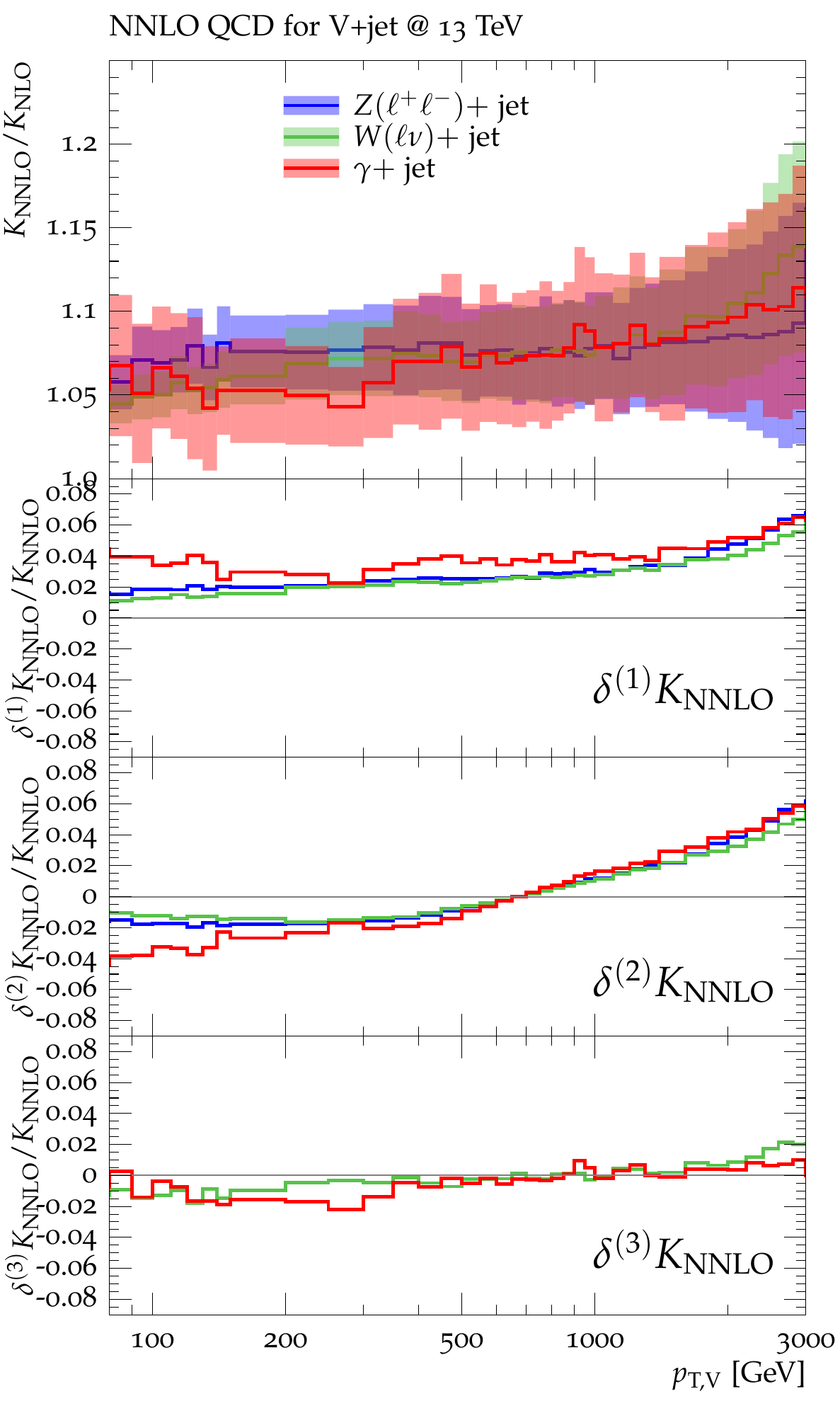}
\caption{QCD $K$-factors at NLO (with respect to LO) on the left and at NNLO (with respect to NLO) on the right for the various $pp\to V+$\,jet processes at 
13\,TeV.
The bands in the upper frame correspond to scale variations, \ie $\delta^{(1)}K_{\rm NLO}$ and 
$\delta^{(1)}K_{\NNLO}$.
The lower frames show the individual uncertainties defined in \refeq{eq:var1Kfact},
\refeq{eq:dKqcd2}, and \refeq{eq:dKqcd3}.
They are displayed as ratios $\delta^{(i)}K_{\NkLO}/K_{\NkLO}$, which 
corresponds to the relative impact of uncertainties on $p_\rT$ distributions at $\NLO$ and $\NNLO$. 
}
\label{fig:QCD_corr_error}
\end{figure*}

\begin{figure*}[t]   
\centering
  \includegraphics[width=\ratiotextwidth\textwidth]{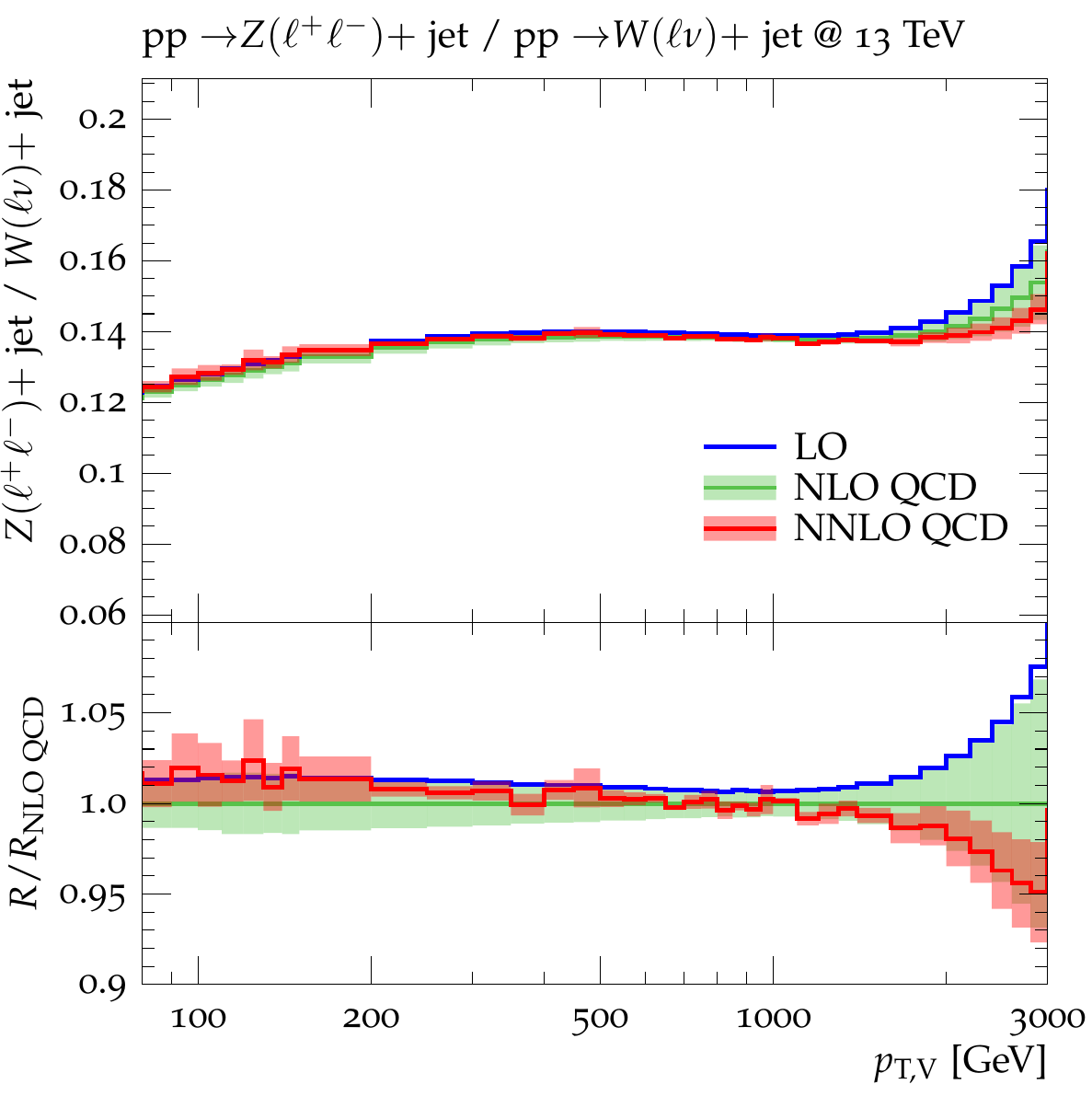}
    \includegraphics[width=\ratiotextwidth\textwidth]{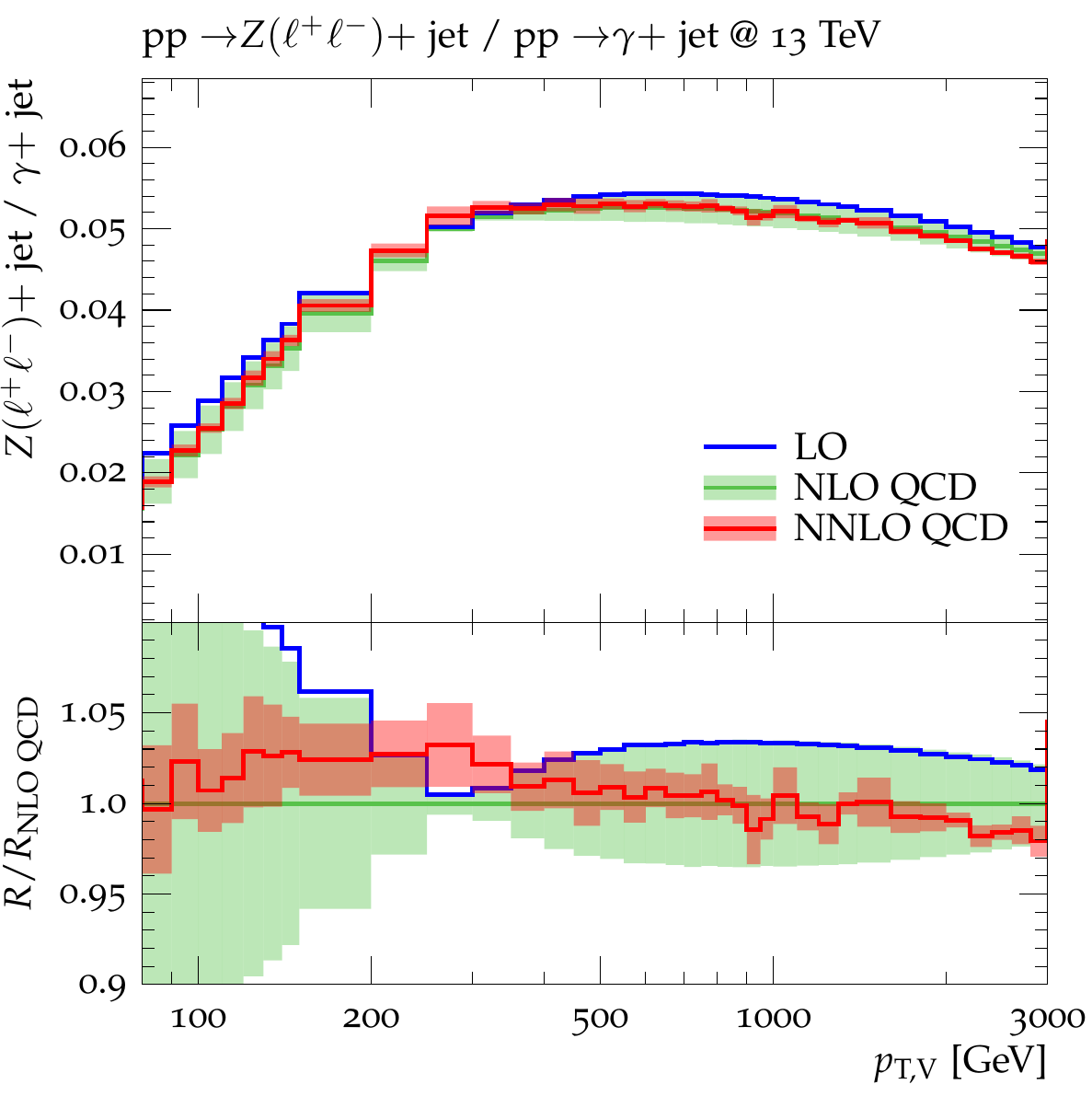}\\[6mm]
        \includegraphics[width=\ratiotextwidth\textwidth]{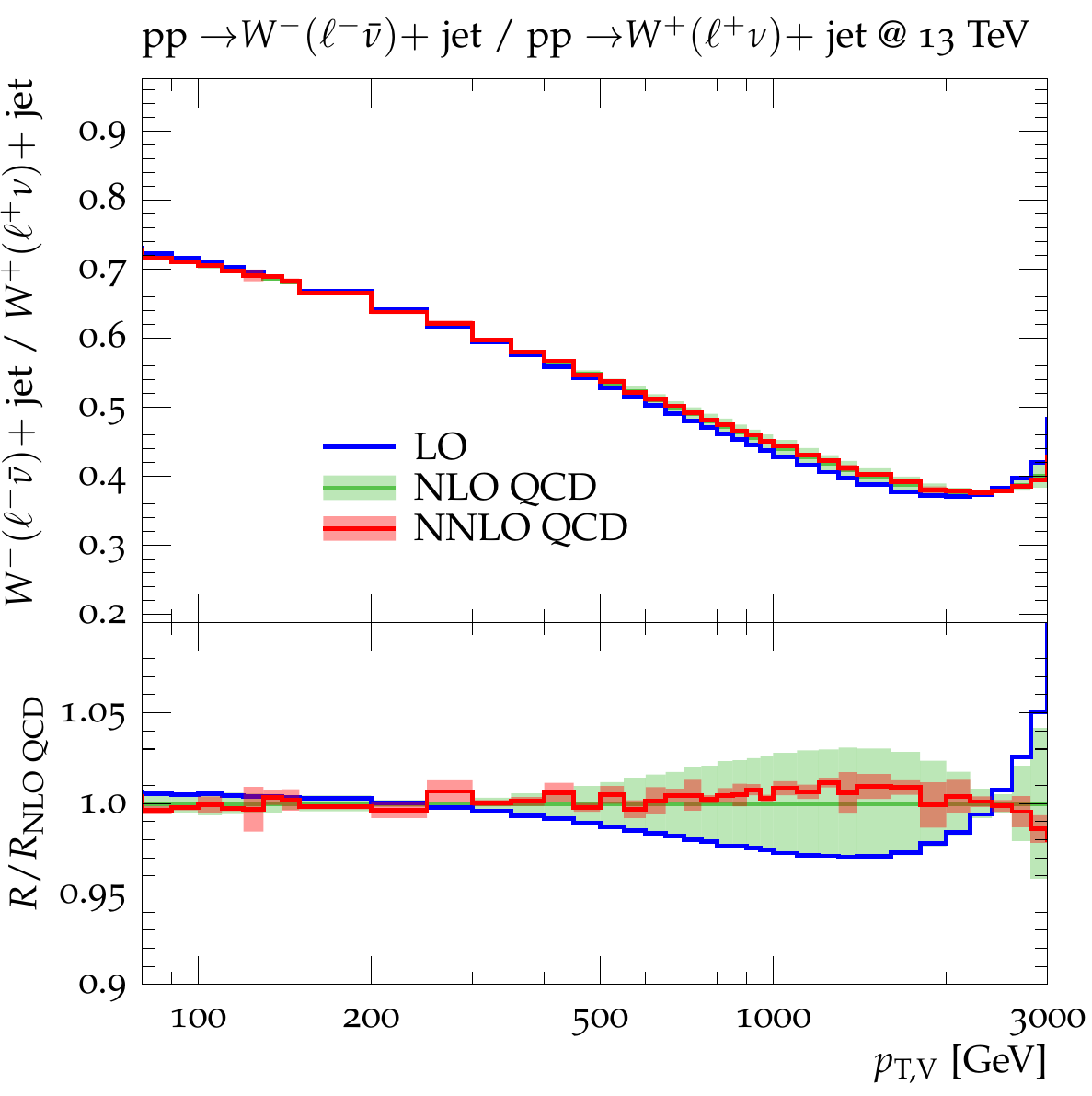}
    \includegraphics[width=\ratiotextwidth\textwidth]{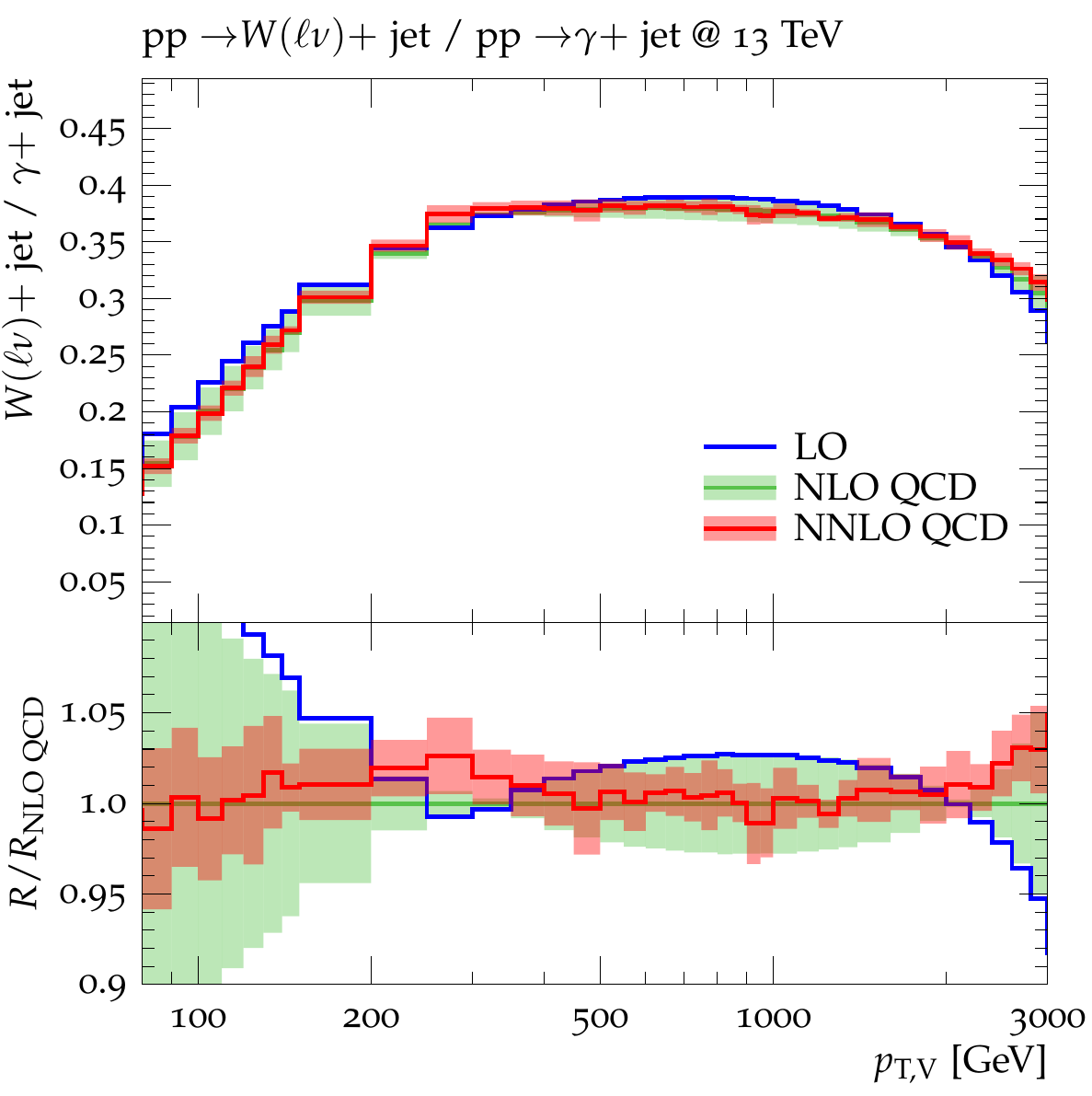}
\caption{
Ratios of $p_\rT$-distributions for various $pp\to V$+jet processes at 
LO, NLO and NNLO QCD. The NLO and NNLO QCD uncertainties, estimated according to \refeq{eq:var1Kfact},
\refeq{eq:dKqcd2}, and \refeq{eq:dKqcd3} are correlated amongst 
processes as described in the text and combined in quadrature. At LO only nominal predictions are shown.
}
\label{fig:ratios_qcd}
\end{figure*}

Predictions for $V$+jet distributions and their ratios at LO, NLO and NNLO
QCD are presented in \reffis{fig:QCD_error}{fig:ratios_qcd} as well as in
\reffis{fig:app_QCD_error}{fig:app_ratios_qcd} (see \refapp{app:unc}).
In Figures~\ref{fig:QCD_corr_error}
and~\ref{fig:app_QCD_error}--\ref{fig:app_ratios_qcd}, scale uncertainties
\refpar{eq:var1Kfact}, shape uncertainties \refpar{eq:dKqcd2}, and
process-correlation uncertainties~\refpar{eq:dKqcd3} are shown separately,
while in Figures \ref{fig:QCD_error} and~\ref{fig:ratios_qcd} the three QCD
uncertainties are combined in quadrature.
Here and in the following $W$ denotes $W^+$ and $W^-$ combined. 

At high transverse momentum, we find that QCD corrections and uncertainties
for the various $V+\,$jet production processes behave in a very similar way. 
At NLO the corrections amount to 40--60\% with residual uncertainties around
10--20\%, while NNLO corrections increase the cross section by 5--10\% and
reduce the combined uncertainty to 3--10\%.  Scale variations
$\delta^{(1)}K_{\NkLO}$ and shape variations $\delta^{(2)}K_{\NkLO}$ are the
dominant sources of uncertainty in $\pT$-distributions.  Their contributions
are very similar across $V+\,$jet processes.  Thus in the ratios scale and
shape variations largely cancel, and the process-correlation uncertainty
$\delta^{(3)}K_{\NkLO}$ tends to dominate.

The ratio plots (\reffi{fig:ratios_qcd}) allow one to appreciate small
differences in the QCD dynamics of the various $V+$\,jet processes.  As
reflected in the $Z/W$ ratio, the NLO and NNLO corrections for the
corresponding proceses are almost identical, with differences below \mbox{1--2\%}
up to one TeV.  Only at very large $\pT$ the NLO and also NNLO corrections to
$W$+jet grow faster than in the case of $Z$+jet.  This results in
an increase of the process-correlation uncertainty $\delta^{(3)}K_{\NLO}$ up
to about $5\%$ beyond $\pT=2~\TeV$.

As can be seen in the $Z/\gamma$ and $W/\gamma$ ratios,  the higher-order QCD corrections to $\gamma$+jet
production behave very similarly as for $Z+$\,jet and $W+$\,jet production at
large $\pT$. This is the result of the dynamic photon isolation~\refpar{eq:dynisolation}, which guarantees
that  the differences in the NLO and NNLO
corrections remain below 3--4\% for $\pT>200\,$\,GeV.
Instead, at lower $\pT$ the behaviour of
$\gamma+$\,jet production changes drastically due to mass effects, which
results in sizeable process-correlation uncertainties.\footnote{In this
regime, which is not the main focus of the present study, the
process-correlation uncertainty~\refpar{eq:dKqcd3} ceases to be a meaningful
uncertainty estimate.} Note that for $\pT\approx 300~\GeV$ the
NLO process-correlation uncertainty in $pp\to\gamma+$jet is accidentally very
small (see \reffi{fig:app_QCD_error}) yielding a pinch in the total QCD
uncertainty for the $Z/\gamma$ and the $W/\gamma$ ratios (see also \reffi{fig:app_ratios_qcd}).  However, one
should keep in mind that an additional analysis-dependent photon-isolation
uncertainty (see \refse{se:objects}) has to be considered for these ratios.

In general, comparing QCD predictions at different orders we observe a good
convergence of the perturbative expansion, and the fact that process ratios
receive very small corrections both at NLO and NNLO provides strong evidence
for the universality of QCD dynamics is all $V+$\,jet processes.  Results at
NNLO provide also a crucial test of the goodness of the proposed approach
for the estimate of QCD uncertainties and their correlations.  In
particular, the remarkable consistency between NNLO and NLO predictions in
\reffi{fig:ratios_qcd} confirms that QCD uncertainties for process ratios
are as small as 1--2\%.

%%%%%%%%%%%%%%%%%%%%
%%%%%%%%%%%%%%%%%%%%
%\clearpage

%%%%%%%%%%%%%%%%%%%%
%%%%%%%%%%%%%%%%%%%%

\subsection{Electroweak corrections}
\label{se:ew}

For EW higher-order corrections we use the notation,
\beqar
\label{eq:ew0}
\parx\siv_{\NLO\,\EW}&=&
\parx\siv_{\LO\,\QCD}+
\parx\Delta\siv_{\NLO\,\EW},
\\\nonumber
\parx\siv_{\nNLO\,\EW}&=&
\parx\siv_{\NLO\,\EW}+
\parx\Delta\siv_{\nnlo\,\Sud},
\eeqar
where $\Delta\siv_{\NLO\,\EW}$ denotes exact $\ord(\alpha^2\alphaS)$
contributions, and `NNLO\,Sud' stands for
$\ord(\alpha^3\alphaS)$ EW Sudakov logarithms in NLL approximation (see below).
Their combination is dubbed nNLO\,EW as it accounts for the 
dominant EW effects at NNLO.
While our power counting does not consider the extra factor $\alpha$ 
associated with vector-boson decays, all 
predictions for $pp\to W/Z+$\,jet at (N)NLO QCD + NLO EW  are at the level of the full
processes, $pp\to \ell\nu/\ell\ell/\nu\nu+$\,jet, including off-shell effects and NLO EW corrections 
in decays.
Since EW Sudakov logarithms do not enter $W$ and $Z$ decays, they
are applied only at the level 
of $pp\to V+$\,jet production, including off-shell decays at LO.

The EW corrections, similarly as for the QCD ones, are also expressed in terms of 
correction factors with respect to LO QCD,
\beqar\label{eq:ewkfactors}
\parx\siv_{\EW}(\vec\mu)&=& \left[1+\kappa^{(V)}_{\EW}(x,\vec\mu)\right]
\parx\siv_{\LO\,\QCD}(\vec\mu),
\eeqar
where EW stands for NLO~EW or nNLO~EW.
At variance with \refeq{eq:kfactors}, here
the EW $\kappa$-factors are defined by taking the factorized LO cross section 
at the same QCD scales, 
$\vec\mu=(\mur,\muf)$, as in the higher-order EW prediction. In this way, 
since QCD scale variations at LO~QCD and (n)NLO~EW have almost identical impact,
the relative EW correction is essentially independent of $\vec\mu$. Thus, in practice,
$\kappa_\EW$ can be computed at the fixed reference scale,
\beqar
\kappa^{(V)}_\EW(x,\vec\mu)\simeq
\kappa^{(V)}_\EW(x,\vec\mu_0)=
\kappa^{(V)}_\EW(x),
\eeqar
while the scale dependence of $\siv_{\EW}$ is generated through 
$\siv_{\LO\,\QCD}(\vec\mu)$ in \refeq{eq:ewkfactors}.
Moreover, the EW correction factor $\kappa^{(V)}_\EW$ is rather
insensitive to the choice of PDF set
as long as it is derived from cross sections that are based on the same PDFs.
Analogously to \refeq{eq:ew0}, nNLO~EW correction factors are split into 
a full NLO part and an NNLO Sudakov part,
\beqar\label{eq:EWcorsplitting}
\kappa^{(V)}_{\nNLO\,\EW}(x)=
\kappa^{(V)}_{\NLO\,\EW}(x)+\kappa^{(V)}_{\NNLO\,\Sud}(x).
\eeqar

At NLO EW, all relevant contributions of $\ord{(\alpha^2\alphaS)}$ are included.
In the $q\bar q$ channel, and in all crossing-related channels, 
they comprise the following types of corrections:
\renewcommand{\labelenumi}{(a.\arabic{enumi})}
\begin{enumerate}
%(a.1) 
\item virtual EW corrections to $q\bar q\to Vg$;
%(a.2) 
\item $q\bar q\to Vg\gamma$ photon bremsstrahlung;
%(a.3) 
\item virtual QCD corrections to $q\bar q\to V\gamma$, which are needed to cancel soft-gluon 
singularities from (a.2) if the final-state QCD partons
are allowed to become unresolved;
%
%(a.4) 
\item $q \bar q\to Vq'\bar q'$ bremsstrahlung, which contributes at
$\ord(\alpha^2\alphaS)$ through the interference of $\ord(eg_S^2)$ and
$\ord(e^3)$ tree amplitudes in the same-flavour case, $q=q'$;

\end{enumerate}
Formally at $\ord(\alpha^2\alphaS)$ in perturbation theory also the following contributions appear and are not included:
\begin{enumerate}
  \setcounter{enumi}{4}
%(a.5) 
\item 
\label{it:gammaindqbrem}
$\gamma q\to V q g$ photon-induced
quark-bremsstrahlung\footnote{Note that, in spite of the fact that we
present them as separate terms in \refeq{eq:th1}, $\gamma$-induced
contributions and NLO EW corrections to $pp\to V+$\,jet are interconnected
at $\ord(\alpha^2\alphaS)$.},
at $\ord(\alpha^2\alphaS)$,
which plays the dual role of NLO EW correction to the $q\bar q\to Vg$ channel 
and NLO QCD correction to the $\gamma q\to Vq$ channel.
As discussed in \refse{se:gammaind}, given the relatively small impact 
of $\gamma q\to V q$ processes at $\ord(\alpha^2)$, photon-induced contributions
of $\ord(\alphaS\alpha^2)$ will not be included in the present study;
%(a.6) 
\item real-boson emission, i.e.~$pp\to VV'j$, 
contributes at $\ord(\alpha^2\alphaS)$. 
As discussed in \refse{se:dibosons}, in order to avoid double
counting with diboson production,
such contributions should be treated as separate background samples and 
not as part of the EW corrections to $pp\to Vj$.
\end{enumerate}

\newcommand\slog[1]{\ln^{#1}\left(\frac{Q^2}{M^2}\right)}
\newcommand\slogij[1]{\ln^{#1}\left(\frac{Q_{ij}^2}{M^2}\right)}
\newcommand\sreslog[1]{\ln^{#1}\left(\frac{Q^2}{(\xi M)^2}\right)}
\newcommand\sreslogij[1]{\ln^{#1}\left(\frac{Q_{ij}^2}{(\xi M)^2}\right)}
\newcommand\sco[2]{C_{#1}^{(#2)}}

At very high transverse momentum, EW corrections are strongly enhanced by Sudakov effects, 
and the inclusion of higher-order Sudakov logarithms 
becomes mandatory in order to achieve few-percent level accuracy.
In the high-$p_\rT$ regime, where
all energy scales are far above the weak-boson mass scale,
higher-order virtual EW corrections to hard scattering cross sections
can be described by means of 
resummation formulas of the  type\footnote{Here, in order to discuss qualitative
features of Sudakov logarithms, we adopt a generic and rather
schematic representation of the asymptotic high-energy limit. In particular,
we do not consider some aspects, such as the helicity dependence of the corrections
or SU(2) soft-correlation effects. However, in the numerical analysis 
 all relevant aspects are consistently included.} \linebreak \cite{Jantzen:2005az,Chiu:2007dg}
\beqar
\label{eq:Sudakov}
\rd\sigma_\EW
= \exp\left\{\int_{M_W^2}^{Q^2}\frac{\rd t}{t}
\left[ \int_{M_W^2}^{t}\rd\tau 
\frac{\gamma(\alpha(\tau))}{\tau} \right.\right.\\ \left.\left.
+\chi(\alpha(t))+\xi\left(\alpha(M_W^2)\right)\right]
\right\}
\rd\sigma_\hard, \nonumber
\eeqar
where $\gamma$, $\chi$ and $\xi$ are anomalous dimensions depending on the EW quantum numbers of the 
scattering particles.
The hard cross section has the form 
\beqar\label{eq:hardME}
\rd\sigma_\hard=\left[
1+\frac{\alpha}{\pi}\delta^{(1)}_\hard
+\left(\frac{\alpha}{\pi}\right)^2\delta^{(2)}_\hard+\dots
\right]\rd\sigma_{\mathrm{Born}},
\eeqar
and the correction factors $\delta^{(k)}_\hard$ are finite in the limit $Q^2/M_W^2\to \infty$, while
EW Sudakov 
and subleading high-energy
logarithms of type $\alpha^m \ln^n\left(Q^2/M_W^2\right)$ 
are factorised in the exponential.
Expanding in $\alpha=\alpha(M^2)$ with 
$\gamma_i(\alpha)=\frac{\alpha}{\pi}\gamma_i^{(1)}+\dots,$ and 
\beqar
\alpha(t)=\alpha\left[1+\frac{\alpha}{\pi}b^{(1)}\ln\left(\frac{t}{M^2}\right)+\dots\right]
\eeqar
yields
\beqar\label{eq:EXPSudakov}
\exp\bigg\{\dots\bigg\}=1+\frac{\alpha}{\pi}\delta^{(1)}_\Sud
+\left(\frac{\alpha}{\pi}\right)^2\delta^{(2)}_\Sud+\dots.
\eeqar
At NLL level, which is the logarithmic accuracy at which NNLO Sudakov effects are known for 
$V+$\,jet  production~\cite{Kuhn:2004em,Kuhn:2005gv,Kuhn:2005az,Kuhn:2007qc,Kuhn:2007cv},
the following types of logarithms are available,
\beqar\label{eq:NLLSudakov}
\delta^{(1)}_\Sud&=&\sum_{i,j}\sco{2,ij}{1}\slogij{2}+\sco{1}{1}\slog{1},
\nonumber\\
\delta^{(2)}_\Sud&=&\sum_{i,j}\sco{4,ij}{2}\slogij{4}
+\sco{3}{2}\slog{3}\nonumber\\&&+\ord\left[\slog{2}\right],\eeqar
where $M=M_W\sim M_Z$, 
$Q^2_{ij}=|(\hat p_i\pm \hat p_j)^2|$ are the various Mandelstam invariants
built from the hard momenta $\hat p_i$ of the $V+$\,jet production process and
$Q^2=Q_{12}^2=\hat s$.

In this work we will employ the explicit NLL Sudakov 
results of~\citeres{Kuhn:2004em,Kuhn:2005gv,Kuhn:2005az,Kuhn:2007qc,Kuhn:2007cv}, which 
have been implemented, in addition to exact NLO QCD+NLO EW amplitudes, in the {\sc
OpenLoops} matrix-element generator~\cite{Kallweit:2014xda,Kallweit:2015dum}.
Let us recall that the results 
of\linebreak \citeres{Kuhn:2004em,Kuhn:2005gv,Kuhn:2005az,Kuhn:2007qc,Kuhn:2007cv} 
are based on the high-energy limit of virtual one- and two-loop corrections 
regularised with a 
fictitious photon mass of order $M_W$. This generates 
logarithms of the form $\alpha^n\ln^k(\hat s/M^2_W)$ that 
correspond to the combination of virtual one- and two-loop EW corrections 
plus corresponding photon radiation contributions 
up to an effective cut-off scale of order $M_W$.
In the case of $V+$\,jet production, for physical observables that are inclusive 
with respect to photon radiation,
this approximation is accurate at the one-percent level~\cite{Kuhn:2005gv,Kuhn:2007cv,Badger:2016bpw}.

In this work we will employ full EW results at NLO and NLL Sudakov logarithms at NNLO. 
In the notation of \refeqs{eq:ewkfactors}{eq:EWcorsplitting}, for fully-differential 
partonic cross sections, this implies
\beqar\label{eq:sudakovkfactors1}
\kappa_{\NLO\,\EW}(\hat s, \hat t)&=&\frac{\alpha}{\pi}\left[\delta^{(1)}_\hard+\delta^{(1)}_\Sud\right],\\
\label{eq:sudakovkfactors2}
\kappa_{\NNLO\,\Sud}(\hat s, \hat t)&=&\left(\frac{\alpha}{\pi}\right)^2\delta^{(2)}_\Sud.
\eeqar

%The quality of the NLL Sudakov approximation for vector-boson $\pT$ distributions
%is studied in~\reffi{fig:EW_error}, where Sudakov logarithms are compared to 
%exact EW corrections at NLO. The observed agreement indicates that the
%Sudakov approximation at NLO works very wells, thereby supporting
%the usage of EW Sudakov logarithms at NNLO.

\subsubsection*{Pure EW uncertainties}
\label{se:pureewunc}

Assuming that the NLL Sudakov approximation at NNLO is comparably 
accurate as at NLO, we can consider unknown Sudakov logarithms beyond NNLO
as the dominant source of EW uncertainty at high $p_\rT$.  
Such Sudakov terms of relative $\ord(\alpha^3)$ can be easily estimated via naive exponentiation,
which implies the following relations between NLO, NNLO and NNNLO terms,
\beqar\label{eq:n3loestimate}
\delta^{(2)}_\Sud&\simeq&\frac{1}{2}\left[\delta^{(1)}_\Sud\right]^2,\nonumber\\
\delta^{(3)}_\Sud&\simeq&\frac{1}{3!}\left[\delta^{(1)}_\Sud\right]^3\simeq\frac{1}{3}\delta^{(1)}_\Sud\,\delta^{(2)}_\Sud.
\eeqar
Based on these relations, we estimate the uncertainty due to unknown high-$p_\rT$ EW effects
beyond NNLO as
\beqar
\label{eq:dkappaEW1}
\delta^{(1)}\kappa^{(V)}_{\nNLO\,\EW}(x) &=& 
\frac{2}{3} \left\vert \kappa^{(V)}_{\NLO\,\EW}(x)\,\kappa^{(V)}_{\NNLO\,\Sud}(x) \right\vert, \nonumber\\
\eeqar
which is an approximate implementation of \refeq{eq:n3loestimate}, obtained 
by  neglecting effects from angular integration,
replacing  $\delta^{(1)}_\Sud$ by the full NLO EW correction,
and multiplying 
the term $\delta^{(3)}_\Sud$ by a factor two, in order to be conservative.

Besides Sudakov exponentiation effects, we introduce 
a second source of uncertainty, defined, at \linebreak nNLO~EW level,  as 5\% of the absolute full NLO EW correction,
\beqar\label{eq:dkappaEW2}
\delta^{(2)} \kappa^{(V)}_{\nNLO\,\EW}(x)=
0.05\, \left\vert \kappa^{(V)}_{\NLO\,\EW}(x)\right\vert.
\eeqar
This type of uncertainty has a twofold motivation.
At high $p_\rT$, 
where Sudakov logarithms dominate,
it accounts for unknown terms of order $\alpha^2\slog{2}$ that
can arise from effects of the form
\beqar\label{eq:dkappaEW2b}
\left(\frac{\alpha}{\pi}\right)^2\delta^{(1)}_\hard\,
\delta^{(1)}_\Sud&=&
\kappa_{\NLO\,\hard}\,
\kappa_{\NLO\,\Sud}\nonumber \\ &\simeq&
\kappa_{\NLO\,\hard}\, \kappa_{\NLO\,\EW}.
\eeqar
In general, the non-Sudakov factor
$\kappa_{\NLO\,\hard}=(\frac{\alpha}{\pi})\delta^{(1)}_\hard$ can amount to
several percent,  e.g.\ due to photon-\linebreak bremsstrahlung effects in highly
exclusive observables.  However, 
for the boson-$p_\rT$ distributions considered in this paper,
where dressed leptons are used,
the quality of the Sudakov approximation
observed in~\reffi{fig:EW_error} 
indicates that $\kappa_{\NLO\,\hard}$ is very small.
Nevertheless, to be conservative, in \refeq{eq:dkappaEW2} 
we choose a prefactor that allows for effects as large as $\kappa_{\NLO\,\hard}=5\%$.

As a second motivation,
the uncertainty~\refpar{eq:dkappaEW2} accounts also for NNLO
effects of type $\left(\frac{\alpha}{\pi}\right)^2\delta^{(2)}_\hard$,
which can become relevant in the case where hard contributions dominate.
In this situation, \refeq{eq:dkappaEW2} amounts to a bound 
on hard NNLO effects,
\beqar\label{eq:dkappaEW2c}
\left(\frac{\alpha}{\pi}\right)^2\delta^{(2)}_\hard\le
0.05\, \kappa_{\NLO\,\EW} \simeq 
0.05 \left(\frac{\alpha}{\pi}\right)\delta^{(1)}_\hard,
\eeqar
which corresponds to $\delta^{(2)}_\hard \le \frac{0.05\pi}{\alpha}\delta^{(1)}_\hard\simeq
20\,\delta^{(1)}_\hard$. This limit should be conservative enough to 
hold also in situations where the NLO hard correction is
accidentally small with respect to its NNLO counterpart.

In order to account for the limitations of the Sudakov approximation at nNLO
in a sufficiently conservative way, we introduce an additional source of uncertainty
defined as the difference between the rigorous NLL Sudakov approximation
\refpar{eq:sudakovkfactors2} and a naive exponentiation of the full NLO EW correction, 
\beqar
\label{eq:dkappaEW3} 
\delta^{(3)}\kappa^{(V)}_{\nNLO\,\EW}(x) &=& 
\left\vert \kappa^{(V)}_{\NNLO\,\Sud}(x)-
\frac{1}{2}[\kappa^{(V)}_{\NLO\,\EW}(x)]^2 \right\vert. \nonumber \\
\eeqar
This expression provides an estimate of the typical size of 
terms of type $\left[\delta^{(1)}_\hard\right]^2$ and $\delta^{(1)}_\hard\times\delta^{(1)}_\Sud$.

In correspondence to the nNLO uncertainties of \linebreak Eqs.~\refpar{eq:dkappaEW1}, \refpar{eq:dkappaEW2} and \refpar{eq:dkappaEW3},
at NLO\,EW we introduce uncertainties $\delta^{(i)}\kappa_{\NLO\,\EW}$, defined as
\beqar
\label{eq:EWuncertainties3}
\delta^{(1)}\kappa^{(V)}_{\NLO\,\EW}(x) &=& 
\frac{2}{2} \left[\kappa^{(V)}_{\NLO\,\EW}(x)\right]^2,\nonumber\\
\delta^{(2)}\kappa^{(V)}_{\NLO\,\EW}(x) &=& 2000\times\left(\frac{\alpha}{\pi}\right)^2 \simeq  1.2\% ,\nonumber\\
\delta^{(3)}\kappa^{(V)}_{\NLO\,\EW}(x) &=& 0. 
\eeqar
Here the first term is the direct transposition of Eq.~\refpar{eq:dkappaEW1} to NLO. It 
accounts for the unknown $\ord{(\alpha^2)}$ Sudakov terms $\delta_\Sud^{(2)}$ 
in~\refeq{eq:n3loestimate} supplemented with an extra factor of two.
As explained in the following, the second uncertainty in~\refeq{eq:EWuncertainties3}
is the NLO counterpart of 
the nNLO\,EW uncertainty \refpar{eq:dkappaEW2}. The latter 
accounts for unknown $\ord{(\alpha^2)}$ terms of
type \refpar{eq:dkappaEW2b} and \refpar{eq:dkappaEW2c}, which correspond 
to the intrinsic uncertainty of the employed Sudakov approximation at nNLO.
At NLO\,EW the situation is different, since the calculations are exact, \ie there are no unknown 
terms of $\ord(\alpha)$. Thus, we assume an uncertainty 
$\delta^{(2)}\kappa^{(V)}_{\NLO\,\EW}(x)$ of type 
$\left(\frac{\alpha}{\pi}\right)^2\delta^{(2)}_\hard$.
We do not consider additional
uncertainties of type $\left(\frac{\alpha}{\pi}\right)^2\delta^{(2)}_\Sud$ 
since they are already covered by the first term in~\refeq{eq:EWuncertainties3}.
As estimate of the size of the unknown $\delta^{(2)}_\hard$ coefficient,
following the discussion of~\refeq{eq:dkappaEW2c}, we impose a very generous upper bound to the ratio 
between $\delta^{(2)}_\hard$ and $\delta^{(1)}_\hard$. To be conservative,
at NLO\,EW we adopt a ten times looser bound as compared to nNLO\,EW, \ie
we require $\delta^{(2)}_\hard\lsim 200\,\delta^{(1)}_\hard$.
Finally, setting $\delta^{(1)}_\hard=10$, which corresponds to the typical size of non-Sudakov enhanced
EW corrections, 
$10\times\left(\frac{\alpha}{\pi}\right)\simeq 2\%$, we arrive at 
$2000\times\left(\frac{\alpha}{\pi}\right)^2$ for the second term in~\refeq{eq:EWuncertainties3}. 
The third uncertainty in~\refeq{eq:EWuncertainties3} 
is set to zero, since there is no counterpart 
of~\refeq{eq:dkappaEW3} at NLO.

%The rough estimate \refpar{eq:dkappaEW1} of higher-order EW effects,
%based on naive exponentiation,
%can be validated at NLO by comparing the corresponding estimate
%$\delta^{(1)}\kappa^{(V)}_{\NLO\,\EW}(x)$ 
%against the known NLL Sudakov results at NNLO.
%This is illustrated in \reffi{fig:EW_error}, which demonstrates that 
%\refeq{eq:EWuncertainties3} (green band)
%provides a fairly realistic estimate of nNLO EW corrections.
%The expected effects beyond nNLO, estimated according to~Eqs.~\refpar{eq:dkappaEW1}, \refpar{eq:dkappaEW2} 
%and \refpar{eq:dkappaEW3},
%turn out to be around $\pm 5\%$ in the multi-TeV tails. 

Similarly as for QCD uncertainties, the EW uncertainties 
in Eqs.~\refpar{eq:dkappaEW1}, \refpar{eq:dkappaEW2}, \refpar{eq:dkappaEW3} and 
\refeq{eq:EWuncertainties3}, can be parametrised in terms of 
nuisance parameters $\vec \eps_\EW$ and combined via
\beqar\label{eq:ewunccomb}
\parx\siv_{\EW}(\vec\eps_\EW,\vec\eps_\QCD)&=& 
\left[\kappa^{(V)}_{\EW}(x)
+\sum_{i=1}^3\eps^{(V)}_{\EW,i}\,\delta^{(i)}\kappa^{(V)}_\EW
(x)\right]\nonumber\\&&{}\times
\parx\siv_{\LO\,\QCD}(\vec\eps_\QCD),
\eeqar
where EW stands for NLO~EW or nNLO~EW.
The nuisance parameters $\eps^{(V)}_{\EW,i}$
should be Gaussian distributed with one standard deviation
corresponding to the range $\eps^{(V)}_{\EW,i}\in[-1,+1]$,
and their variations should be applied in a correlated way across $p_\rT$-bins.
Since the first uncertainty \refpar{eq:dkappaEW1} reflects the universal exponentiation 
properties of Sudakov EW corrections, which permits to predict the magnitude and size of the
dominant higher-order corrections for each individual processes,
this variation should be correlated across processes,
i.e.~a single  nuisance parameter should be used,
\beqar
\eps^{(W^\pm)}_{\EW,1}=
\eps^{(Z)}_{\EW,1}=
\eps^{(\gamma)}_{\EW,1}=
\eps_{\EW,1}.
\eeqar
In contrast, the 
remaining EW uncertainties \refpar{eq:dkappaEW2} and \refpar{eq:dkappaEW3}
describe subleading NNLO effects whose sign, magnitude and 
process dependence are unknown. Thus these uncertainties should be treated as uncorrelated, i.e.~independent 
nuisance parameters $\eps^{(V)}_{\EW,2}$ and $\eps^{(V)}_{\EW,3}$ should be used for each process. 

\subsubsection*{Numerical results}

%%%%%%%%%%%%%%%%%%%%
\begin{figure*}[tp] %[htbp!]   
\centering
  \includegraphics[width=\ratiotextwidthbig\textwidth]{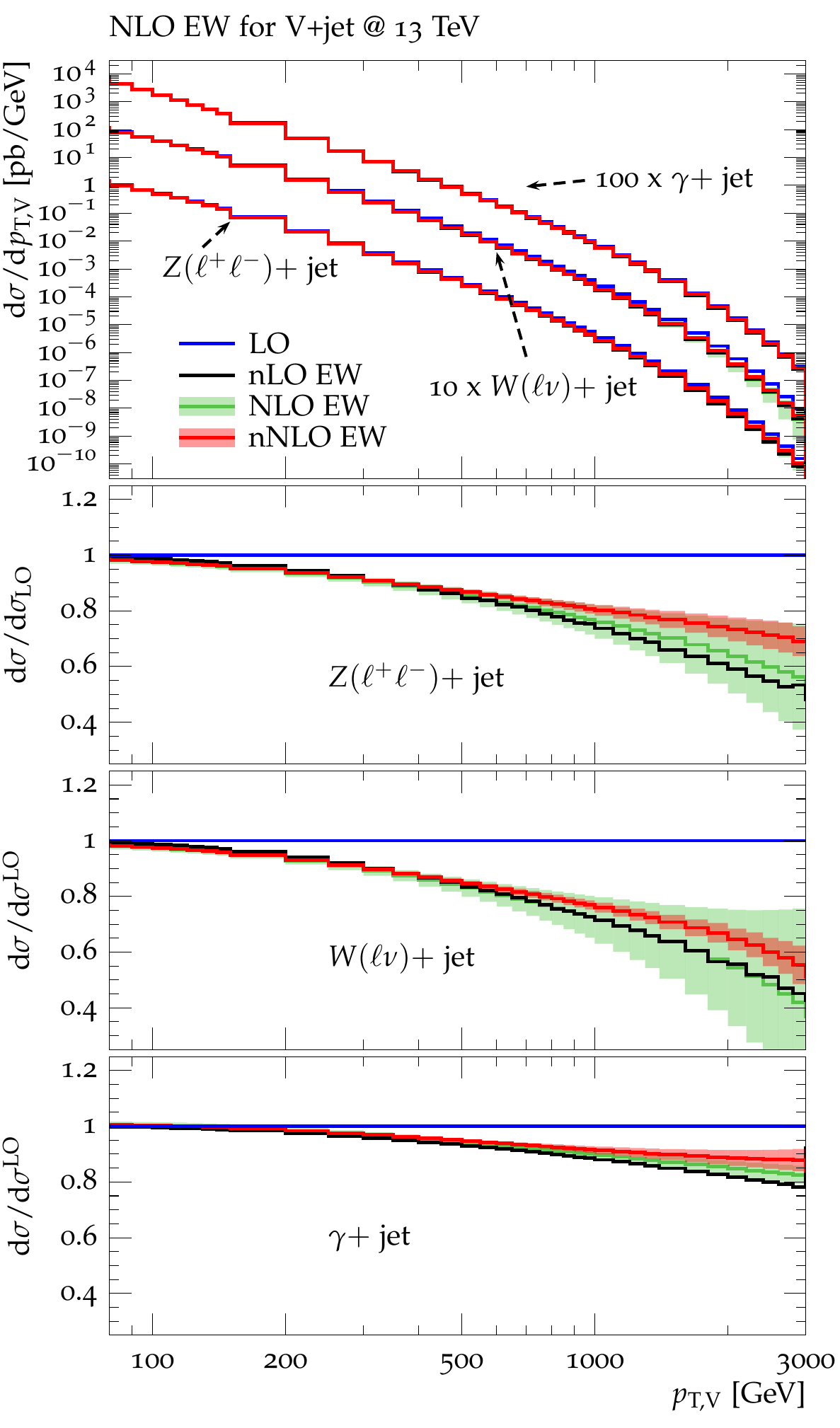}
\caption{
Higher-order EW predictions and uncertainties for different $pp\to V+$\,jet 
processes at 13\,TeV.
The main frame displays absolute predictions at LO (blue), NLO EW (green) and nNLO EW (red), 
as well as NLL Sudakov logarithms at NLO (black), which are denoted as nLO EW.
In the ratio plots all results are normalised to LO.
Uncertainties at nNLO EW (red band) 
are evaluated by combining in quadrature the corresponding 
variations $\delta^{(i)}\kappa^{(V)}_{\nNLO\,\EW}$ 
as defined
in Eqs.~\refpar{eq:dkappaEW1}, \refpar{eq:dkappaEW2} and \refpar{eq:dkappaEW3}
and for  $\delta^{(i)}\kappa^{(V)}_{\NLO\,\EW}$  in \refeq{eq:EWuncertainties3}.
}
\label{fig:EW_error}
\end{figure*}
%%%%%%%%%%%%%%%%%%%%
%\clearpage

%%%%%%%%%%%%%%%%%%%%
\begin{figure*}[t]   
\centering
\vspace{20mm}
 \includegraphics[width=\ratiotextwidth\textwidth]{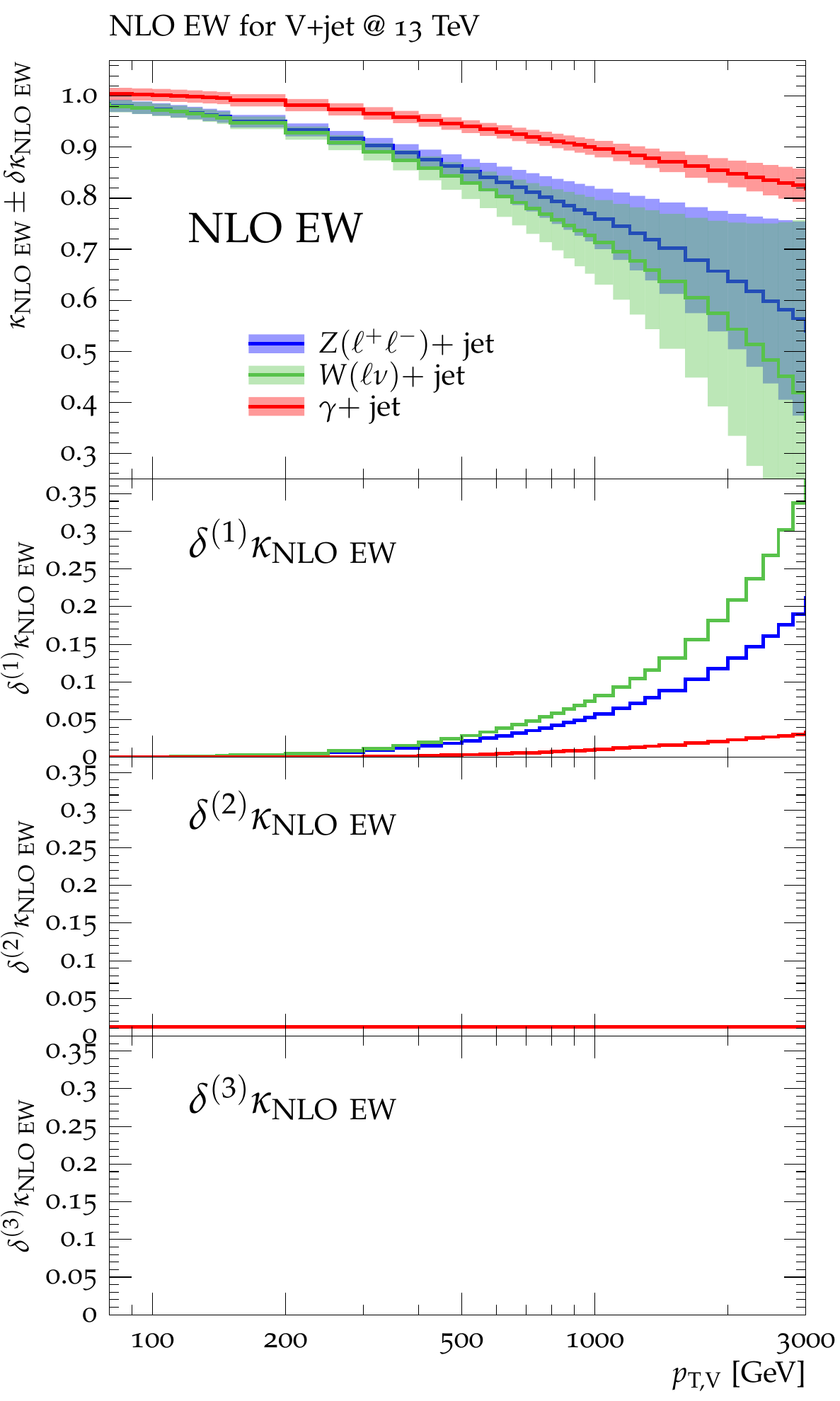}
  \includegraphics[width=\ratiotextwidth\textwidth]{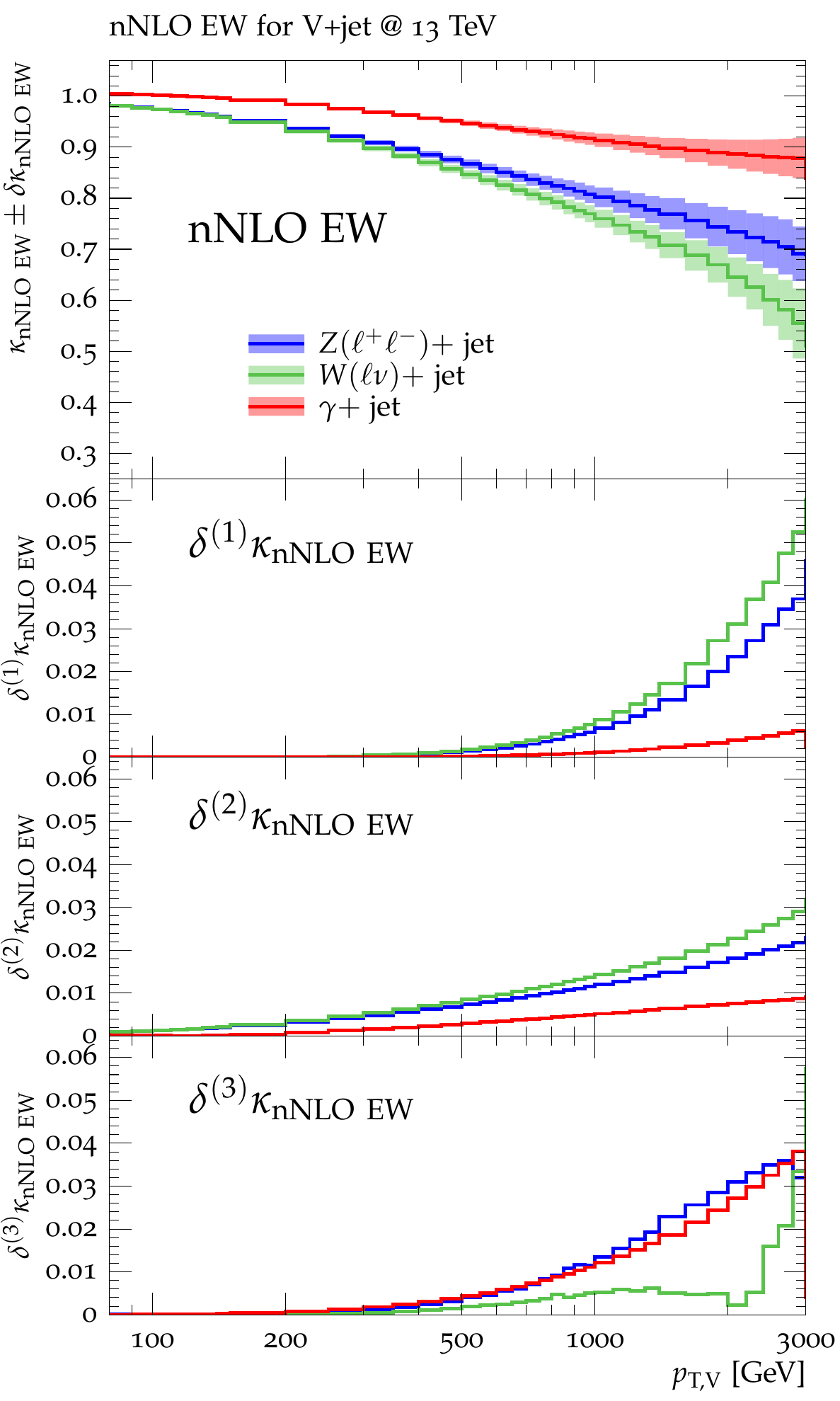}
\caption{
NLO EW (left) and nNLO EW (right) $\kappa$-factors for the various $pp\to V+$\,jet processes at 
13\,TeV. 
The individual uncertainties $\delta^{(i)}\kappa^{(V)}_{\EW}$
are defined in Eqs.~\refpar{eq:dkappaEW1}, \refpar{eq:dkappaEW2} and \refpar{eq:dkappaEW3},
at nNLO and in \refeq{eq:EWuncertainties3} at NLO.
The bands in the main frame correspond to their combination in quadrature.
}
\label{fig:EW_corr_error}
\end{figure*}
%%%%%%%%%%%%%%%%%%%%

\begin{figure*}[tp]   
\centering
  \includegraphics[width=\ratiotextwidth\textwidth]{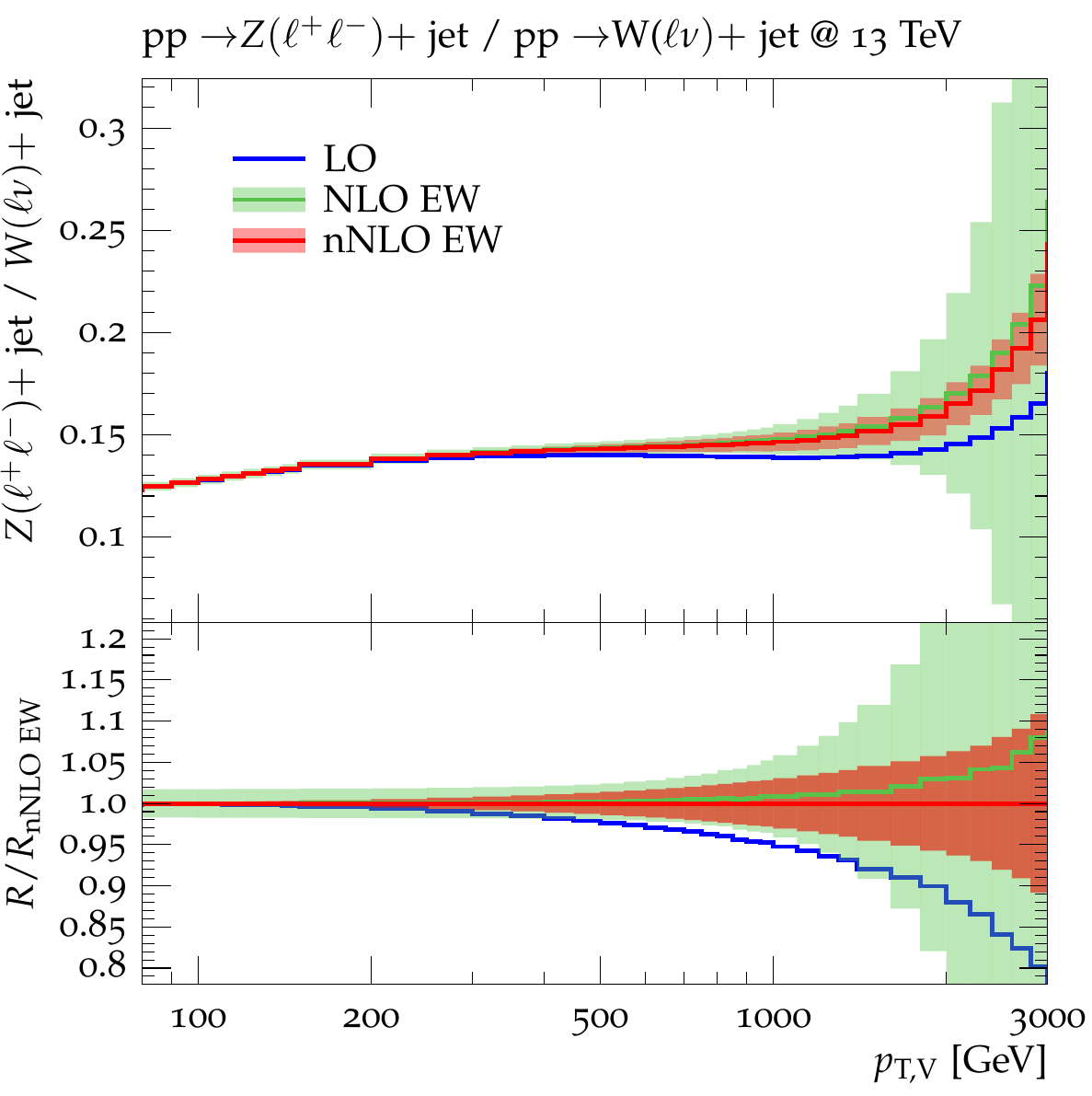}
    \includegraphics[width=\ratiotextwidth\textwidth]{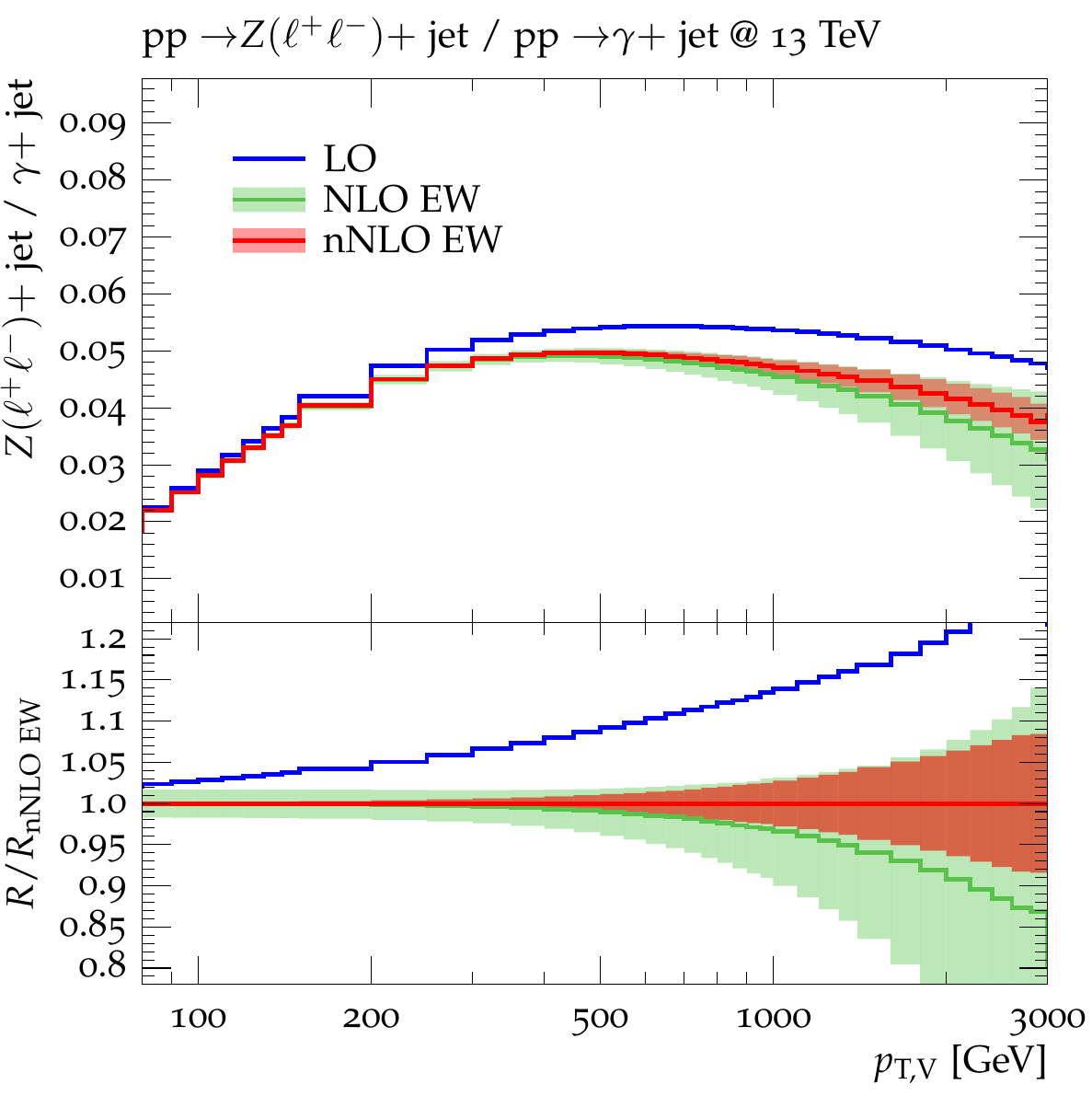}\\[6mm]
        \includegraphics[width=\ratiotextwidth\textwidth]{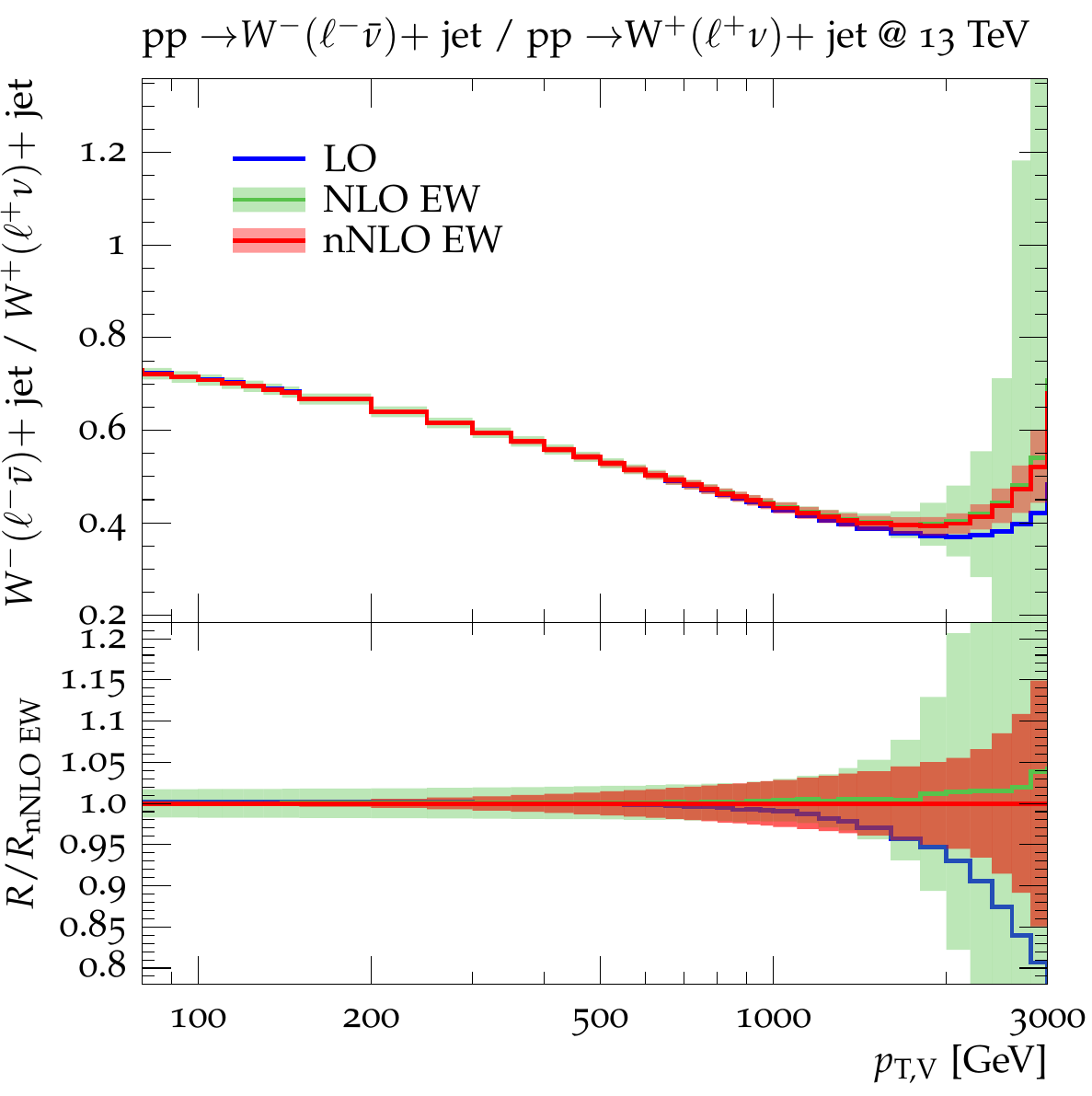}
    \includegraphics[width=\ratiotextwidth\textwidth]{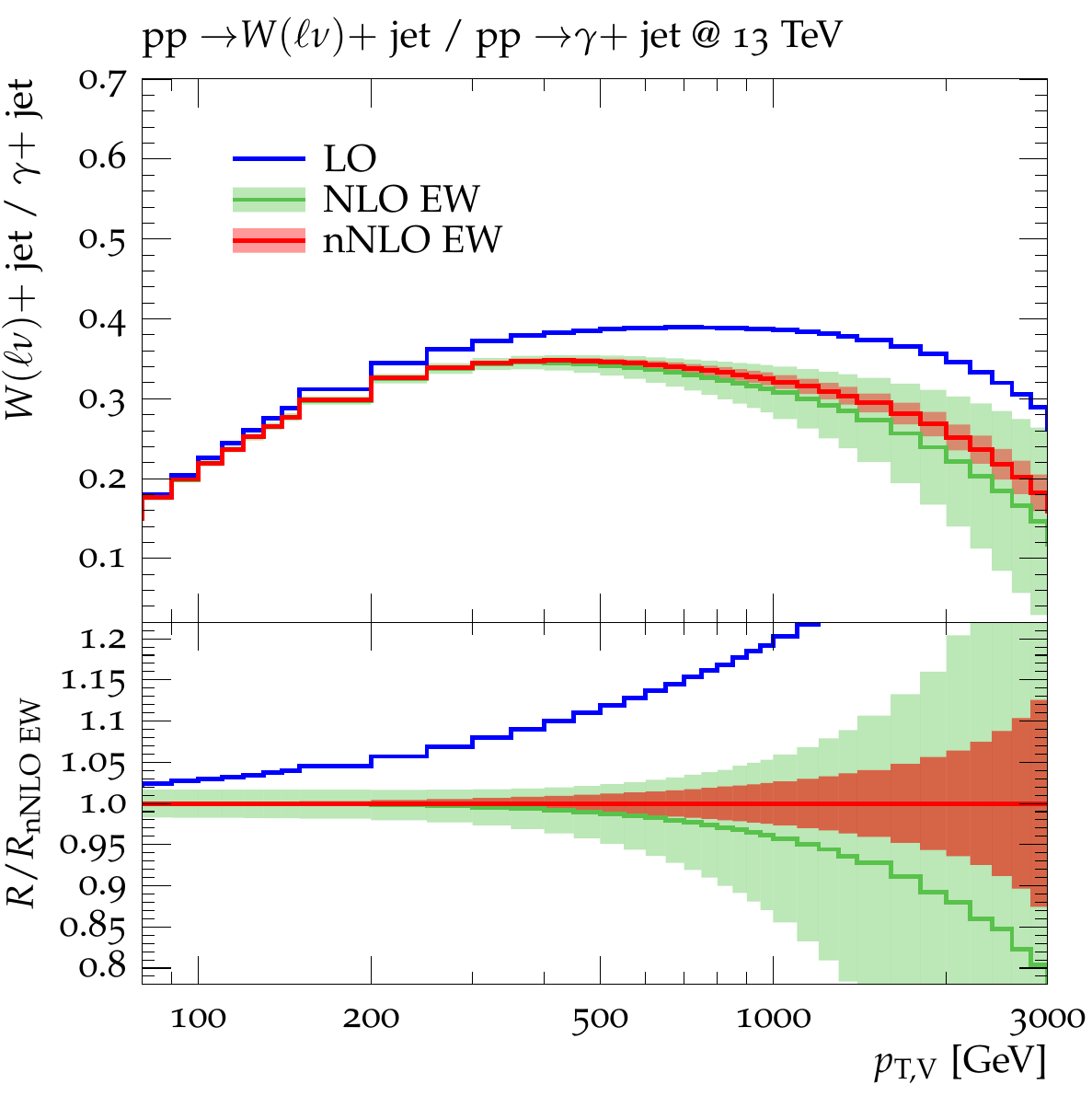}\\[6mm]
        \includegraphics[width=\ratiotextwidth\textwidth]{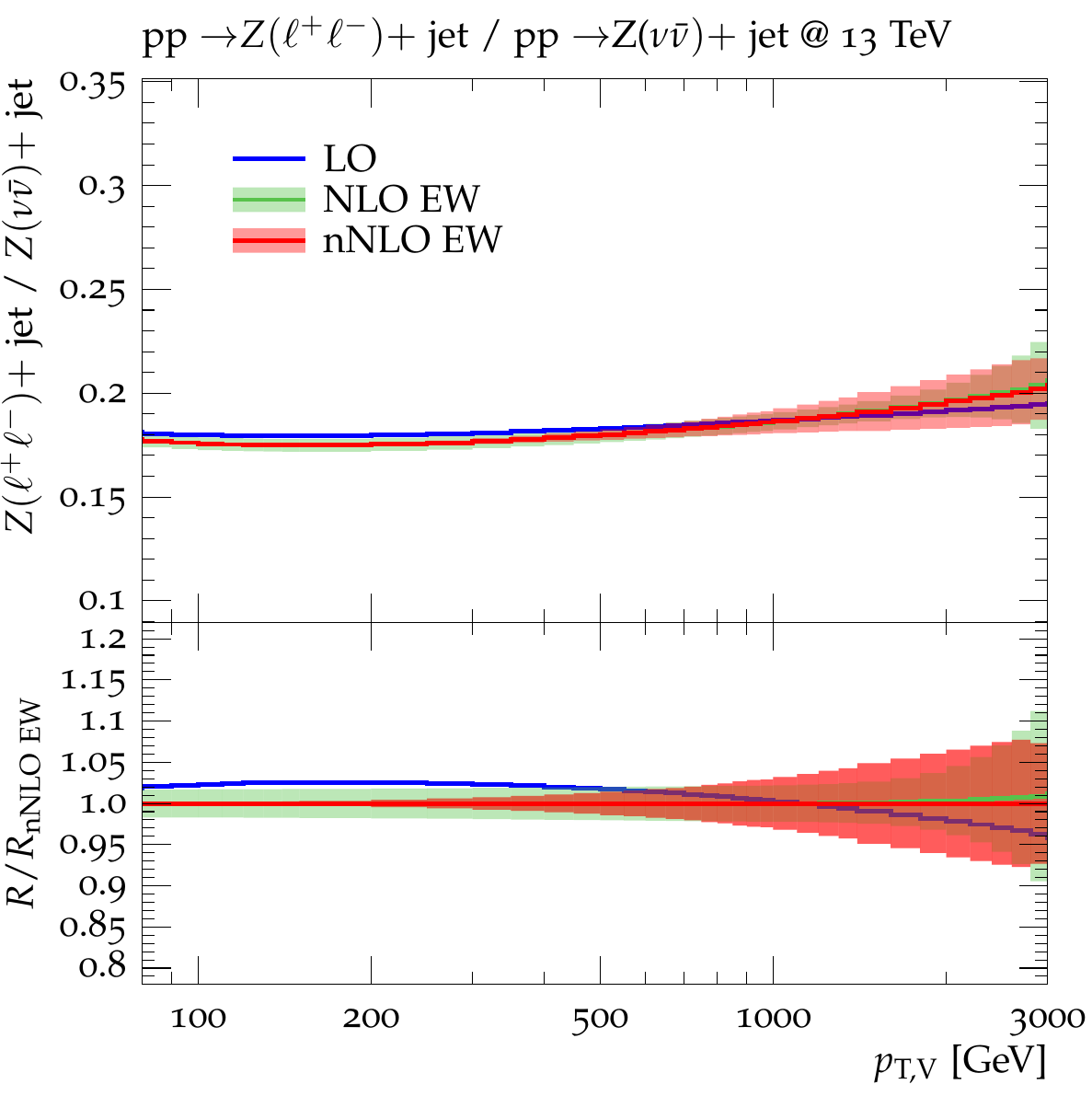}\\[6mm]
\caption{
Ratios of $p_\rT$-distributions for various $pp\to V$+jet processes at 
LO, NLO EW and nNLO EW accuracy. 
Relative uncertainties normalised to nNLO EW
are illustrated in the lower frames.
The bands correspond to a combination (in quadrature) 
of the three EW uncertainties $\delta^{(i)}\kappa^{(V)}_{\EW}$
defined in Eqs.~\refpar{eq:dkappaEW1}, \refpar{eq:dkappaEW2} and \refpar{eq:dkappaEW3}
at nNLO and in \refeq{eq:EWuncertainties3} at NLO.
As discussed in the text, the uncertainty $\delta^{(1)}\kappa^{(V)}_{\EW}$
is correlated amongst processes, while the effect of $\delta^{(2)}\kappa^{(V)}_{\EW}$
and $\delta^{(3)}\kappa^{(V)}_{\EW}$ in the numerator and denominator of ratios
is kept uncorrelated, \ie
added in quadrature.
}
\label{fig:ratios_ew}
\end{figure*}

Predictions for $V$+jet distributions and their ratios at LO,  $\NLO\,\EW$ and $\nNLO\,\EW$
are presented in \linebreak \reffis{fig:EW_error}{fig:ratios_ew} as well as in
\reffis{fig:app_EW_error}{fig:app_ratios_ew} (see \linebreak \refapp{app:unc}).
In Figures~\ref{fig:EW_corr_error} and \mbox{\ref{fig:app_EW_error}--\ref{fig:app_ratios_ew}}, 
the EW uncertainties defined in~\refeq{eq:dkappaEW1}, \refeq{eq:dkappaEW2}, and \refeq{eq:dkappaEW3}
are shown separately,
while in Figures \ref{fig:EW_error} and \ref{fig:ratios_ew}
they are combined in quadrature.

Contrary to the case of QCD corrections, higher-order EW effects have a
significant impact on the shapes of $\pT$ distributions as well as a
pronounced dependence on the scattering process.  This behaviour is mainly due to
the $\pT$ dependence of  EW Sudakov logarithms and their dependence on the SU(2) charges of the
produced vector bosons.

As can be seen in \reffi{fig:EW_error}, the vector-boson $\pT$ spectra
receive negative EW corrections that grow with $\pT$ and become very sizable
in the tails.  At the TeV scale, NLO EW effects reach 20--50\% for $Z$+jet
and $W$+jet production, and 10--15\% for $\gamma$+jet production.  As
expected from exponentiation, NNLO Sudakov logarithms have positive sign. 
Thus they compensate in part the impact of NLO EW corrections.

In \reffi{fig:EW_error} exact NLO EW results are also compared to the NLL
Sudakov approximation at the same order, denoted as $\nLO\,\EW$.  The
observed agreement indicates that the Sudakov approximation at NLO works
very well, thereby supporting the usage of EW Sudakov logarithms at $\NNLO$. 
Moreover, the fact that $\nNLO\,\EW$ results are well consistent with \NLO
predictions supplemented by the corresponding uncertainties \refpar{eq:EWuncertainties3} 
provides an
important confirmation of the goodness of the proposed approach for the
estimate of EW uncertainties.

The importance of NLO and $\nNLO\,\EW$ corrections for different processes 
and the role of individual uncertainties is shown in more detail in 
\reffi{fig:EW_corr_error}. Regarding the size of EW uncertainties we observe that 
the inclusion of  $\nNLO\,\EW$ corrections is crucial in order to achieve few-percent 
accuracy in the tails, while uncertainties at $\NLO\,\EW$ can be as large as 10\% or beyond.

%%%%%%%%%%%%%%%%%%%%
%%%%%%%%%%%%%%%%%%%%

As shown in \reffi{fig:ratios_ew},
the various ratios of $\pT$ distributions and their shape 
receive significant EW corrections, with the
largest effects observed in the \linebreak $Z(\ell^+\ell^-)/\gamma$ and $W/\gamma$
ratios.  In these ratios the remaining combined EW uncertainties are at the
level of few percent in the TeV range, reaching about $5\%$ for $p_{{\rm T},
V}\simeq 2$ TeV.  
Interestingly, also the $Z(\ell^+\ell^-)/Z(\nu\bar\nu)$ 
and \linebreak  $W^-/W^+$ ratios receive non-negligible EW corrections.  
%
%This is due,
%respectively, to corrections of QED type and to
%mixed QCD-EW interference
%contributions. At very high
%$\pT$, the latter give relevant (negative) contributions in $W^+$+jet production 
%but less in $W^-$+jet production.  
%
In the case of the $W^-/W^+$ ratio
this is due to the behaviour of mixed QCD--EW interference contributions at
high $\pT$, which yield relevant (negative) contributions in $W^+$+jet
production but less in $W^-$+jet production.  As for the
$Z(\ell^+\ell^-)/Z(\nu\bar\nu)$ ratio, the observed EW effects can be
attributed to $\pT$-migration effects induced by QED radiation off leptons. 
At moderate $p_{\rT,Z}$, the invariant mass of photon-lepton pairs 
that lie inside the recombination cone $\Delta R_{\ell\gamma}<0.1$ 
is well below $M_Z$. Thus a significant fraction of the
$Z\to \ell^+\ell^-\gamma$ phase space does not undergo 
photon-lepton recombination, and photon radiation results in 
a negative mass and momentum shift for the $\ell^+\ell^-$
system. The $Z$-mass shift 
is typically not sufficient to push 
$Z\to \ell^+\ell^-\gamma$  events outside the inclusive  $m_{\ell\ell}$
window defined in \refse{se:cutsnadobs}. However,
the reduction of the reconstructed $p_{\rT,\ell\ell}$ results in a negative
correction to the $Z(\ell^+\ell^-)/Z(\nu\bar\nu)$ ratio.  Vice versa, 
for $p_{\rT,Z}\gsim 1$\,TeV the recombination cone $\Delta R_{\ell\gamma}<0.1$ 
covers photon-lepton invariant masses up to $p_{\rT,Z}\Delta_{\ell\gamma}> M_Z$, \ie beyond the
$Z\to \ell^+\ell^-\gamma$ phase space. As a result, $p_{\rT,\ell\ell}$
starts capturing a non-negligible
amount of ISR QED radiation, which results in a positive shift of 
$p_{\rT,\ell\ell}$ and
thus in a positive correction to the $Z(\ell^+\ell^-)/Z(\nu\bar\nu)$
ratio.  Note, that the quantitative
impact of such corrections depends on the choice of the $m_{\ell\ell}$ mass window. 
Thus, for a consistent implementation of the predictions presented in this
study it is crucial to reweight MC samples using the $m_{\ell\ell}$ window defined in
\refse{se:cutsnadobs}.  Moreover, in order to guarantee a consistent extrapolation of QED
radiative effects to the $m_{\ell\ell}$  window employed in experimental analyses, it
is mandatory to employ MC samples that account for QED radiation off
leptons.

\subsection{Photon-induced production and QED effects on PDFs}
\label{se:gammaind}

Higher-order QCD and EW calculations for $pp\to V+$\,jet require PDFs at a
corresponding accuracy level, \ie including also QED corrections.
The effect of QED interactions on parton densities is twofold.
Firstly they introduce a photon parton distribution and so open up
partonic channels such as $\gamma q \to V q'$.
Secondly they modify the quark (and even gluon) PDFs both through QED
effects in the initial conditions and especially in the DGLAP
evolution.

Photon-induced $V+$\,jet production is accounted for by the term $\parx\siv_{\gamma-{\rm ind.}}$
in eq.~(\ref{eq:th1}). It might become relevant in the TeV range, especially in the case
of $W+$\,jet production~\cite{Denner:2009gj,Kallweit:2015dum}, where the initial-state photon directly couples to a 
virtual $W$ boson in the $t$-channel.
Such contributions are 
suppressed by a relative factor $\alpha/\alpha_S$ and can be treated at LO, which corresponds to 
$\gamma q\to Vq$  at  $\ord(\alpha^2)$
or, if necessary, at NLO QCD, \ie up to order $\ord(\alpha^2\alphaS)$.
This order comprises:
\renewcommand{\labelenumi}{(b.\arabic{enumi})}
\begin{enumerate}
% (b.1) 
\item virtual QCD corrections to $\gamma q\to Vq$;
% (b.2) 
\item $\gamma g\to Vq\bar q$ quark bremsstrahlung;
% (b.3) 
\item $\gamma q\to Vqg$ gluon bremsstrahlung. 
\end{enumerate}
The latter can also be understood as photon-induced quark-bremsstrahlung NLO EW contribution to the dominant $q\bar q$ channel.
See the contributions of type (a.\ref{it:gammaindqbrem}) in \refse{se:ew}.

%%%%%%%%%%%%%%%%%%%%
\begin{figure*}[t]   
\centering
  \includegraphics[width=\ratiotextwidth\textwidth]{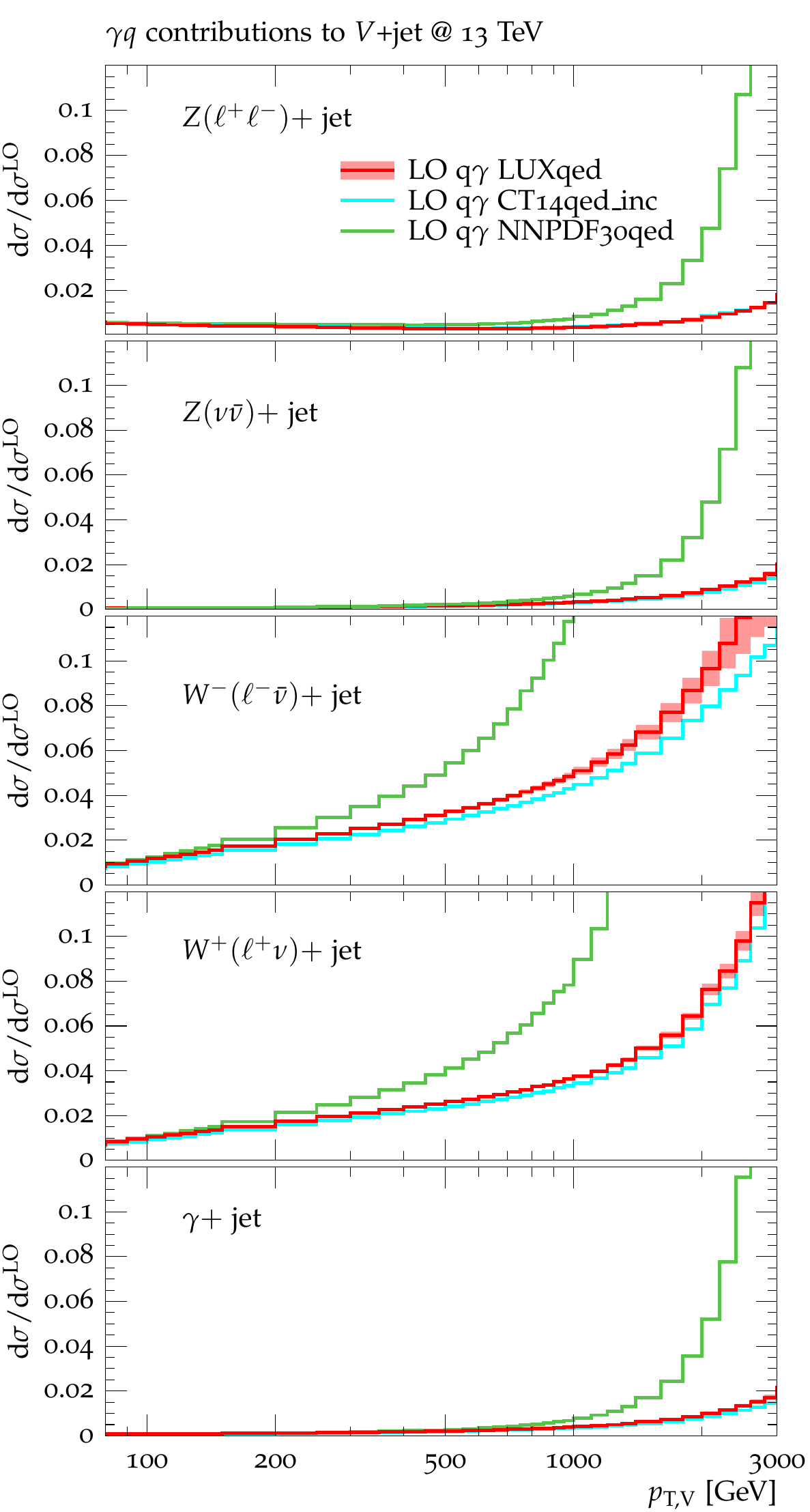}
  \includegraphics[width=\ratiotextwidth\textwidth]{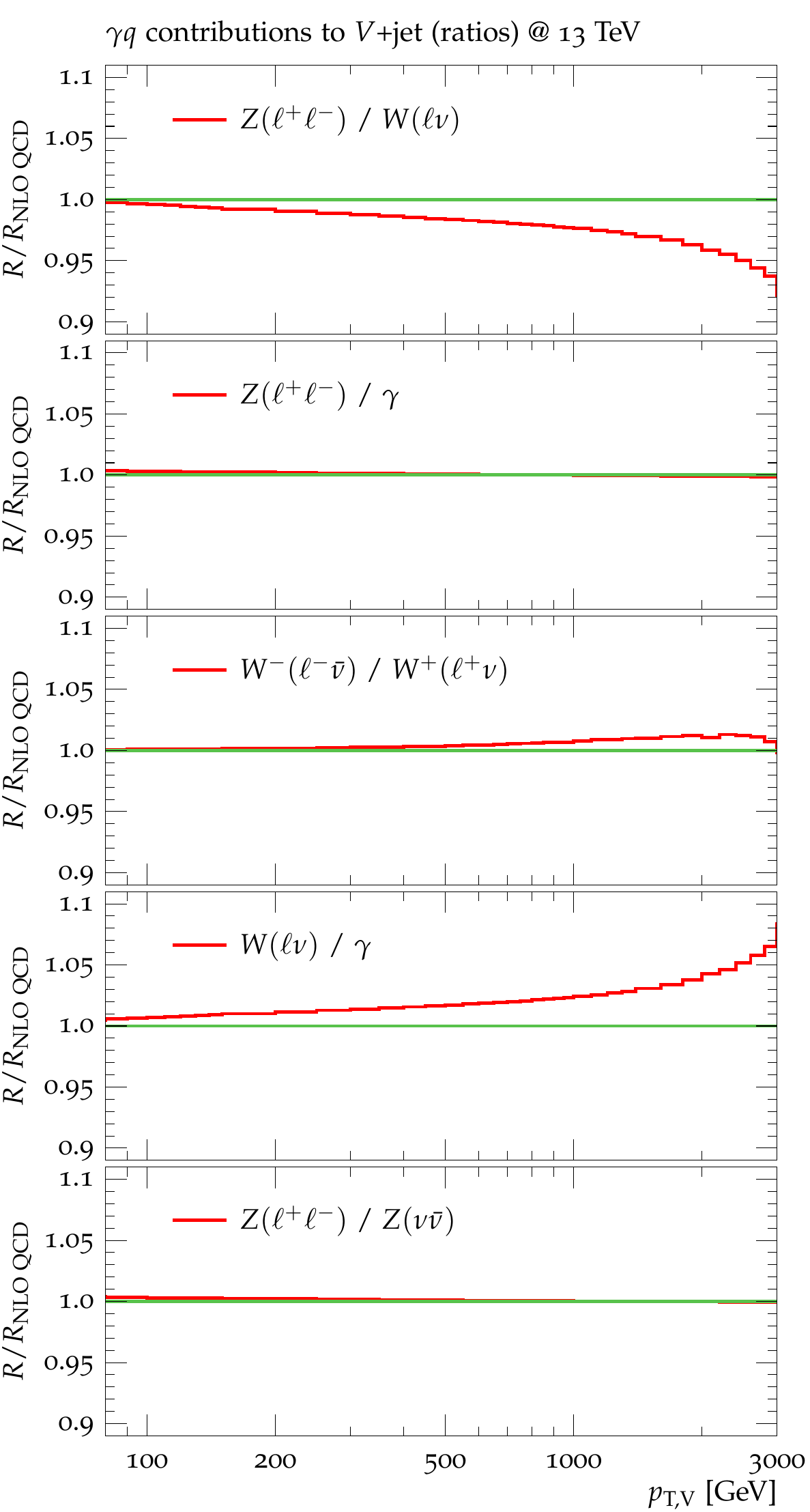}
\caption{
The left plot illustrates the impact of photon-induced contributions at LO, \ie 
$\gamma p\to V+$\,jet at $\ord(\alpha^2)$,  relative to $pp\to V+$\,jet at LO QCD
for different $V+\,$jet processes. Predictions obtained with 
\texttt{LUXqed\_plus\_PDF4LHC15\_nnlo\_100}, \texttt{CT14qed\_inc} and \texttt{NNPDF30qed} 
PDFs are compared.
The error band, shown only for the \texttt{LUXqed\_plus\_PDF4LHC15\_nnlo\_100} prediction, 
reflects PDF uncertainties. 
The right plot shows ratios of 
$V+$jet distributions at NLO QCD with (red) and without (green) $\gamma$-induced contributions 
based on \texttt{LUXqed\_plus\_PDF4LHC15\_nnlo\_100} PDFs.
}
\label{fig:LO_gamma_ind}
\end{figure*}
%%%%%%%%%%%%%%%%%%%%

\reffi{fig:LO_gamma_ind} illustrates the impact of photon-induced
$V+$\,jet production at LO according to three recent PDF sets that 
implement QED corrections.
Effects of the order of
5--10\% for $W$+jet can be observed in the TeV region if
\texttt{CT14qed\_inc}~\cite{Schmidt:2015zda} or \texttt{LUXqed}
PDFs~\cite{Manohar:2016nzj} are used. 
Much larger effects are found with
\texttt{NNPDF30qed}~\cite{Ball:2013hta,Bertone:2016men}.
The impact of photon-induced production to $Z$+jet (and also
$\gamma$+jet) processes  on the other hand is
negligible~\cite{Denner:2011vu,Denner:2012ts}.  

For the description of PDFs and their uncertainties 
we will use the \texttt{LUXqed} PDFs and their intrinsic uncertainties, given that this
set of parton distributions implements a model-independent, data-driven determination of the photon
distribution.
From~\reffi{fig:LO_gamma_ind} one sees that the 
\texttt{LUXqed} uncertainties for $\gamma p\to V+$\,jet
are small.
Using the \texttt{CT14qed\_inc} PDFs, based on a non-perturbative model with limited data-based
constraints for the inelastic contribution, would result in fairly similar photon-induced 
cross sections but somewhat larger
uncertainties (not shown) as compared to \texttt{LUXqed} PDFs. 
The \texttt{NNPDF30qed} parton distributions are model independent and data driven, but are based 
on a different approach from \texttt{LUXqed} for deducing the photon distribution from
data, which results in large uncertainties in the photon-induced
component, of the order of $100\%$ for $pp\to \ell^+\nu_\ell+$\,jet at
$p_{\rT,\ell}=1\,$TeV~\cite{Denner:2009gj}.

We have verified that the NLO QCD corrections to photon-induced production have an impact at the 
percent level relative to $\ord(\alpha^2)$
and can safely be omitted. This implies that $\gamma p\to V+$\,jet can be 
regarded as independent processes. Thus photon-induced $V+$\,jet production can be 
either included through the parton-level predictions provided in this study
or handled as separate background processes through dedicated MC simulations.

\begin{figure*}[t]   
\centering
  \includegraphics[page=1,width=\ratiotextwidth\textwidth]{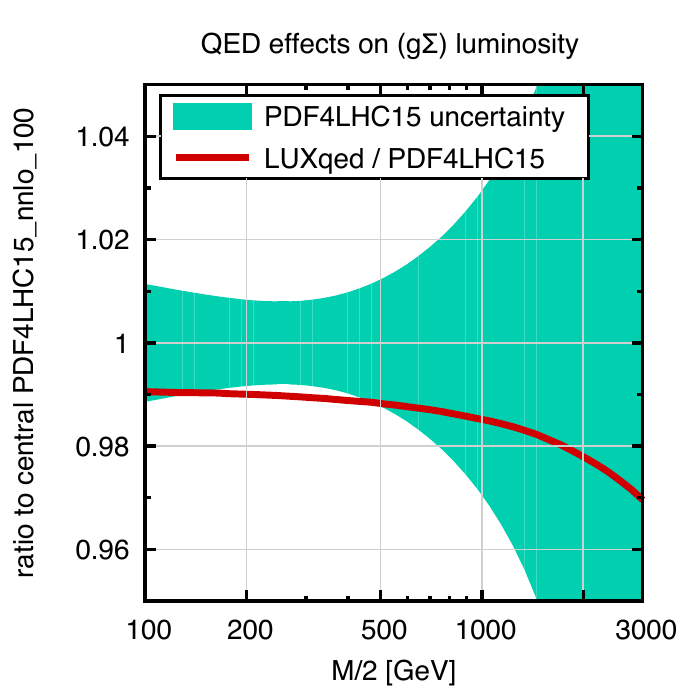}\quad
  \includegraphics[page=2,width=\ratiotextwidth\textwidth]{qed-lumi-effects.pdf}
\caption{
Impact of QED effects on the two partonic luminosities ($g\Sigma$ and
$q\bar q$) that contribute
dominantly to the $Z$+jet cross section. 
The luminosity for producing a system of mass $M$ from two flavours
$a$ and $b$ is defined as
${\cal L}_{ab} = \int_{M^2\!/\!s}^1 \frac{dx}{x} f_{a/p}(x, M^2)
f_{b/p}(\frac{M^2}{xs}, M^2)$ and the $g\Sigma$ luminosity corresponds
to $2 \sum_i ({\cal L}_{gq_i} + {\cal L}_{g\bar q_i})$, while the
$q\bar q$ luminosity corresponds to $2\sum_i {\cal L}_{q_i\bar q_i}$,
where $i$ runs over quark flavours.
The solid red lines correspond to the ratio of luminosities obtained with the
\texttt{LUXqed\_plus\_PDF4LHC15\_nnlo\_100}~\cite{Manohar:2016nzj} and
\texttt{PDF4LHC15\_nnlo\_100}~\cite{Butterworth:2015oua} sets, where a
given $M/2$ value corresponds roughly to the same $p_{\rT,Z}$.
The bands represent the \texttt{PDF4LHC15\_nnlo\_100} uncertainty, shown for
comparison.  }
\label{fig:lumis_PDFQED}
\end{figure*}

Concerning the size of the QED effects on the QCD partons,~\reffi{fig:lumis_PDFQED} examines the two main parton luminosities that
contribute to the $Z+$jet process, i.e.\ $g\Sigma=2 \sum_i ({\cal L}_{gq_i} + {\cal L}_{g\bar q_i})$ (which dominates)
and $q\bar q=2\sum_i {\cal L}_{q_i\bar q_i}$ (which accounts for the remaining $15\%{-}30\%$).
It shows the ratio of these luminosities in
\texttt{LUXqed\_plus\_\linebreak PDF4LHC15\_nnlo} relative to the
\texttt{PDF4LHC15\_nnlo} set on which it is based.
The ratio is given as a function of half the partonic invariant mass, $M/2$,
which is commensurate with the $p_{\rT}$ of the $Z$.

Most of the difference between the \texttt{LUXqed} set and
\texttt{PDF4LHC15\_nnlo} results in~\reffi{fig:lumis_PDFQED} comes
from the QED effects in the DGLAP evolution~\cite{deFlorian:2015ujt},
with photon emission during the evolution reducing the momentum in the
quarks.
This effect reaches about $2\%$ at $2\,$TeV for the $g\Sigma$
luminosity.
There is also a part of the correction associated with the impact of
QED effects on the initial partons.
In the \texttt{LUXqed} set this has been approximated by absorbing
the photon momentum from the gluon distribution in
\texttt{PDF4LHC15\_nnlo} and keeping the quarks unchanged at a scale
of $10\,\GeV$.
This is an ad-hoc procedure, however, insofar as the photon carries
only $\simeq 0.3\%$ of the proton momentum (at a scale of $10\,\GeV$),
the uncertainty associated with the arbitrariness of this choice
should be below $1\%$.

%%%%%%%%%%%%%%%%%%%%
%%%%%%%%%%%%%%%%%%%%

%%%%%%%%%%%%%%%%%%%%
\subsection{PDF uncertainties}
\label{se:pdfuncert}
%%%%%%%%%%%%%%%%%%%%

\begin{figure*}[t]   
\centering
  \includegraphics[width=\ratiotextwidth\textwidth,valign=t]{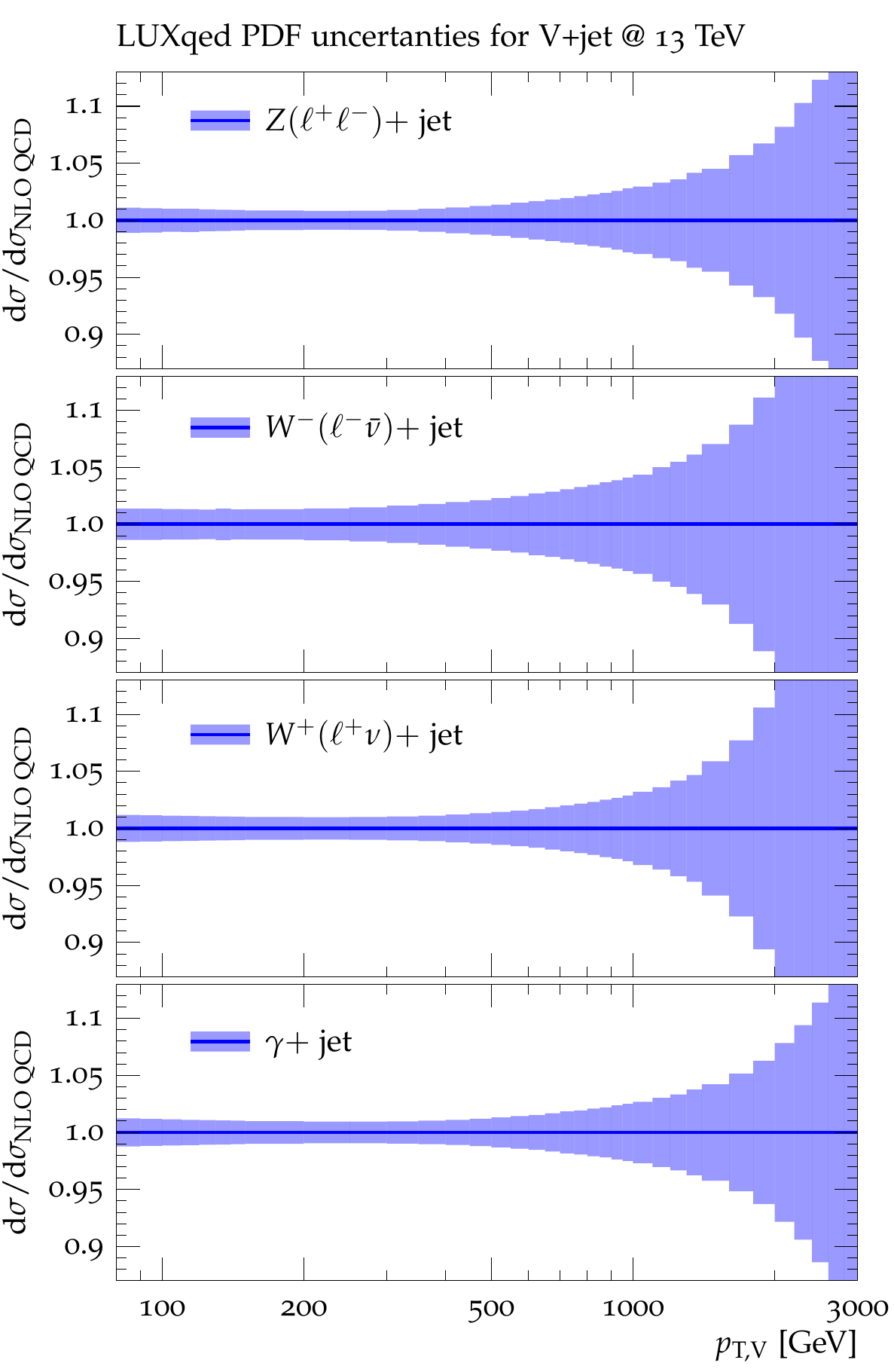}
    \includegraphics[width=\ratiotextwidth\textwidth,valign=t]{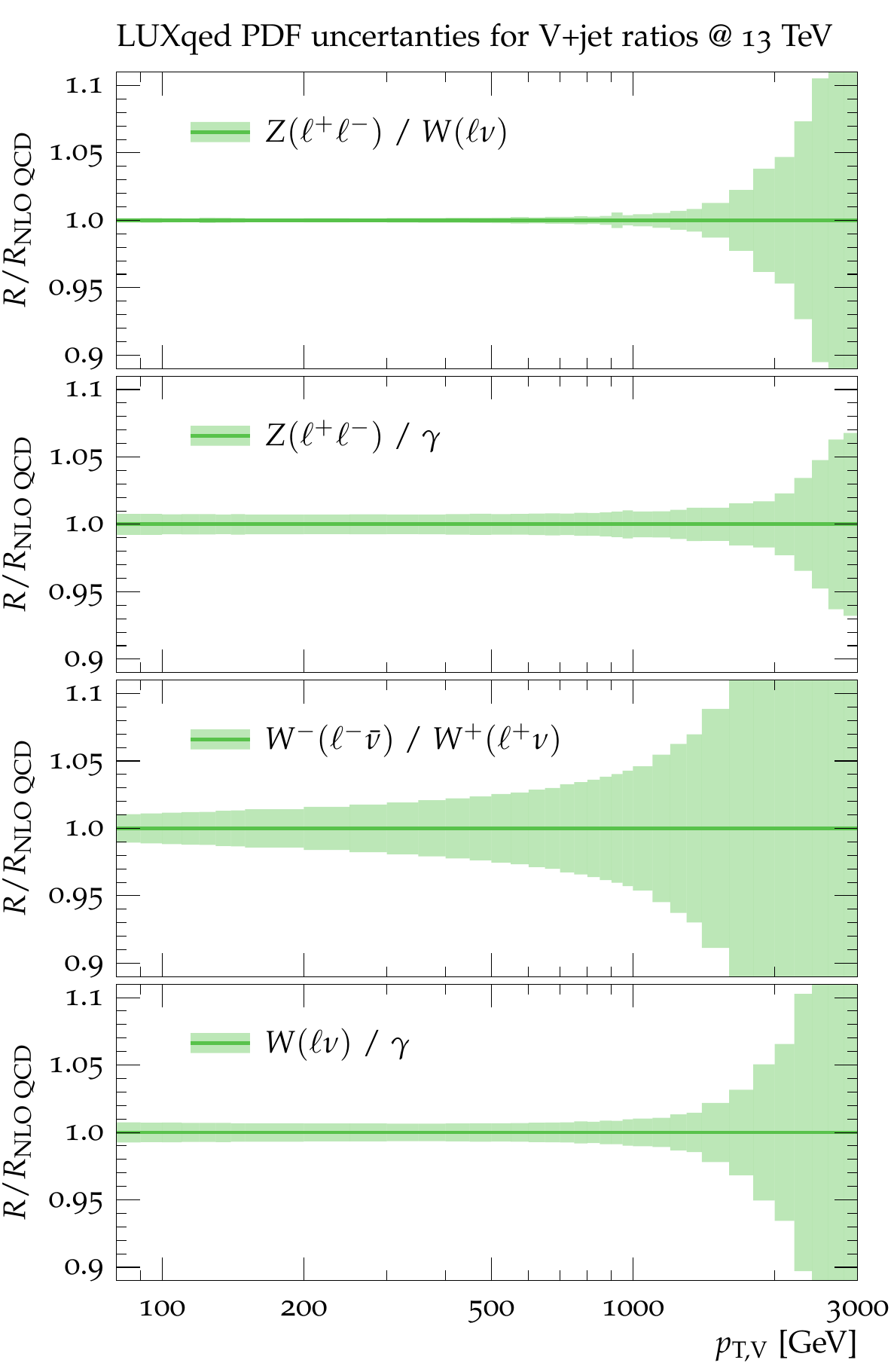}
\caption{Relative \texttt{LUXqed\_plus\_PDF4LHC15\_nnlo\_100} PDF uncertainties on the nominal $\pT$ distributions
 for the different $pp\to V+$jet processes at 13\,TeV evaluated to NLO QCD are shown on the left.
Corresponding PDF uncertainties for ratios of $V+$\,jet distributions
are shown on the right. In the ratios different PDF replicas are correlated across processes and the resulting errors on the respective ratio are combined in quadrature.
}
\label{fig:pdfvariations}
\end{figure*}

The role of PDF uncertainties can be significant especially  at
high-$p_\rT$, where PDFs tend to be less precise.
In \reffi{fig:pdfvariations} we illustrate the effect of PDF uncertainties
within \texttt{LUXqed} (for the quark and gluon uncertainties based on
\texttt{PDF4LHC15\_nnlo\_100}) for the different $V+$jets processes and
process ratios at NLO QCD.  Up to about $800$ GeV the PDF uncertainties on
the nominal $\pT$ distributions remain below $2\%$.  In the tails of the
distributions the PDF uncertainties significantly increase.  They grow
beyond $5\%$ for $\pT \gtrsim 1.5$ TeV.  In the Z/W ratio the PDF
uncertainties cancel almost completely and remain below $0.5(2)\%$ up to
$\pT \approx 800(1500)$~GeV.  In the $Z/\gamma$ and $W/\gamma$ ratios the
PDF uncertainties are at the level of $1-2\%$ up to $\pT \approx 1300$ GeV,
while the $W^-/W^+$ ratio is subject to PDF uncertainties beyond $5\%$
already for $\pT \gtrsim 1$ TeV, driven by uncertainties on the $u/d$-ratio
at large Bjorken-$x$~\cite{Malik:2013kba}.

%%%%%%%%%%%%%%%%%%%%
%%%%%%%%%%%%%%%%%%%%

To keep track of PDF uncertainties 
in the combination of QCD and EW corrections we introduce 
a generalised set of QCD nuisance parameters,
\beqar
\label{se:epsQCDPDF}
\vec\eps_\QCD=(\eps_{\QCD,1},\eps_{\QCD,2},\eps_{\QCD,3},\eps_{\PDF,1},\eps_{\PDF,2},\dots), \nonumber \\
\eeqar
which comprises QCD scale and shape variations, as well as process-correlation and PDF uncertainties.
To this end we extent Eq.~\eqref{ew:QCDcomb},
\begin{align}\label{ew:QCDcombPDF}
\nonumber
&\parx\siv_{\NkLO\,\QCD}(\vec\eps_{\QCD})= 
\left[K^{(V)}_{\NkLO}(x) \right. \\  \nonumber
&\left. +\sum_{i=1}^3\eps_{\QCD,i}\,\delta^{(i)} K^{(V)}_{\NkLO}(x)
+  \sum_{i=1}^{107}\eps_{\PDF,i}\,\delta^{(i)} K^{(V)}_{\PDF}(x) \right]\\
&{}\times\,\parx\siv_{\LO\,\QCD}(\vec\mu_0)\,,
\end{align}
introducing a sum over the 107 independent Hessian PDF replicas provided by the PDF set \texttt{LUXqed\_plus\_\linebreak PDF4LHC15\_nnlo}.
Such a combination corresponds to the PDF4LHC recommendation as detailed in Eq. (20) of Ref.~\cite{Butterworth:2015oua}.
These PDF variations should be applied in a fully correlated way across processes and $p_\rT$ bins. 
As specified in more detail in \refse{se:ewqcd}, 
the various uncertainties parametrised through \refeq{se:epsQCDPDF} should be
applied at the level of QCD calculations and treated on the same 
footing in the combination of QCD and EW corrections.
%Rather than giving an explicit definition of 
%the $\eps_{\PDF,i}$'s, 
%for details of the implementation
%we simply refer to the PDF4LHC recommendation~\cite{Butterworth:2015oua}, see alternatively~\cite{Buckley:2014ana}.

%
%More details on PDF variations in photon-induced processes are given in \refse{se:gammaind} and~\ref{se:inputs}.

\subsection{Real-boson emission}
\label{se:dibosons}
Inclusive diboson production (in particular \linebreak $pp\to VV'+$jets) can be understood as the real-emission counterpart 
to NLO EW corrections to $pp\to V+$\,jet. Both contributions are separately finite and well defined if $V'=W,Z$. 
Although they are expected to cancel against each other to a certain (typically small) extent,
in practice one should only make sure that both types of processes, $pp\to V+$\,jet and $pp\to VV'$(+jets) with leptonic and hadronic decays of the $V'$, are included in the analysis, and, in order to avoid double counting,
contributions of type $VV'$(+jets) should be included in separate diboson MC samples and not as EW correction effects in $V+$\,jets samples.
Unless a very strong cancellation is observed (which is typically not the case), 
there is no reason to worry about the possible correlation of 
uncertainties in $V+$\,jets and $VV'$(+jets) production, i.e.~one can treat the respective uncertainties as uncorrelated.

As concerns the accuracy of MC simulations of 
$pp\to VV'$(+jets), it is important to notice that 
a large diboson background to inclusive vector-boson production at high $p_\rT$ is expected to arise from 
$pp\to VV'j$ topologies with a hard back-to-back $Vj$ system
accompanied by a relatively soft extra vector boson.
This calls for a reliable description of $VV'+$\,jet including QCD
(and possibly EW) corrections. Thus we recommend the use of merged diboson samples that include at least one extra jet at matrix-element level.
At the TeV scale, the EW corrections to $pp\to VV'+$\,jet can become quite large~\cite{Li:2015ura,Yong:2016njr} 
and should ultimately be included, together with the corresponding 
QCD corrections~\cite{Campbell:2007ev,Dittmaier:2007th,Dittmaier:2009un,Binoth:2009wk,Campanario:2009um,
Campanario:2010hp,Cascioli:2013gfa,Campbell:2015hya}.

\subsection{Combination of QCD and electroweak corrections}
\label{se:ewqcd}

The combination \refpar{eq:th1} of higher-order predictions presented in the previous sections
can be cast in the form, 
\beqar\label{eq:QCDEWcomb1}
\parx\siv_{\TH}(\vec\mu)
&=& 
K^{(V)}_{\TH}(x,\vec\mu)\,
\parx\siv_{\LO\,\QCD}(\vec\mu_0) \\ 
&&+\parx\siv_{\gamma-{\rm ind.}} \nonumber
(x,\vec\mu),
\eeqar
where
\beqar\label{eq:QCDEWcomb2}
K^{(V)}_{\TH} = K^{(V)}_{\TH,\oplus}(x,\vec\mu) &= &
K_{\NkLO}^{(V)}(x,\vec\mu)  \\ &&+ \kappa_{\EW}^{(V)}(x)\, \nonumber
K_{\LO}^{(V)}(x,\vec\mu)
\eeqar 
corresponds to the standard additive combination of QCD and EW corrections as defined, respectively,
 in \refeq{eq:kfactors} and \refeqs{eq:ewkfactors}{eq:EWcorsplitting}.
Note that the scale-\linebreak dependent LO 
QCD \mbox{$K$-factor}
in \refeq{eq:QCDEWcomb2} is due to the fact that QCD and EW correction factors 
are normalised to  $\sigma_{\LO\,\QCD}^{(V)}(\vec\mu_0)$ and
$\sigma_{\LO\,\QCD}^{(V)}(\vec\mu)$, respectively.

Mixed QCD--EW corrections of relative $\ord(\alpha\alphaS)$
are not known to date. However, it is possible to obtain an improved prediction 
that partially includes such mixed effects
by combining higher-order EW and QCD corrections through a factorised 
prescription\footnote{See, e.g.~\citeres{Dittmaier:2012vm,Heinemeyer:2013tqa,deFlorian:2016spz}
for a factorised treatment of QCD and EW corrections for Higgs-strahlung and vector-boson fusion processes.},
\beqar\label{eq:QCDEWcomb3}
K^{(V)}_{\TH}=
K^{(V)}_{\TH,\otimes}(x,\vec\mu) =
K_{\NkLO}^{(V)}(x,\vec\mu)\left[1  + \kappa_{\EW}^{(V)}(x)\right].\nonumber \\
\eeqar 
The higher-order terms induced by this factorised formula can be written as
\beqar\label{eq:QCDEWcomb5}
K^{(V)}_{\TH,\otimes}(x,\vec\mu)-K^{(V)}_{\TH,\oplus}(x,\vec\mu) &=&
\kappa_{\NkLO}^{(V)}(x,\vec\mu)\,\kappa_{\EW}^{(V)}(x), \nonumber \\
\eeqar 
where 
$\kappa_{\NkLO}^{(V)}$ denotes the 
pure higher-order contribution to the QCD $K$-factor, \ie
\beqar\label{eq:QCDEWcomb6}
K_{\NkLO}^{(V)}(x,\vec\mu)
=
K_{\LO}^{(V)}(x,\vec\mu)
+
\kappa_{\NkLO}^{(V)}(x,\vec\mu),
\eeqar 
in analogy with the definition of the $\kappa_{\EW}$ correction factor~\refpar{eq:ewkfactors}.

The prescription \refpar{eq:QCDEWcomb3}
is motivated by the factorisation of QCD corrections from
the large Sudakov-\linebreak enhanced EW corrections at high energies~\cite{Chiu:2007dg}
and by the observation that in cases where the multiplicative and additive approach are far apart from
each other, such as in the presence of giant
$K$-factors~\cite{Rubin:2010xp,Kallweit:2015dum}, the former turns out to be 
much more reliable.
In general, when QCD and EW corrections are simultaneously enhanced,
the $\ord(\alpha\alphaS)$  mixed terms
that are controlled by the multiplicative prescription can become quite significant.
We also note that, thanks to the fact that the relative EW correction factors $\kappa_{\EW}^{(V)}(x)$ are 
essentially insensitive to QCD scale variations,
the scale dependence of the multiplicative combination~\refpar{eq:QCDEWcomb3}
is similar as for pure $\NkLO$ QCD predictions. 
In contrast,  the additive approach~\refpar{eq:QCDEWcomb2} 
can suffer from sizable scale uncertainties 
when EW corrections become large.

In order to estimate the typical size of higher-order effects that are not captured 
by the factorised 
prescription~\refpar{eq:QCDEWcomb3}, we cast 
mixed QCD--EW corrections of $\ord(\alpha\alphaS)$ in the form 
\beqar\label{eq:QCDEWcomb7}
K^{(V)}_{\mix}(x,\vec\mu) &=&\frac{\parx\Delta\sigma_{\mix}^{(V)}(x,\vec\mu)}
{\parx\sigma_{\LO}^{(V)}(x,\vec\mu_0)}
\nonumber \\ &=&
\kappa_{\NkLO}^{(V)}(x,\vec\mu)\left[
\kappa_{\EW}^{(V)}(x)+
\delta \kappa^{(V)}_{\mix}(x)
\right],
\eeqar 
and to model the non-factorising term we use the simple Ansatz\footnote{%
As discussed below, the goodness of this naive Ansatz 
will be justified by fitting it to a realistic estimator of
$\delta \kappa^{(V)}_{\mix}(x)$.
} 
\beqar\label{eq:QCDEWcomb8}
\delta\kappa^{(V)}_{\mix}(x) &=& \xi^{(V)}\,
\kappa_{\EW}^{(V)}(x).
\eeqar 
The expectation that the bulk of QCD and EW corrections factorise
implies that the absolute value of the free process-dependent factors
$\xi^{(V)}$ should be well below one.
Note that \refeq{eq:QCDEWcomb8} is equivalent to
\beqar\label{eq:QCDEWcomb9}
\delta K^{(V)}_{\mix}(x,\vec\mu) &=& 
\xi^{(V)} 
\left[
K^{(V)}_{\TH,\otimes}(x,\vec\mu) -
K^{(V)}_{\TH,\oplus}(x,\vec\mu)
\right],\nonumber\\
\eeqar
\ie we assume that non-factorising EW--QCD mixed terms are proportional to the difference 
between the additive and multiplicative combination of QCD and EW corrections.

The NLO EW corrections to
$pp\to V+2$\,jets~\cite{Denner:2014ina,Kallweit:2015dum}, which represent
a real--virtual contribution to the unknown mixed EW--QCD NNLO corrections to 
$V+$\,jet production,
can provide useful insights into the typical size of the $\xi^{(V)}$ factors and the goodness of the
Ansatz~\refpar{eq:QCDEWcomb7}--\refpar{eq:QCDEWcomb8}.
In particular, starting from the $\ord(\alpha\alphaS)$ contributions to \refeq{eq:QCDEWcomb7}, 
\beqar\label{eq:dkmix1}
K^{(V)}_{\NNLO\, \mix}(x,\vec\mu)=
&\kappa_{\NLO}^{(V)}(x,\vec\mu)
\left[
\kappa_{\NLO\, \EW}^{(V)}(x)
\right. \nonumber \\ &+ \left.
\delta \kappa^{(V)}_{\NNLO\,\mix}(x)\right],
\eeqar
it is possible to establish a relation between
non-\linebreak factorising  NNLO mixed corrections
and the differences between NLO EW $K$-factors for $V+2$\,jet and  $V+1$\,jet production.
To this end, we consider the identity
\beqar\label{eq:dkmix2}
&\parx\sigma^{V+2\,\jets}_{\NLO\, \EW}(x,\taucut)
=
\parx\sigma^{V+2\,\jets}_{\LO\, \QCD}(x,\taucut)
\nonumber \\ &\times \left[
\kappa_{\NLO\, \EW}^{V+1\,\jet}(x)
+
\delta \kappa^{(V)}_{\NNLO\,\mix}(x,\taucut)\right],
\eeqar
which is obtained by multiplying both sides of~\refeq{eq:dkmix1} 
by the LO QCD cross section for $pp\to V+1$\,jet and restricting the phase space to 
real--virtual contributions with $V+2$\,jet final states.
This restriction is implemented by means of an $N$-jettiness~\cite{Stewart:2010tn} 
resolution parameter $\taucut$, as described in more detail below, and
the above equation should be understood as definition of
$\delta \kappa^{(V)}_{\NNLO\,\mix}(x,\taucut)$, which will be used as estimator of 
$\delta \kappa^{(V)}_{\NNLO\,\mix}(x)$ in~\refeq {eq:dkmix1}.
In~\refeq{eq:dkmix2} we use the notation $\kappa_{\NLO\, \EW}^{V+1\,\jet}(x)=\kappa_{\NLO\, \EW}^{(V)}(x)$,
and  we keep the \mbox{$\mu$-dependence} as implicitly understood, since
the term $\delta \kappa^{(V)}_{\NNLO\,\mix}(x,\taucut)$ is expected to be quite 
stable with respect to scale variations.
Instead, the $\taucut$ parameter plays an important role since it
acts as a cutoff of infrared QCD singularities in the regions 
where the second jet becomes soft or collinear.
Based on the universal behaviour of IR QCD effects, 
such singularities are expected to factorise into
identical singular factors on the left- and the right-hand side 
of \refeq{eq:dkmix2}. Thus, while the $\delta \kappa^{(V)}_{\NNLO\,\mix}(x,\taucut)$
term on the right-hand side depends on $\taucut$, this dependence is expected to be 
free from large $\taucut$-logarithms and thus reasonably mild.
%
%Note that~\refeq{eq:dkmix2} should be understood as definition of
%$\delta \kappa^{(V)}_{\NNLO\,\mix}(x,\taucut)$, which will be used as estimator of 
%$\delta \kappa^{(V)}_{\NNLO\,\mix}(x)$ in~\refeq {eq:dkmix1}.

As anticipated above, solving for $\delta \kappa^{(V)}_{\NNLO\,\mix}$ 
we obtain the relation
\beqar\label{eq:dkmix3}
\delta \kappa^{(V)}_{\NNLO\,\mix}(x,\taucut)&=&
\kappa^{V+2\,\jets}_{\NLO\, \EW}(x,\taucut)
-\kappa_{\NLO\, \EW}^{V+1\,\jet}(x),\nonumber \\
\eeqar
which allows us to estimate non-factorising mixed effects in terms of the difference between
the $V+2$-jet and $V+1$-jet EW $\kappa$-factors. To this end, we will match the estimator \refpar{eq:dkmix3} 
to the Ansatz~\refpar{eq:QCDEWcomb8}.
More precisely, we will fix the free coefficients $\xi^{(V)}$ in \refeq{eq:QCDEWcomb8}
in such a way that 
\beqar
\label{eq:dkmix4}
\xi^{(V)}\,
\kappa_{\NLO\,\EW}^{V+1\,\jet}(x)&\gsim &
%\delta \kappa^{(V)}_{\NNLO\,\mix}(x,\taucut)=
\kappa^{V+2\,\jets}_{\NLO\, \EW}(x,\taucut)
-\kappa_{\NLO\, \EW}^{V+1\,\jet}(x)\nonumber \\
\eeqar
for the whole $x$-spectrum and within an appropriately chosen $\taucut$ range.  
Thanks to the cancellation of IR QCD singularities in \refeq{eq:dkmix3}, the 
resulting $\xi^{(V)}$ coefficients should be reasonably stable with respect to 
the choice of the resolution parameter. Thus, $\taucut$ can be varied in a 
rather wide range. In principle one could even consider the $\taucut\to 0$ limit of \refeq{eq:dkmix4}.
However, given that two-loop mixed EW--QCD contributions are
not taken into account, this limit does not converge towards the full NNLO result
corresponding to $\taucut=0$.  Moreover, for very small values of $\taucut$ the
numerator and denominator of $\kappa^{V+2\,\jets}_{\NLO\, \EW}(x,\taucut)$
are dominated by universal $\taucut$-logarithms  that should 
cancel against virtual two-loop terms, 
and since such logarithms factorise, their dominance can result in 
an underestimation of non-factorising effects.
Vice versa, excessively large values of $\taucut$ can lead to  
an overestimation of non-factorising effects. 
This is due to the fact that increasing $\taucut$ enhances the difference
between EW $\kappa$-factors in \refeq{eq:dkmix4} but also 
suppresses the cross section of the $V+2$-jet subprocess, rendering it a
less and less significant estimator of the behaviour of mixed corrections
for inclusive $V+$\,jet production.
Thus, excessively small or large values of $\taucut$ should be
avoided.

Based on the above considerations, for the fit of the $\xi^{(V)}$ 
coefficients we require that \refeq{eq:dkmix4} is
fullfilled in a wide $\taucut$-range while keeping the
$\sigma^{V+2\,\jet}/\sigma^{V+1\,\jet}$ ratio
at order one, in such a way that
the $V+2\,$jet cross section is neither too suppressed nor too enhanced.
This procedure is implemented using an~\mbox{$N$-jettiness} 
cut parameter~\cite{Stewart:2010tn}. 
%of~\citere{Boughezal:2015dva}.
More precisely, we use the dimensionless one-jettiness parameter
\begin{equation}
\label{eq:taudef}
{\tau}_1 = \sum_k \text{min}_i \left\{ \frac{2 p_i \cdot q_k}{Q_i\,\sqrt{\hat{s}}}\right\},
%{\tau}_1 = \sum_k \text{min}_i \left\{ \frac{2 p_i \cdot q_k}{Q_i\right\},
\end{equation}
where the $p_i$ are light-like vectors for each of the initial beams and the
hardest final-state jet, and the $Q_i$ 
characterise their respective hardness, which we set as $Q_i = 2 E_i$.
The hardest final-state jet is defined by applying an anti-$k_{\rm{T}}$ algorithm with R=1
to all final-state partons.\footnote{In order to guarantee a proper cancellation of 
QCD and EW singularities, 
the jet algorithm is 
applied to all QCD partons and photons, excluding photons that
are recombined with leptons, as well as the leading identified photon in case of
the $\gamma+$jets process.}
The $q_k$ denote the four-momenta of any 
such  final-state parton, and $\sqrt{\hat{s}}$ is the partonic centre-of-mass energy. 
All quantities are defined in the hadronic
centre-of-mass system.

%=======

%%%%%%%%%%%%%%%%%%%%
\begin{figure*}[t]   
\centering
  \includegraphics[width=\ratiotextwidth\textwidth]{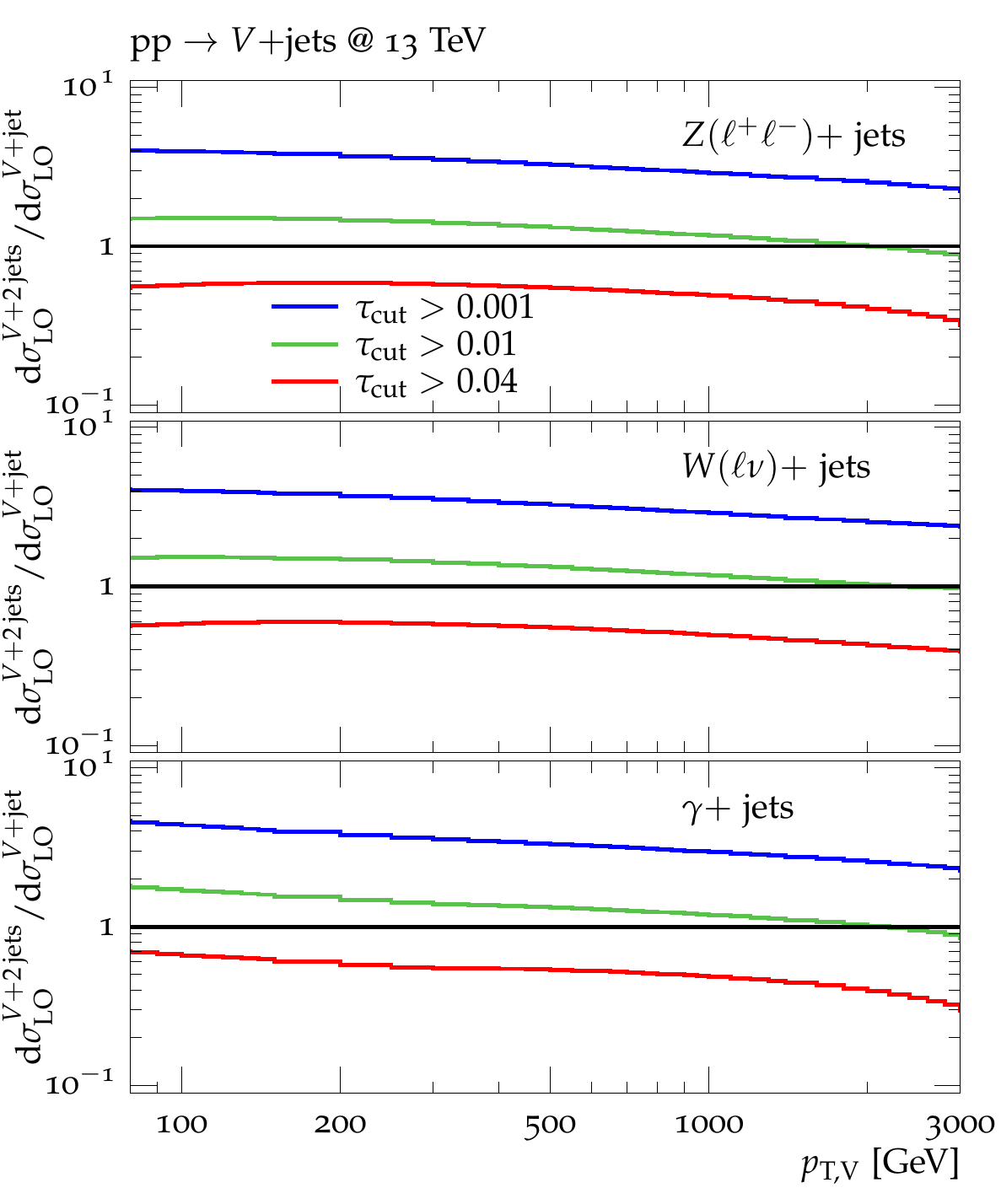}
 \includegraphics[width=\ratiotextwidth\textwidth]{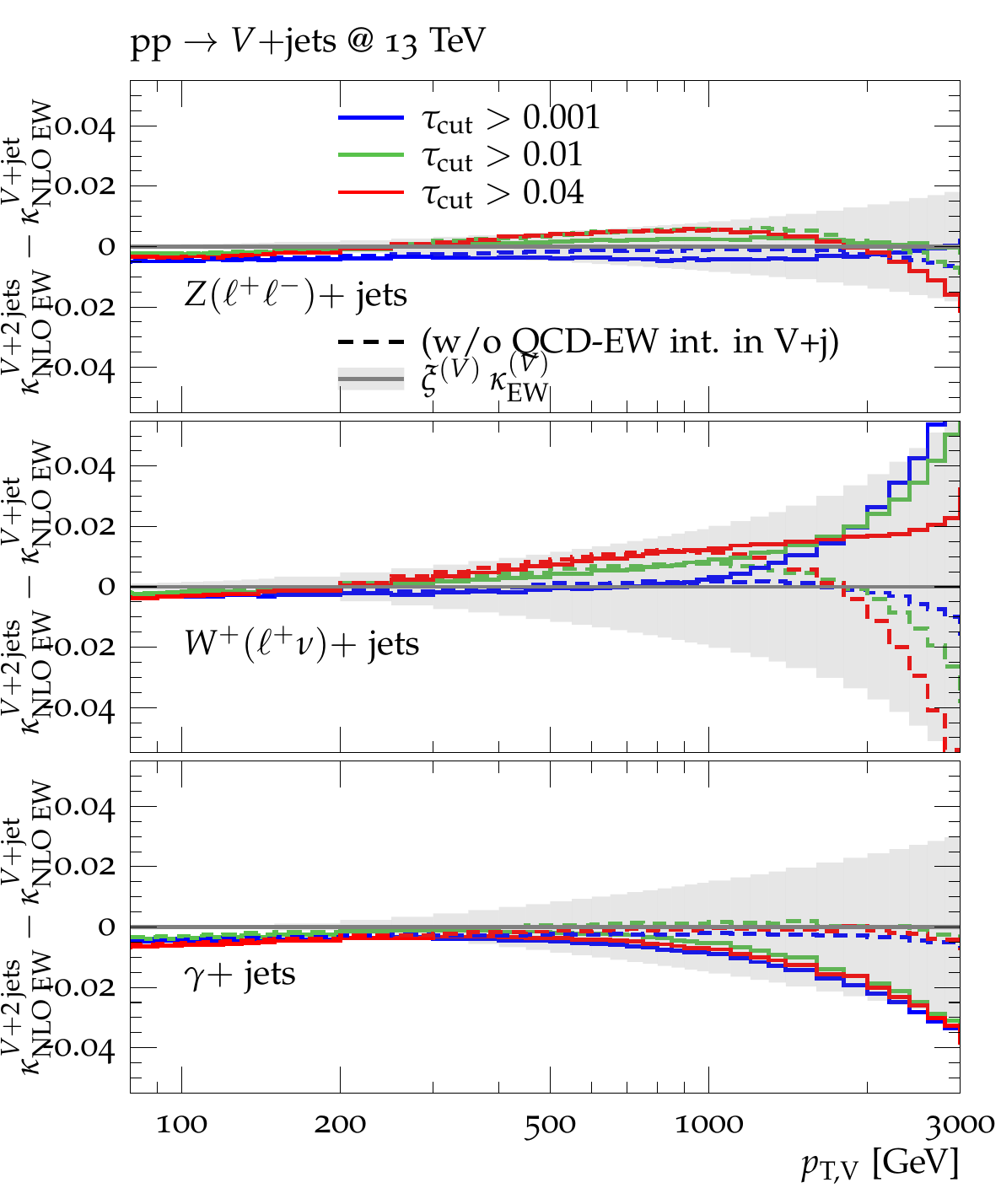}
\caption{
The left plot shows ratios of the different $V+2\, $jets over $V + 1\, $jet predictions at LO for three 
values of the jettiness resolution parameter $\tau_{\rm cut}$.
The right plot shows the estimator of non-factorising mixed EW--QCD effects \refpar{eq:dkmix3}, \ie
the  difference between the EW $K$-factors for one- and
two-jet processes. Results with full EW corrections (solid lines) are compared to the case where
QCD--EW bremsstrahlung interference contributions to $pp\to V+1$\,jet are not included
(dashed lines).
The gray band corresponds to the Ansatz~\refpar{eq:QCDEWcomb8} 
with the $\xi^{(V)}$ coefficients specified in \refeq{eq:dkmix5}.
}
\label{fig:Vjj_NLOEW}
\end{figure*}
%%%%%%%%%%%%%%%%%%%%

To isolate two-jet
configurations against one-jet configurations 
we require $\tau_1>\taucut$, and the cut is 
varied in the range 
%
%$0.1\,\GeV \le  \taucut \le 10\,\GeV$.
%
$0.001 \le  \taucut \le 0.04$.
As demonstrated in \reffi{fig:Vjj_NLOEW}, this choice 
keeps the $\sigma^{V+2\,\jet}/\sigma^{V+1\,\jet}$ ratio around order one, as desired.
Moreover, we observe that the estimator \refpar{eq:dkmix4} remains quite stable with respect to
$\taucut$ variations (see the solid lines in the right plot).
Non-factorising effects turn out to be generally very small. They exceed the percent level only in the TeV tails
of the distributions. As illustrated by the gray band in
\reffi{fig:Vjj_NLOEW} (right), setting
\beqar
\label{eq:dkmix5}
\xi^{Z}=0.1, \qquad
\xi^{W}=0.2, \qquad
\xi^{\gamma}=0.4,
\eeqar
guarantees an acceptable matching of the Ansatz~\refpar{eq:QCDEWcomb8} to the estimator 
\refpar{eq:dkmix4}.
More precisely, for $W+$\,jet production the shape of the
Ansatz~\refpar{eq:QCDEWcomb8} tends to overestimate the uncertainty
in the $\pT$ range between one and two TeV.
However,
we have checked that the Ansatz becomes
much less adequate if
the full EW correction in~\refeq{eq:QCDEWcomb7} is replaced 
by its non-Sudakov part.

The rather small values of the $\xi^{(V)}$ coefficients confirm that the bulk of the EW 
and QCD corrections factorise. However, 
in the case of $W+$\,jet and  $\gamma+$\,jet production,
the relative size of
non-factorising corrections appears to be rather significant.
This is due to the behaviour of the EW $\kappa$-factors in the 
multi-TeV region, where the difference 
between the EW $\kappa$-factors for $pp\to V+1$\,jet and $pp\to V+2$\,jet
is enhanced by the presence of mixed EW--QCD interference contributions 
in channels of 
type $qq\to qq V$ (see the contributions of type a.\ref{it:gammaindqbrem} in \refse{se:ew}).
More precisely, EW--QCD interference 
effects of $\ord(\alphaS\alpha^2)$  enhance the EW corrections to $pp\to V+1$\,jet
as a result of the opening of the $qq$ channel at NLO EW,
while in $pp\to V+2$\,jet the EW K-factor is not enhanced since 
the $qq$ channel is already open at LO.
Based on this observation, and also due to the 
fact that the main effect of the opening of the 
$qq$ channel is already reflected in the NLO QCD 
$K$-factor for $V+1\,$jet production, 
the above mentioned EW--QCD interference effects could be 
excluded from the factorisation prescription~\refpar{eq:QCDEWcomb3}
and treated as a separate contribution.
As illustrated by the dashed curves in \reffi{fig:Vjj_NLOEW}, this approach would lead to a drastic reduction of 
non-factorising effects, especially for $\gamma+$\,jet production.
Nevertheless, given that the effects observed in \reffi{fig:Vjj_NLOEW}
are subdominant with respect to current PDF and statistical uncertainties,
in the present study we refrain from implementing such a splitting.

%%%%%%%%%%%%%%%%%%%%%
%\begin{figure*}[t]   
%\centering
%  \includegraphics[width=\ratiotextwidth\textwidth]{./eejj_NLOEW.pdf}
% \includegraphics[width=\ratiotextwidth\textwidth]{./evjj_NLOEW.pdf}
%\caption{
%NLO EW predictions for the production of $Z(\ell^+\ell^-)$+jets 
%(left) and $W^\pm(\ell\nu)$+jets (right) at 13\,TeV.
%The NLO EW corrections for $pp\to V+1$\,jet 
%(blue) and  $pp\to V+2$\,jets are compared  (green). 
%In the $V+2$\,jet predictions we require, besides the inclusive event selection detailed in 
%\refse{se:setup}, at least two anti-$\kT$ jets with $R=0.4$ and $p_{T}>30~\GeV$ (without any $\eta$ cuts). 
%The lower ratio plot shows the difference in the EW corrections between the one- and
%two-jet processes,
%$\Delta \kappa_{\NLO\,\EW}=\kappa^{Vjj}_{\NLO\,\EW}-\kappa^{Vj}_{\NLO\,\EW}$ for the
%full NLO EW corrections (red) and excluding the finite mixed QCD-EW bremsstrahlung interference contributions 
%to $pp\to V+1$\,jet (magenta). 
%}
%\label{fig:Vjj_NLOEW}
%\end{figure*}
%%%%%%%%%%%%%%%%%%%%%

\subsection*{Combination of QCD and EW corrections with related uncertainties}
\label{se:MIXunc}

Based on the above analysis, we recommend to combine QCD and EW corrections
according to the multiplicative prescription~\refpar{eq:QCDEWcomb7},
treating the non-factorising term \refpar{eq:QCDEWcomb8} as uncertainty and
using the estimated $\xi^{(V)}$ factors given in \refeq{eq:dkmix5}.
Including QCD and EW uncertainties as specified in~\refeq{ew:QCDcomb} and \refeq{eq:ewunccomb},
this leads to the combination formula
%\begin{widetext}
\begin{align}\label{eq:QCDEWcomb4}
&K^{(V)}_{\TH}(x,\vec\eps_\QCD,\vec\eps_\EW,\eps_\mix) \nonumber \\ 
&= K^{(V)}_{\TH,\otimes}(x,\vec\eps_\QCD,\vec\eps_\EW) + \eps_\mix \, \delta K^{(V)}_{\mix}(x) \nonumber
\\ \nonumber &=
\left[K_{\NkLO}^{(V)}(x)
+\sum_{i=1}^3\eps_{\QCD,i}\,\delta^{(i)} K^{(V)}_{\NkLO}(x) \right. \\
&\quad\quad \left. + \sum_{i=1}^{107}\eps_{\PDF,i}\,\delta^{(i)} K^{(V)}_{\PDF}(x) \right]
\nonumber \\
 &\quad \times
\left[1  + \kappa_{\EW}^{(V)}(x)
+\sum_{i=1}^3\eps^{(V)}_{\EW,i}\,\delta^{(i)}\kappa^{(V)}_\EW(x)
\right] \nonumber \\ 
&\quad + \eps_\mix \, \delta K^{(V)}_{\mix}(x),
\end{align}
%\end{widetext}
where the uncertainty associated with non-factorising mixed 
EW--QCD terms reads
\beqar
\label{eq:uncert_mix}
\delta K^{(V)}_{\mix}(x) &=& 
\xi^{(V)} \left[K_{\NkLO}^{(V)}(x)-1\right] \kappa_{\EW}^{(V)}(x)\nonumber \\
&=& 
\xi^{(V)} 
\left[K^{(V)}_{\TH,\oplus}(x)- K^{(V)}_{\TH,\otimes}(x) \right].
\eeqar
The related nuisance parameter, $\eps_\mix$, 
should be Gaussian distributed 
with one standard deviation
corresponding to the range  $\eps_\mix\in[-1,+1]$. 
Given that mixed uncertainties have been estimated using a proxy of
the full NNLO QCD--EW calculation, it would be reasonable to assume some 
degree of correlation across different $V+$\,jet processes.
However, for simplicity 
in this study we keep $\eps_\mix$ variations 
fully uncorrelated, bearing in mind that this approach is probably too conservative.

%%%%%%%%%%%%%%%%%%%%
\begin{figure*}[t]   
\centering
  \includegraphics[width=\ratiotextwidth\textwidth]{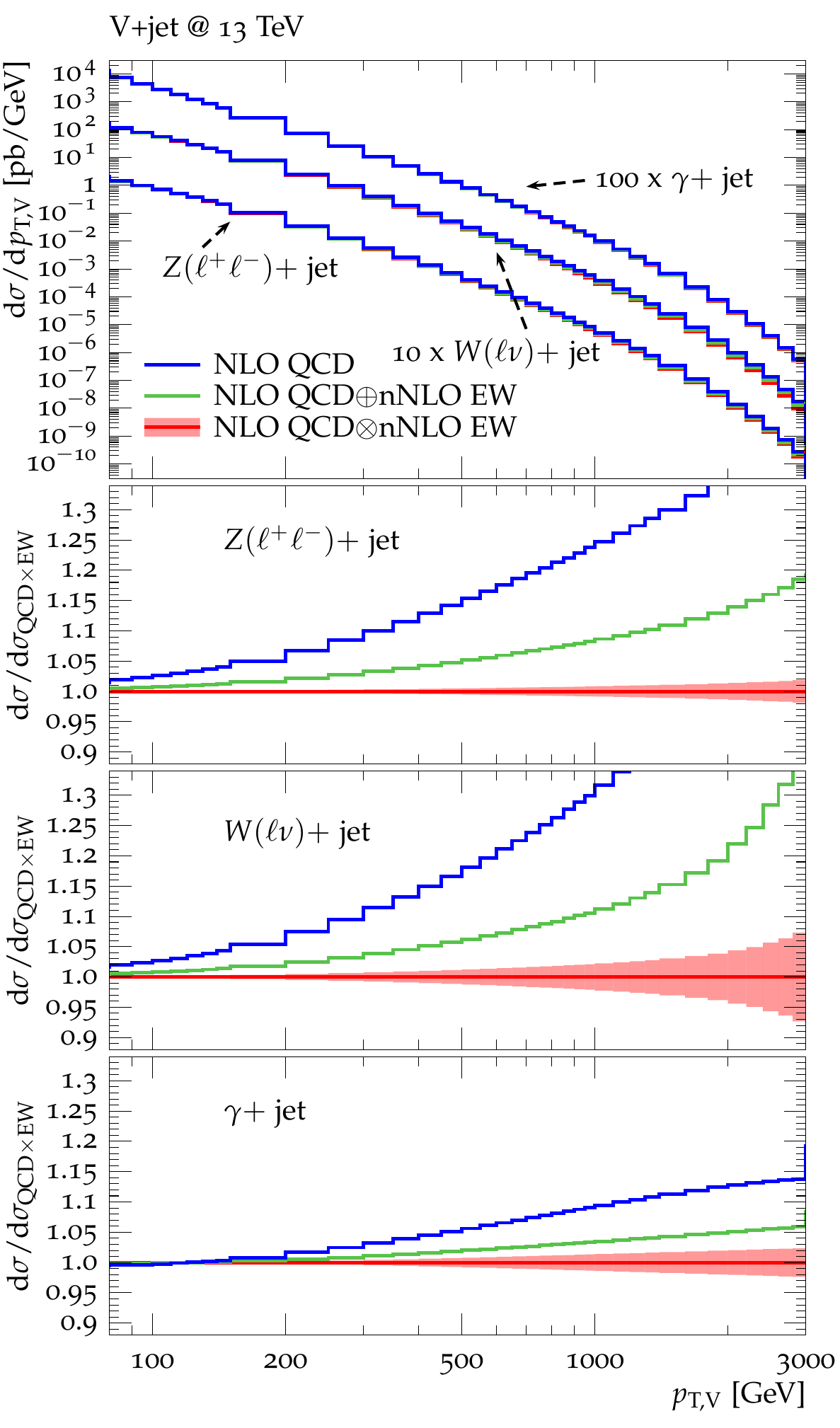}
    \includegraphics[width=\ratiotextwidth\textwidth]{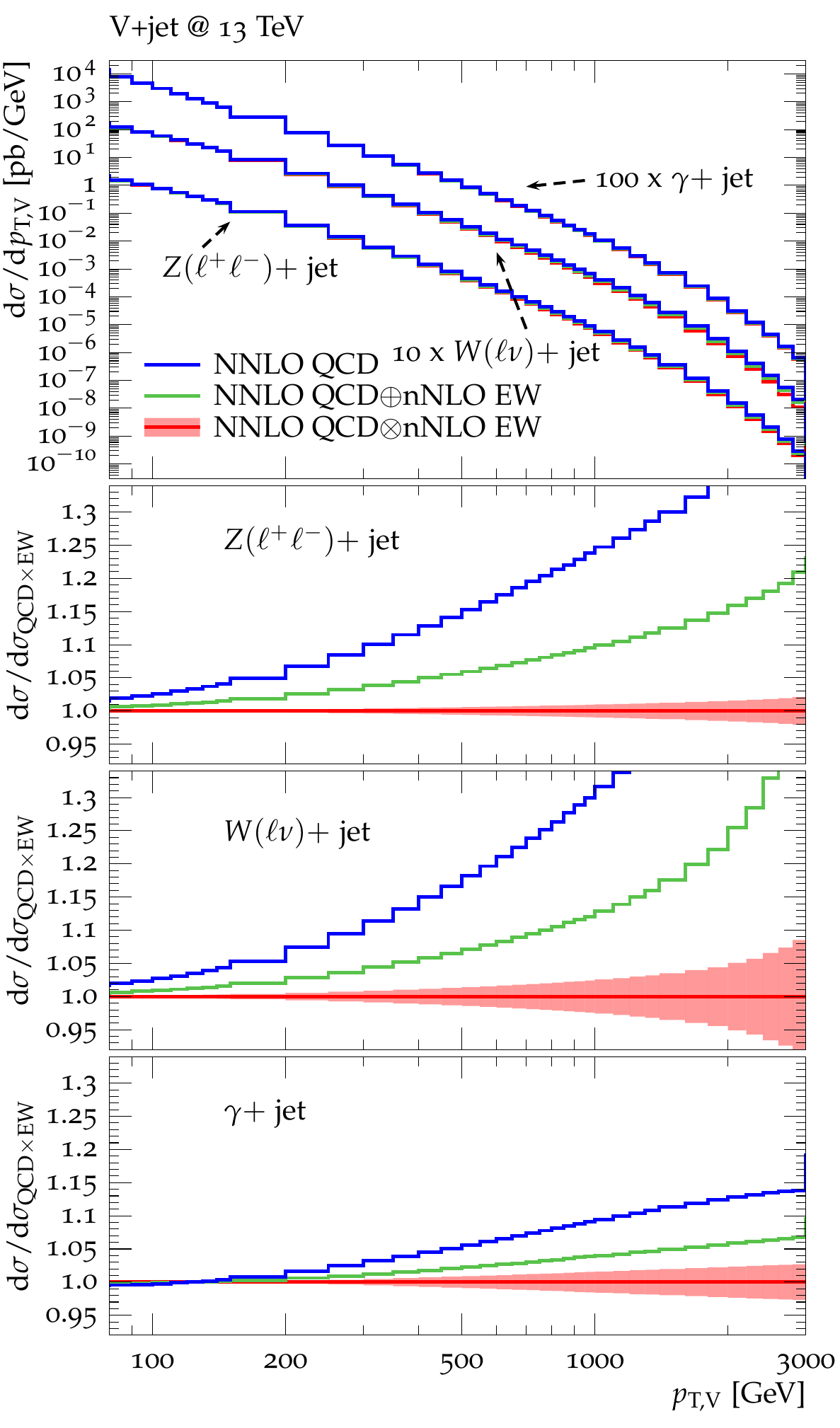}
\caption{
Comparison of additive (green) and multiplicative (red) combination of (N)NLO QCD and 
nNLO EW corrections for 
various $pp\to V$+jet 
processes at 13\,TeV.
The red band corresponds to the mixed QCD--EW uncertainty~\refpar{eq:uncert_mix}. The (N)NLO QCD result without EW corrections is shown in blue. The combination at NLO QCD is shown on the left and at NNLO QCD on the right.
}
\label{fig:QCD_EW}
\end{figure*}
%%%%%%%%%%%%%%%%%%%%

In \reffi{fig:QCD_EW} we compare the additive and multiplicative
combinations of QCD and EW corrections showing also the corresponding uncertainty estimate 
\refpar{eq:uncert_mix} for various $V$+jet processes.

%%%%%%%%%%%%%%%%%%%%%%%%%%%%%%%%%%%%%%%%%%%%
\section{Summary and conclusions}
\label{se:conclusions}

The precise control of SM backgrounds, and notably of 
$pp \to Z(\nu\bar\nu)+$\,jets, is crucial in order to maximise
the potential of MET+jets searches at the LHC.
Such backgrounds can be predicted directly using QCD and EW
calculations.
Alternatively, QCD and EW calculations can be used to relate them to
experimental data for similar $V+$\,jet production  processes, \ie $pp\to \gamma+$\,jets,
$pp \to W(\ell\nu)+$\,jets and $pp \to Z(\ell^+\ell^-)+$\,jets.

In this article we have presented predictions for inclusive vector-boson
$p_\rT$ distributions based on the most advanced calculations available
today, bringing together results from a number of groups so as to have
perturbative QCD to NNLO accuracy, EW corrections to NLO accuracy and
additionally the inclusion of 2-loop EW Sudakov logarithms.

A substantial part of our study concerned uncertainty estimates.
In particular we proposed and applied various new approaches for
uncertainty estimates and correlations across processes and $p_\rT$
regions.

We defined the uncertainties due to normal QCD scale variations
in a way that gives a strong correlation across different $p_\rT$
regions,~\refeq{eq:var1Kfact}.
We then supplemented it with a shape uncertainty that is anti-correlated
across $p_\rT$, \refeqs{eq:var2Kfact}{eq:shapedist}.
To address the long-standing problem of evaluating the correlations
between uncertainties for different processes, we separated the
uncertainty into process-independent and process-dependent components.
The universal component was taken to be composed of the overall scale
and shape uncertainties for the reference $Z+\jet$ process.
The process-dependent component, which is generally small, was
determined by considering the difference between suitably normalised
$K$-factors for the different processes,~\refeq{eq:dKqcd3}.
This amounts to a conservative choice of taking the uncertainty on
ratios as the difference between the best available
prediction and the one at one order lower.

Special attention was devoted to the correlation of $Z/W+$\,jet
and $\gamma+$\,jet production.
In that case a substantial non-universal contribution is associated
with the masslessness of the photon and the need to control
collinear divergent $q\to q\gamma$ radiation through a
photon-isolation prescription.
We introduced a novel photon-isolation prescription with a dynamically
chosen isolation radius,~\refeq{eq:dynrad}, designed to suppress
$q\to \gamma q$ radiative effects in a way that is similar to the
effect of the masses of the $Z$ and $W$ bosons in the case of
$q\to V q$ splittings at large $p_\rT$.
Such a dynamic isolation allows one to split $\gamma$+jet production
into a quasi-universal part, which can be treated on the same footing
as $Z+\jet$ and $W+\jet$ production, and a non-universal part which is
kept uncorrelated.
The non-universal part is given by the difference between the cross
sections with conventional and dynamic photon isolation prescriptions.

For pure EW corrections we considered three uncertainty sources for
unknown higher order contributions.
These address unknown Sudakov logarithms beyond NNLO and/or NLL
accuracy, as well as unknown hard (non-Sudakov) EW corrections beyond
NLO and process-correlation effects.

One potentially large source of uncertainty arises from mixed QCD and
EW corrections, given that both $\ord(\alphaS)$ and  $\ord(\alpha)$ NLO
corrections can be large and that the $\ord(\alpha \alphaS)$ NNLO
corrections are not currently known.
We chose a multiplicative scheme for combining EW and QCD corrections.
To obtain an estimate of unknown $\ord(\alpha\alphaS)$ corrections
not captured by this factorised ansatz, we studied the NLO EW
corrections to $V+2$\,jet production, which represent the real--virtual part
of a full $\ord(\alpha\alphaS)$ calculation for $V+$\,jet production.
Based on this analysis, we concluded that it is reasonable 
to assume that the multiplicative combination of QCD and EW corrections
describes the full $\ord(\alpha\alphaS)$ correction  with a 
relative uncertainty that varies between 10--20\% for $pp\to W/Z+$\,jet
and 40\% for  $pp\to\gamma+$\,jet.

Overall, QCD corrections are substantial, a few tens of percent at
NLO, and up to $10\%$ at NNLO.
The NNLO results are consistent with the NLO predictions within our
prescription for the uncertainty bands of the latter.
This is true not just for absolute cross sections and their shapes, but
also for ratios of cross sections.
These ratios are remarkably stable across LO, NLO and NNLO QCD
corrections, see~\reffi{fig:ratios_qcd}.
Using dynamic photon isolation, this statement holds true also for the
$\gamma+\jet$ process at $p_\rT \gtrsim 300~\GeV$.

The EW corrections to $V$+jet cross sections amount to a few tens of percent
in the TeV region, see \reffi{fig:EW_error}. 
In the ratios they cancel only in part, due to the sensitivity of EW
effects to the SU(2) charges of the produced vector bosons.
At the TeV scale, the NNLO Sudakov logarithms can reach the several
percent level and their systematic inclusion is an important
ingredient in order to achieve percent precision at very high $p_\rT$.

\begin{figure*}[t]   
\centering
  \includegraphics[width=.43\textwidth]{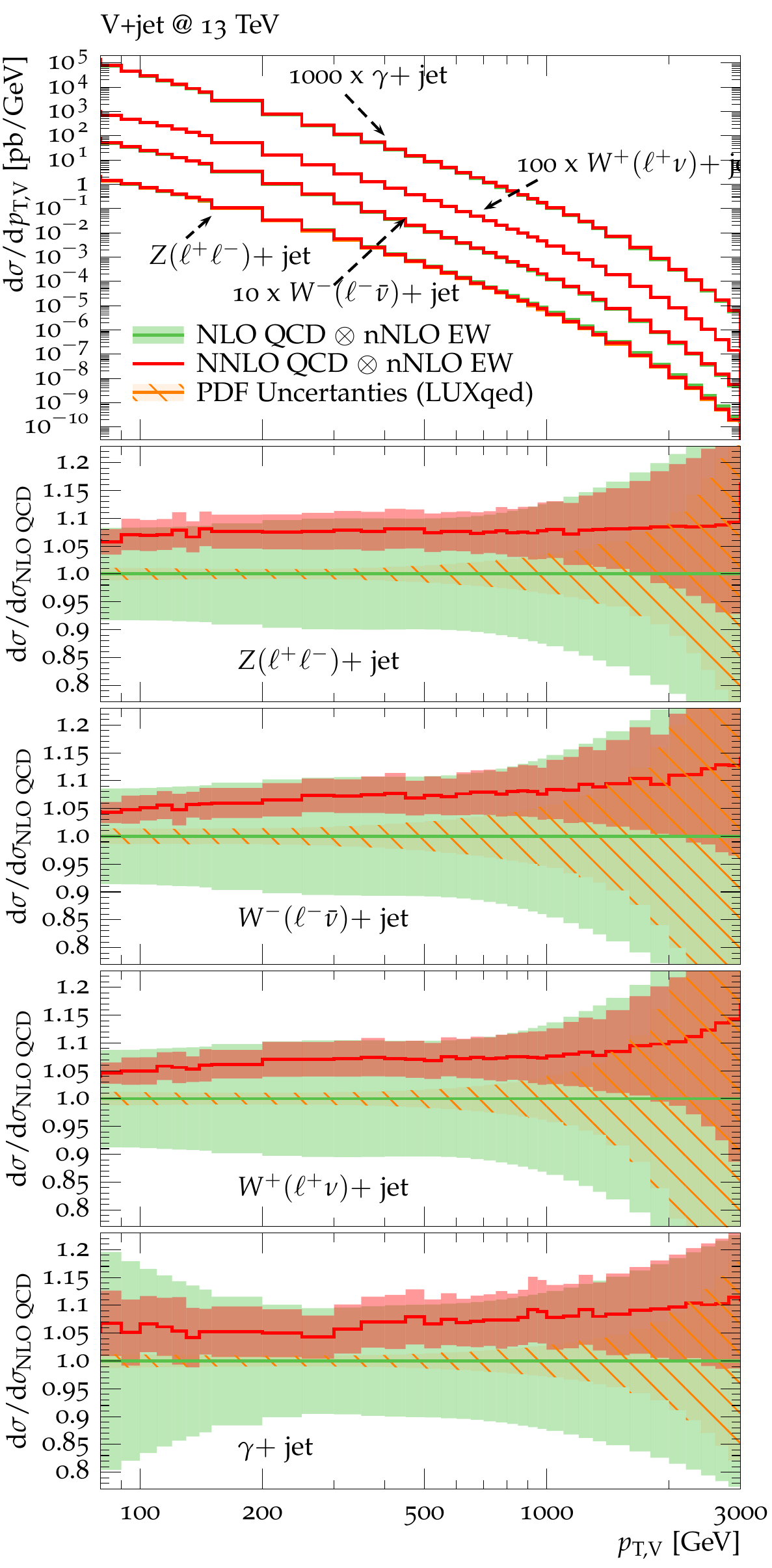}
    \includegraphics[width=.43\textwidth]{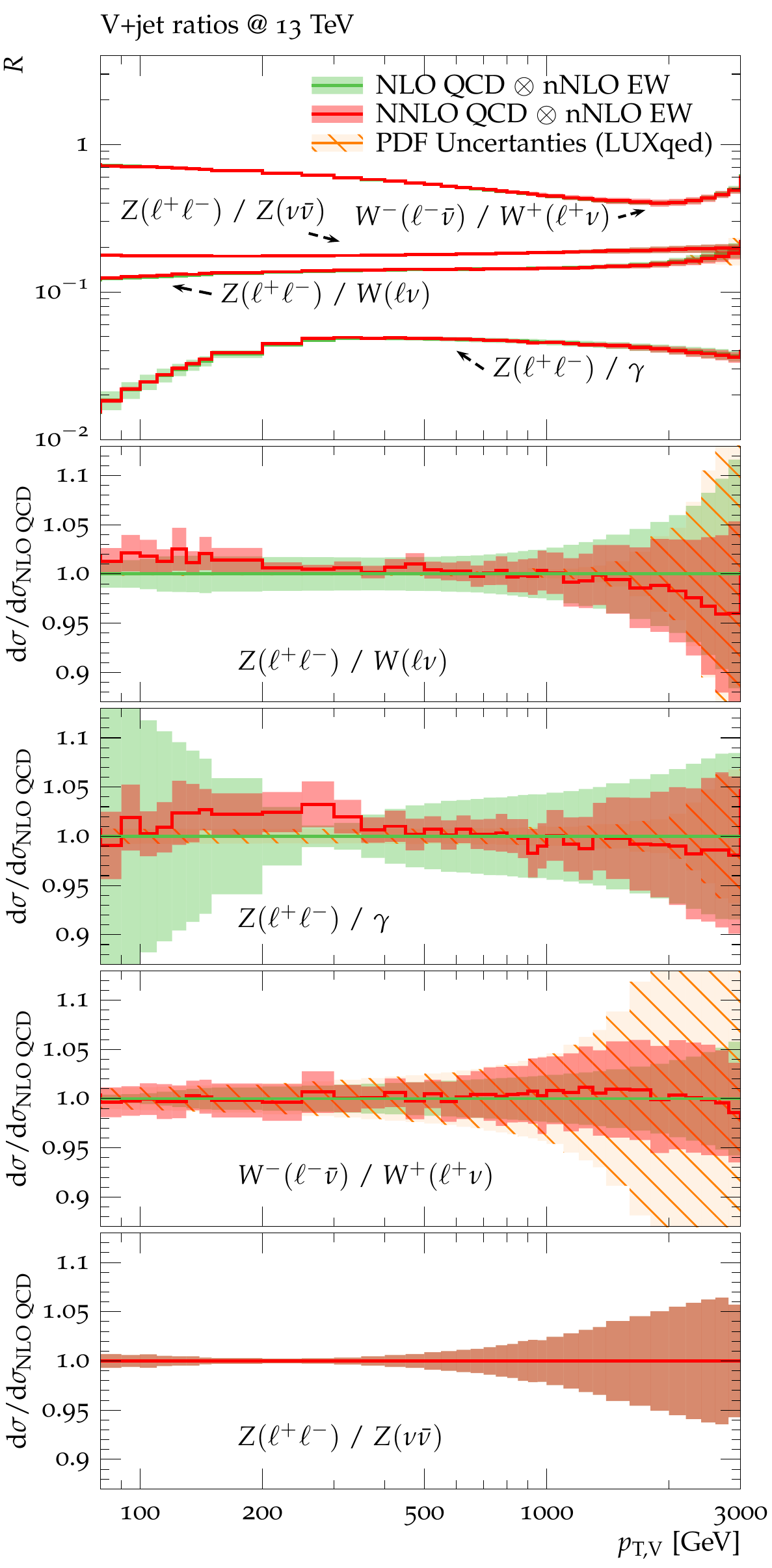}
\caption{Predictions at NLO QCD\,$\otimes$\,nNLO EW and NNLO QCD\,$\otimes$\,nNLO EW
for  $V+$\,jet spectra (left) and ratios (right) at 13\,TeV.
The lower frames show the relative impact of NNLO corrections and theory uncertainties 
normalised to NLO QCD\,$\otimes$\,nNLO EW.  
The green and red bands correspond to the combination (in quadrature) of the perturbative QCD, EW
 and mixed QCD-EW uncertainties, according to~\refeq{eq:QCDEWcomb4}
  at NLO QCD\,$\otimes$\,nNLO EW and NNLO QCD\,$\otimes$\,nNLO EW respectively.
PDF uncertainties based on LUXqed\_plus\_PDF4LHC15\_nnlo are shown at NLO QCD as separate hashed orange bands.
}
\label{fig:QCDxEW_error_combined}
\end{figure*}

In \reffi{fig:QCDxEW_error_combined} we summarize our uncertainty
estimates for the different $V$+jet processes and process ratios. Here we combine
in quadrature all sources of perturbative uncertainties at N(N)LO QCD\,$\otimes$\,nNLO EW
and we overlay the remaining PDF uncertainties.
For convenience, PDF variations have been assessed using NNLO PDFs in combination
with NLO QCD calculations, but can be safely applied to the NNLO QCD results.
The nominal $\pT$ distributions at \linebreak  N(N)LO QCD\,$\otimes$\,nNLO EW  
are constrained at the \linebreak 10(5)\% level up to about 1 TeV and at the 20(10)\% level up to about 2 TeV.
In the process ratios these uncertainties cancel to a large extent. In particular, in the $Z/W$ ratio
remaining uncertainties are at the level of only 1--2\% up to 1\,TeV and below 5\% up to 2\,TeV.
Similarly, the $Z/\gamma$ ratio is constrained at the 5\% level up to 2\,TeV.  Noteworthy,   
including the NNLO QCD corrections the process ratios remain very stable and in particular within the 
uncertainty estimates based on NLO QCD.
This reflects the fact that QCD uncertainties are very well
under control: taking at face value the NNLO QCD 
systematics we are at the level of few percent all the
way up to the multi-TeV scale (see \reffi{fig:ratios_qcd}), 
and at large $\pT$ we are dominated
by EW and PDF uncertainties.
The latter
%
%PDF uncertainties 
%
are below the perturbative 
uncertainties in all nominal distributions 
and all but the $W^-/W^+$ ratio, 
where a precise measurement at high $\pT$
could help to improve PDF fits.
In this respect, we note that the theoretical 
uncertainty for the $W^-/W^+$ ratio is entirely dominated 
by mixed QCD--EW effects and is most likely 
overestimated due to our conservative assumption
of keeping such uncertainties uncorrelated across processes
(see \refse{se:MIXunc}).

%%%%%%%%%%%%%%%%%%%%
%%%%%%%%%%%%%%%%%%%%

We also discussed photon-induced contributions
and  QED corrections to PDFs.
In this context, for a
precise prediction of the $\gamma$-PDF
we have advocated the use of the \texttt{LUXqed\_plus\_PDF4LHC15\_nnlo} PDFs, which
implement a data driven determination of the $\gamma$-PDF.
For a consistency treatment of $\ord(\alpha)$ effects in the PDFs, the
\texttt{LUXqed\_plus\_PDF4LHC15\_nnlo} distributions should be used in all photon-, quark-, and gluon-induced channels.\footnote{%
This is automatically achieved by reweighting MC samples generated with arbitrary PDFs
with our complete $\NkLO\,\QCD\times\nNLO\,\EW$ predictions based on 
\texttt{LUXqed\_plus\_PDF4LHC15\_nnlo} PDFs. Vice versa, restricting the reweighting to 
pure EW corrections and using MC samples based on different PDFs 
can lead to inconsistencies at $\ord(\alpha)$.%
}
Photon-induced effects are negligible in  
$Z+$\,jet and $\gamma+$\,jet production, 
but their impact on \linebreak $pp\to W+$\,jet, and thus on the $W/Z$ and $W/\gamma$ ratios,
can reach the 5\% level at the TeV scale\footnote{%
Note that photon-induced contributions 
are not included in 
the summary plots of \reffi{fig:QCDxEW_error_combined}.
}
 (see \reffi{fig:LO_gamma_ind}).

Our predictions are provided in the form of tables for the central
predictions and for the different uncertainty sources.
Each uncertainty source is to be treated as a $1$-standard deviation
uncertainty and pragmatically associated with a Gaussian-distributed nuisance
parameter.

The predictions are given at parton level as distributions of the
vector boson $p_\rT$, with loose cuts and inclusively over other
radiation.
They are intended to be propagated to an experimental analysis using
Monte Carlo parton shower samples whose inclusive vector-boson $p_\rT$
distribution has been reweighted to agree with our parton-level
predictions.
The impact of additional cuts, non-perturbative effects on lepton
isolation, etc., can then be deduced from the Monte Carlo samples.
The additional uncertainties associated with the Monte Carlo
simulation are expected to be relatively small, insofar as the
vector-boson $p_\rT$ distribution that we calculate is closely connected
to the main experimental observables used in MET+jets searches.

Some caution is needed in implementing the results of this paper: for
example the uncertainty prescriptions are tied to the use of the
central values that we provide.
If an experiment relies on central values that differ, e.g.\ through
the use of MC samples that are not reweighted to our nominal predictions,
then the uncertainty scheme that
we provide may no longer be directly applicable.
Furthermore, for searches that rely on features of the event other
than missing transverse momentum, one should be aware that our
approach might need to be extended.
This would be the case notably for any observable that relies directly
on jet observables, whether related to the recoiling jet or vetoes on
additional jets.

Overall, it is possible to obtain precise theoretical control both for
vector-boson $p_\rT$ distributions, and for their ratios, at the level
of a few percent.
We expect this precision, across a wide range of $p_\rT$, to be of
significant benefit in MET+jets searches, notably enabling reliable
identification or exclusion of substantially smaller BSM signals than was
possible so far.
In fact, since the release of the first version of this paper, the
background estimates we propose here have been adopted in analyses by ATLAS
\cite{ATLAS-CONF-2017-060} and CMS \cite{CMS-PAS-EXO-16-048}.

%\subsubsection*{Note added}
%
%This preprint was released as a reference for dark matter  searches 
%presented at LHCP\,2017.
%While the effect of the recently computed NNLO QCD corrections
%is illustrated in various figures, 
%in order to reflect the usage of our calculations in present experimental analysis, 
%only NLO QCD and nNLO EW effects have been propagated
%through our framework of uncertainty estimates.
%An update of this study, with a full
%NNLO treatment of uncertainties, will be released in the near future.

\begin{acknowledgements}
%\subsubsection*{Acknowledgments}
We wish to thank Frank Krauss, Keith Ellis, Christian G\"utschow, Sarah Malik, Fabio Maltoni, Holger Schulz and Graeme Watt 
for valuable discussions.
This research was supported in part by
the UK Science and Technology Facilities Council,
the Swiss National Science Foundation (SNF) under contracts 200020-162487, CRSII2-160814,
and BSCGI0-157722, 
and by 
the Research Executive Agency (REA) of the European Union under the Grant Agreements 
PITN-GA-2012-316704 (''HiggsTools''),
PITN--GA--2012--315877 (''MCnet''),
and the ERC Advanced Grants MC@NNLO (340983) and LHCtheory (291377).
R.B.~is supported by the DOE contract DE-AC02-06CH11357. 
F.P.~is supported by the DOE grants DE-FG02- 91ER40684 and DE-AC02-06CH11357.  
C.W.~is supported by the National Science Foundation through award number
PHY-1619877.
The research of J.M.C.~is supported by the US DOE under contract DE-AC02-07CH11359.
The work of S.D.~is supported by the German Federal Ministry for Education and Research (BMBF).
This research used resources of the Argonne Leadership Computing Facility,
which is a DOE Office of Science User Facility supported under Contract
DE-AC02-06CH11357.  We also acknowledge support provided by the Center for Computational
Research at the University at Buffalo and the Wilson HPC Computing Facility
at Fermilab.
\end{acknowledgements}

%\clearpage

\begin{appendix}

\section{Theoretical predictions and uncertainties}
\label{se:numpredictions}

Predictions for the various $pp\to V+$\,jet processes 
listed in \refta{Tab:availabe_processes} with $\sqrt{s}=13~\TeV$ are provided  at\\ 
%\url{http://lpcc.web.cern.ch/LPCC/index.php?page=dm_wg_docs}
{\small\url{http://lpcc.web.cern.ch/content/dark-matter-wg-documents}}

The various predictions and related uncertainties at the highest available perturbative order,
\ie NNLO\,QCD and nNLO\,EW, 
as well as the labels of the corresponding histograms are listed in~\refta{Tab:histonames}.
Predictions with uncertainties  at NLO
and additional  
building blocks for the construction of the uncertainties at the various perturbative orders
can also be found in~\refta{Tab:histonames}.

\begin{table} [h]
\begin{center}
\begin{tabular}{c|c|c|c}
process & QCD order & EW order & label 
\\\hline
$pp\to \ell^+\nu_\ell/\ell^-\bar\nu_\ell+$\,jet  & NNLO QCD & nNLO EW & evj \\[2mm]
$pp\to \nu_\ell\bar\nu_\ell+$\,jet & NLO QCD &  nNLO EW& vvj \\[2mm] 
$pp\to \ell^+\ell^-+$\,jet & NNLO QCD &  nNLO EW & eej \\[2mm] 
$pp\to \gamma+$\,jet & NNLO QCD &  nNLO EW & aj \\[2mm]
\end{tabular}
\caption{List of processes, highest available QCD and EW order,
and process labels used in data files (see \refta{Tab:histonames}).
Predictions for $pp\to \nu_\ell\bar\nu_\ell+$\,jet are available only at NLO QCD,
but corresponding NNLO QCD corrections and uncertainties can be taken from 
$pp\to \ell^+\ell^-+$\,jet.}
\label{Tab:availabe_processes}
\end{center}
\end{table}

All ingredients and related uncertainties should be combined as
indicated in~\refeq{eq:QCDEWcomb1} and \refeq{eq:QCDEWcomb4}, and we recall that all 
nuisance parameters in \refeq{eq:QCDEWcomb4} should be 
Gaussian distributed 
with one standard deviation
corresponding to the range $[-1,+1]$ for all
$\eps_{\QCD,i}$, $\eps_{\PDF,i}$, $\eps_{\EW,i}$ and $\eps_\mix$.
In the implementation of the various relative uncertainties, \ie $\delta^{(i)} K_{\NkLO}$,
$\delta^{(i)} K_{\PDF}$, and \linebreak
$\delta^{(i)} K_{\EW}$,
it is crucial to take into account
their correct normalisation according to Eqs.~\refpar{eq:QCDEWcomb1} and 
\refpar{eq:QCDEWcomb4}. For instance, at $\NNLO\,\QCD\otimes\nNLO\,\EW$ the relative 
impact of QCD and EW uncertainties should be \linebreak
$\delta^{(i)} K_{\NNLO}/K_{\NNLO}$ and
$\delta^{(i)} K_{\EW}/(1+\kappa_\EW)$, respectively.

Concerning QCD contributions, predictions at \linebreak NNLO\,QCD should be combined with uncertainties at the same order.
However, before higher-order QCD calculations are thoroughly validated against high-statistics measurements
at moderate transverse momenta, theory uncertainties should be assessed in a 
more conservative way. To this end, we advocate the usage of 
NNLO QCD nominal predictions in combination with NLO QCD uncertainties, while keeping 
all EW effects at nNLO\,EW level.

All predictions and uncertainties for $pp\to \gamma+$\,jet are based on the 
dynamic photon isolation prescription introduced in~\refse{se:photons}. 
As explained therein, this requires an extra $\gamma+$\,jet specific 
uncertainty, which needs to be evaluated by means of a separate reweighting
in a standard Frixione isolation setup with fixed cone. Corresponding 
theoretical predictions at NLO QCD are denoted as 
$K^{(\gamma,\fix)}_{\NLO}(x)$ in~\refta{Tab:histonames}.

\begin{table*} [t]
\begin{center}
\begin{tabular}{l|c|l|c}
prediction &  equation &label & correlation\\\hline\hline
$\parx\siv_{\LO\,\QCD}(\vec\mu_0)$  [pb/GeV]  & \refpar{ew:QCDcomb} & proc\_x\_\LO &  - \\
$K^{(V)}_{\NNLO}(x)$ & \refpar{ew:QCDcomb},\refpar{eq:meanKfact} & proc\_x\_K\_NNLO & - \\
$\delta^{(1)}K^{(V)}_{\NNLO}(x)$ &  \refpar{ew:QCDcomb},\refpar{eq:var1Kfact} & proc\_x\_d1K\_NNLO & yes\\
$\delta^{(2)}K^{(V)}_{\NNLO}(x)$ &  \refpar{ew:QCDcomb}, \refpar{eq:dKqcd2} & proc\_x\_d2K\_NNLO & yes \\
$\delta^{(3)}K^{(V)}_{\NNLO}(x)$ &  \refpar{ew:QCDcomb}, \refpar{eq:dKqcd3} & proc\_x\_d3K\_NNLO & yes \\
$K^{(\gamma,\dyn)}_{\NLO}(x)$ & \refpar{ew:QCDcomb},\refpar{eq:meanKfact},\refpar{eq:dynisolation} & aj\_x\_K\_NLO & - \\
$K^{(\gamma,\fix)}_{\NLO}(x)$ & \refpar{ew:QCDcomb},\refpar{eq:meanKfact},\refpar{eq:standard_isolation_parameters} & aj\_x\_K\_NLO\_fix & - \\
$\kappa^{(V)}_{\EW}(x)=\kappa^{(V)}_{\nNLO\,\EW}(x)$ & \refpar{eq:ew0}--\refpar{eq:EWcorsplitting}, \refpar{eq:ewunccomb} & proc\_x\_kappa\_EW & - \\
$\delta^{(1)}\kappa^{(V)}_{\nNLO\,\EW}(x)$ & \refpar{eq:dkappaEW1}, \refpar{eq:ewunccomb}   & proc\_x\_d1kappa\_EW & yes\\
$\delta^{(2)}\kappa^{(V)}_{\nNLO\,\EW}(x)$ & \refpar{eq:dkappaEW2}, \refpar{eq:ewunccomb}   & proc\_x\_d2kappa\_EW & no\\
$\delta^{(3)}\kappa^{(V)}_{\nNLO\,\EW}(x)$ & \refpar{eq:dkappaEW3}, \refpar{eq:ewunccomb}   &  proc\_x\_d3kappa\_EW & no\\
$\delta K^{(V)}_{\mix}(x)$ &  \refpar{eq:dkmix5}, \refpar{eq:uncert_mix} & proc\_x\_dK\_NLO\_mix & yes\\
$\parx\siv_{\LO\,\gamma-{\rm ind.}}$   [pb/GeV]  & \refpar{eq:th1} & proc\_x\_gammaind\_\LO & - \\ 

$\delta^{(i)}K^{(V)}_{\PDF}(x)$ &  \refpar{ew:QCDcombPDF} & proc\_x\_dK\_PDF\_i & yes\\
\hline
$K^{(V)}_{\LO}(x)$  &  \refpar{ew:QCDcomb}& proc\_x\_K\_LO & - \\
$K^{(V)}_{\NLO}(x)$ & \refpar{ew:QCDcomb},\refpar{eq:meanKfact} & proc\_x\_K\_NLO & - \\
$\delta^{(1)}K^{(V)}_{\NLO}(x)$ &  \refpar{ew:QCDcomb},\refpar{eq:var1Kfact} & proc\_x\_d1K\_NLO & yes\\
$\delta^{(2)}K^{(V)}_{\NLO}(x)$ &  \refpar{ew:QCDcomb}, \refpar{eq:dKqcd2} & proc\_x\_d2K\_NLO & yes \\
$\delta^{(3)}K^{(V)}_{\NLO}(x)$ &  \refpar{ew:QCDcomb}, \refpar{eq:dKqcd3} & proc\_x\_d3K\_NLO & yes \\
$\delta^{(1)}K^{(V)}_{\LO}(x)$  &  \refpar{ew:QCDcomb},\refpar{eq:var1Kfact} & proc\_x\_d1K\_LO & yes \\
$\delta^{(2)}K^{(V)}_{\LO}(x)$  &  \refpar{ew:QCDcomb}, \refpar{eq:dKqcd2} & proc\_x\_d2K\_LO & yes \\
$\kappa^{(V)}_{\NLO\,\EW}(x)$ & \refpar{eq:ew0}--\refpar{eq:EWcorsplitting},\refpar{eq:ewunccomb} & proc\_x\_kappa\_NLO\_EW & - \\
$\kappa^{(V)}_{\NNLO\,\Sud}(x)$ & \refpar{eq:ew0}--\refpar{eq:EWcorsplitting},\refpar{eq:ewunccomb} & proc\_x\_kappa\_NNLO\_Sud & -\\
\end{tabular}
\caption{Naming scheme for the theoretical predictions and uncertainties described in \refse{se:ho}.
The upper part lists the highest available perturbation order, while the predictions in the lower part are included for completeness. 
The last column indicates the correlation of the uncertainties across different $V$+jets processes.
The actual distribution names are x=pTV and the individual processes are available in the files proc.dat 
with process names proc=eej,vvj,evj,aj, as defined in \refta{Tab:availabe_processes}.
Absolute predictions for $p_\rT$ distributions are in pb/GeV. }
\label{Tab:histonames}
\end{center}
\end{table*}

%\newpage

\section{QCD and EW uncertainties}
\label{app:unc}

\begin{figure*}[p]   
\centering
  \includegraphics[width=\ratiotextwidthapp\textwidth]{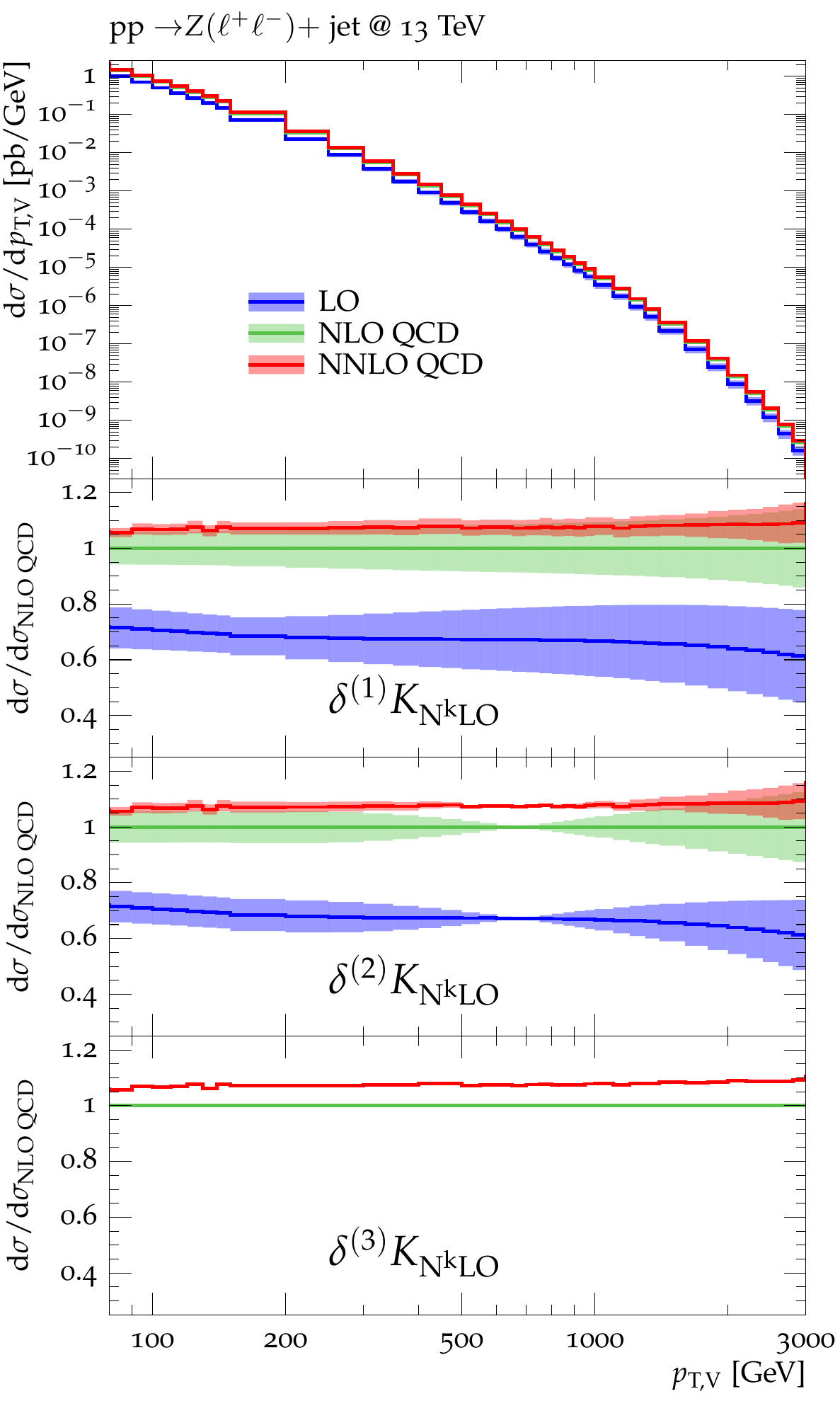}
  \includegraphics[width=\ratiotextwidthapp\textwidth]{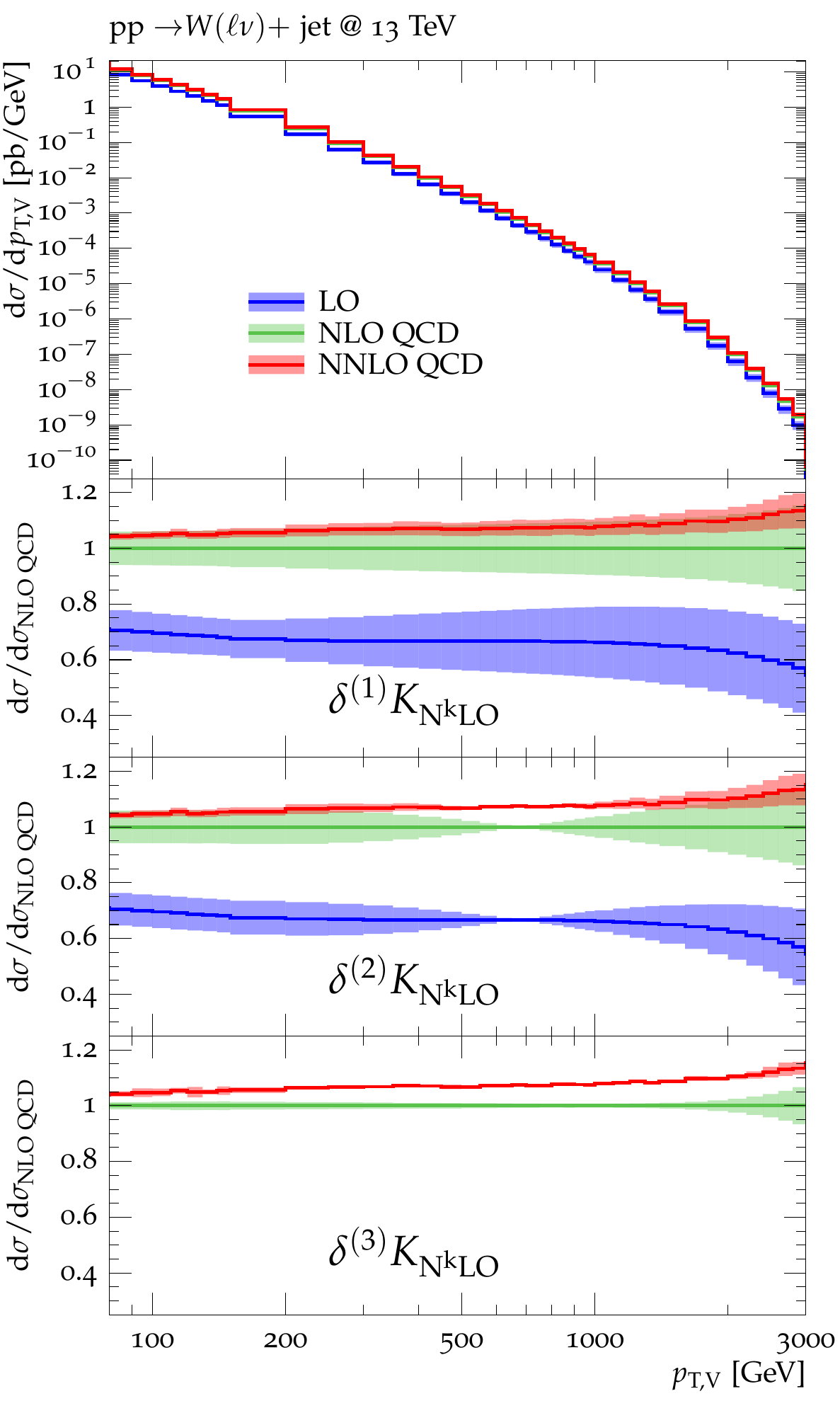}\\[3mm]
   \includegraphics[width=\ratiotextwidthapp\textwidth]{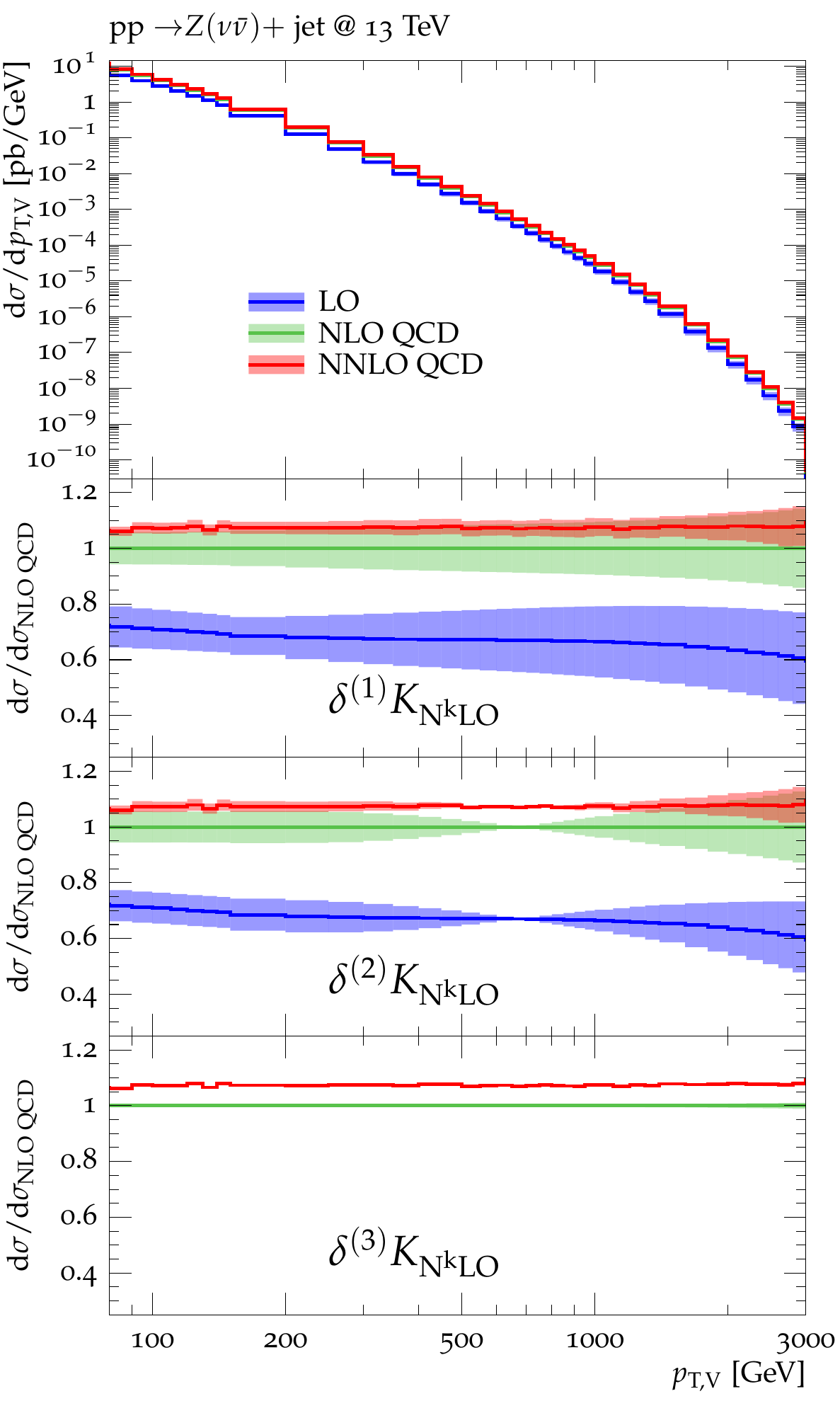}
  \includegraphics[width=\ratiotextwidthapp\textwidth]{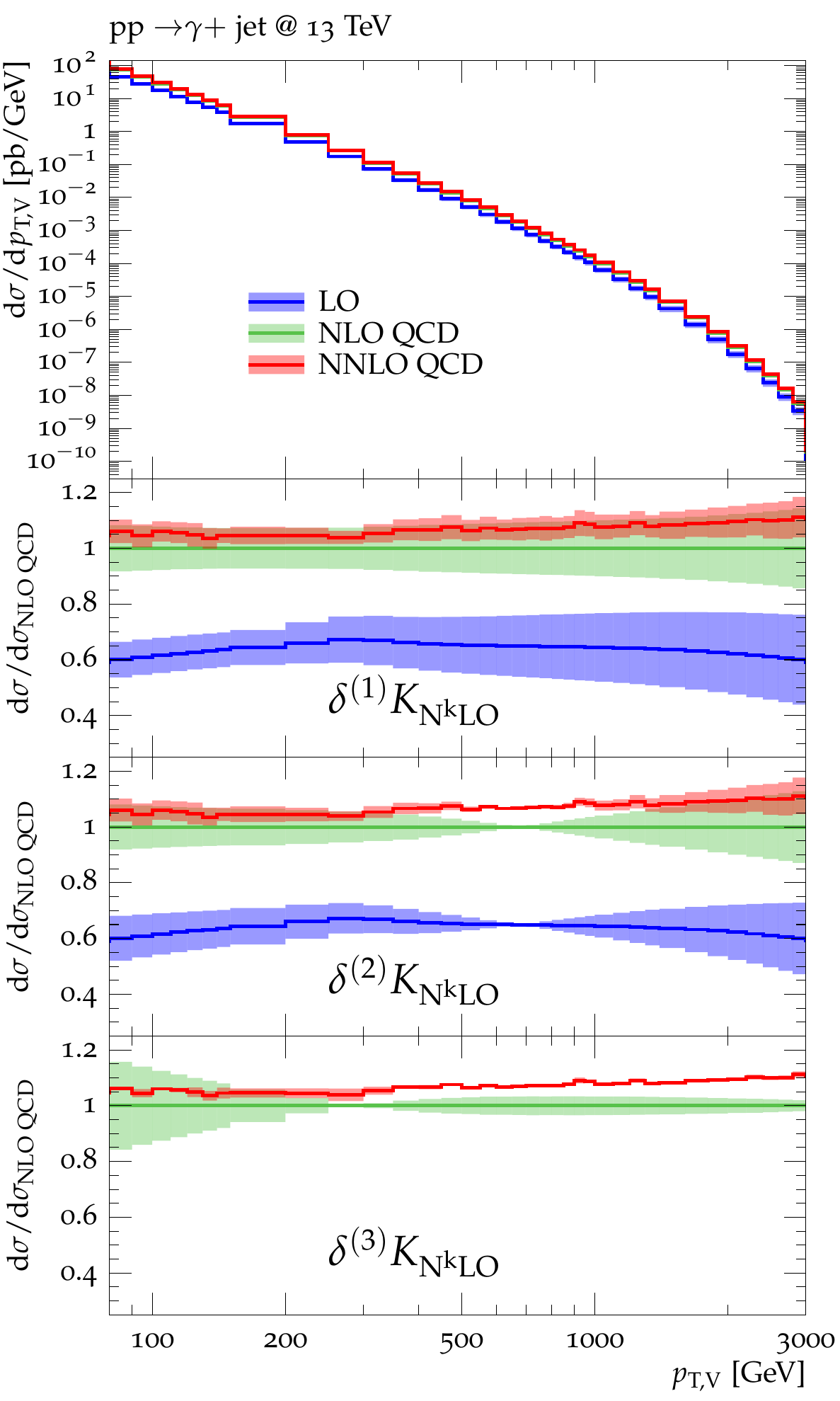}

\caption{Higher-order QCD predictions and uncertainties for
various $pp\to V$+jet processes at
13\,TeV.  Absolute predictions at LO and NLO
%
%LO, NLO and NNLO 
%
QCD are displayed in the
main frame.  In the ratio plots all results are normalised to NLO QCD, and
the bands correspond to the three types of QCD uncertainties,
$\delta^{(i)}K_{\NkLO}$, i.e.~scale uncertainties \refpar{eq:var1Kfact}, shape uncertainties 
\refpar{eq:dKqcd2} and process-correlation uncertainties \refpar{eq:dKqcd3}.  
%
%Note: f2/f4 denotes factor-2 and factor-4 scale variations at NNLO respectively.  
%
} 
\label{fig:app_QCD_error} \end{figure*}

In this appendix we present a series of technical plots that illustrate the relative importance of the
various sources of QCD and EW uncertainties discussed in \linebreak \refses{se:qcd}{se:ew}.
The impact of individual QCD uncertainties, $\delta^{(i)}K_{\NkLO}$, in 
$p_\rT$ spectra and ratios is illustrated in 
\reffis{fig:app_QCD_error}{fig:app_ratios_qcd}.
Similar plots for the three types of EW uncertainties,
$\delta^{(i)}\kappa^{(V)}_{\EW}$, are shown in \linebreak \reffis{fig:app_EW_error}{fig:app_ratios_ew}.

%%%%%%%%%%%%%%%%%%%%
%%%%%%%%%%%%%%%%%%%%

%%%%%%%%%%%%%%%%%%%%
\begin{figure*}[p]   
\centering
  \includegraphics[width=\ratiotextwidthapp\textwidth]{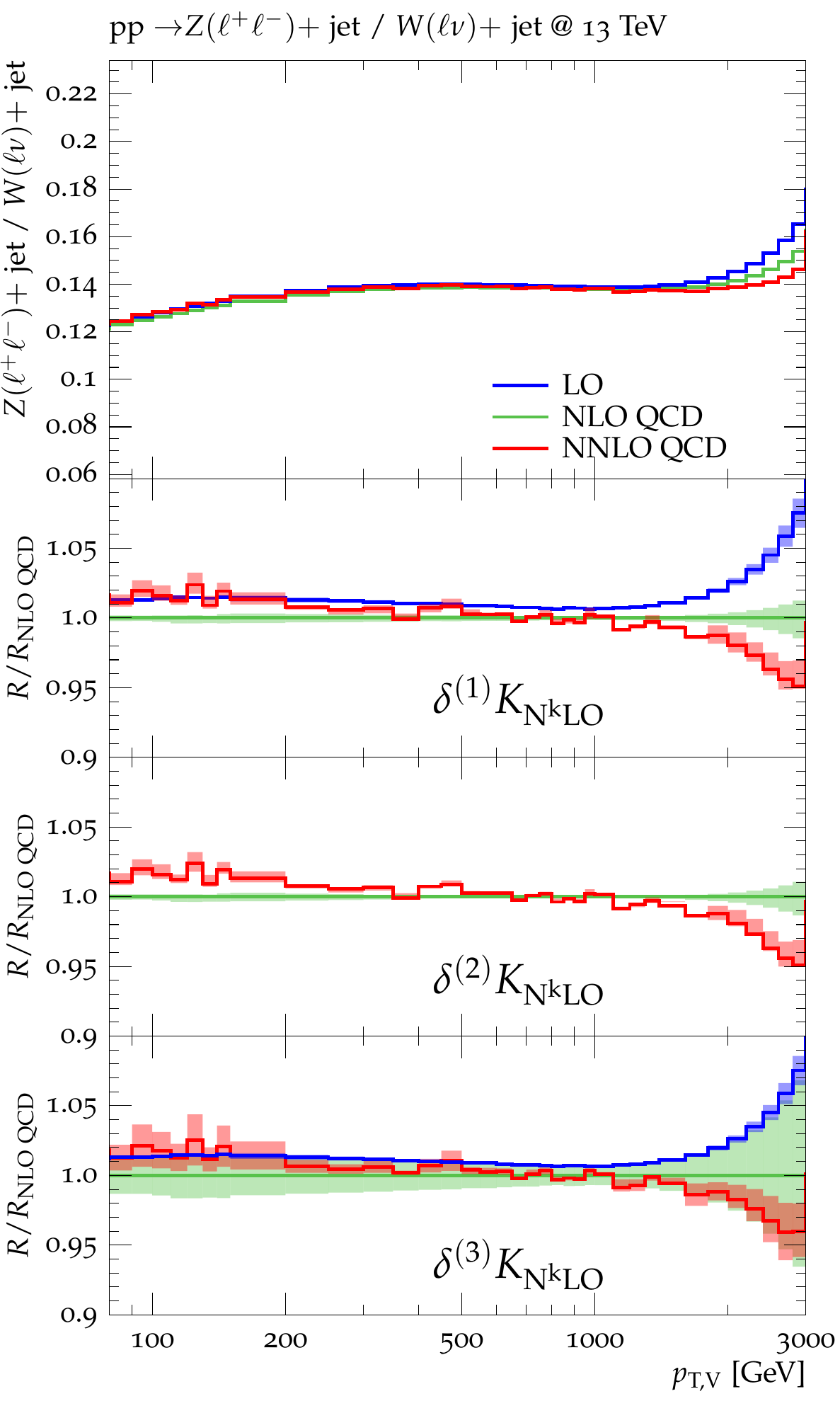}
    \includegraphics[width=\ratiotextwidthapp\textwidth]{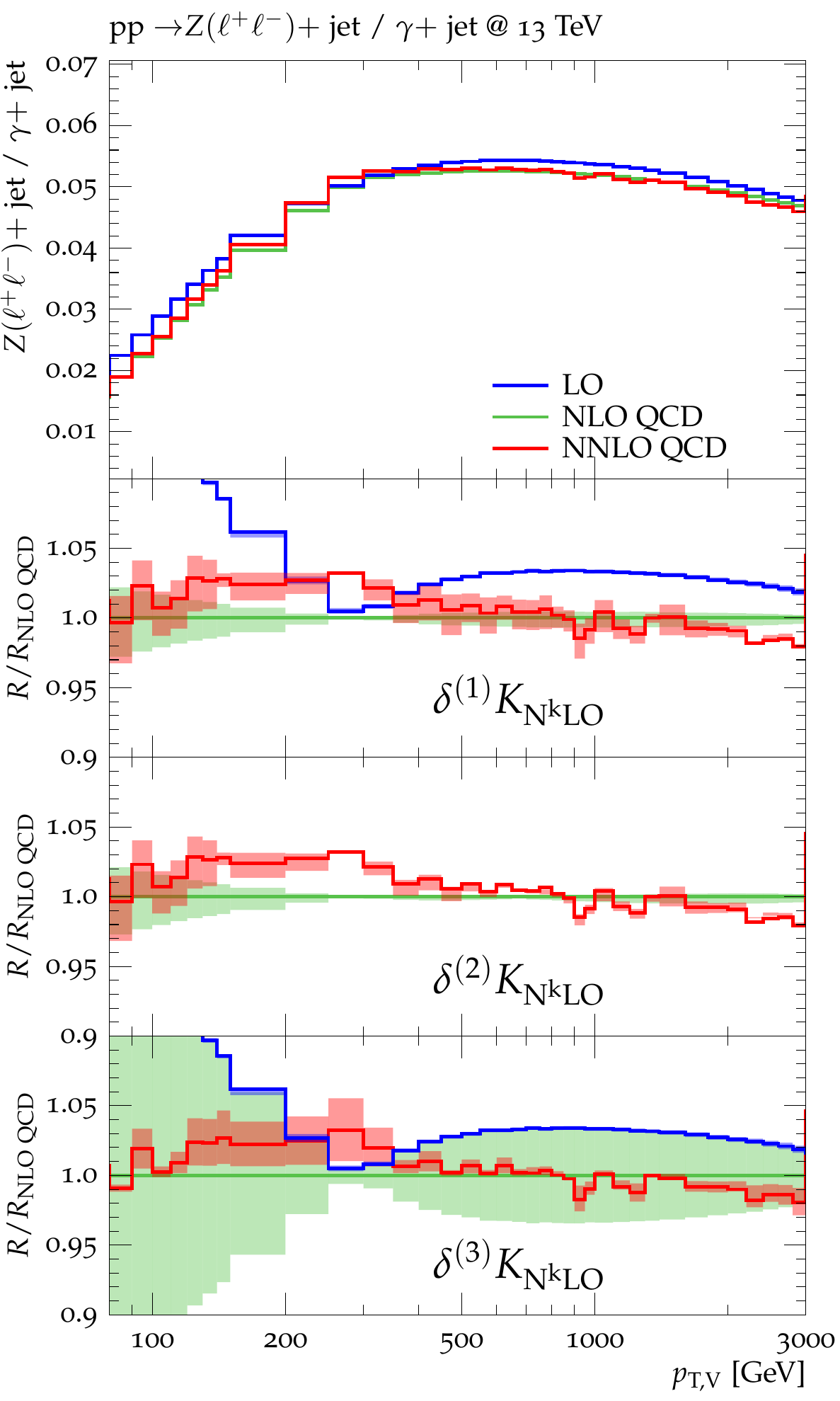}\\[6mm]
        \includegraphics[width=\ratiotextwidthapp\textwidth]{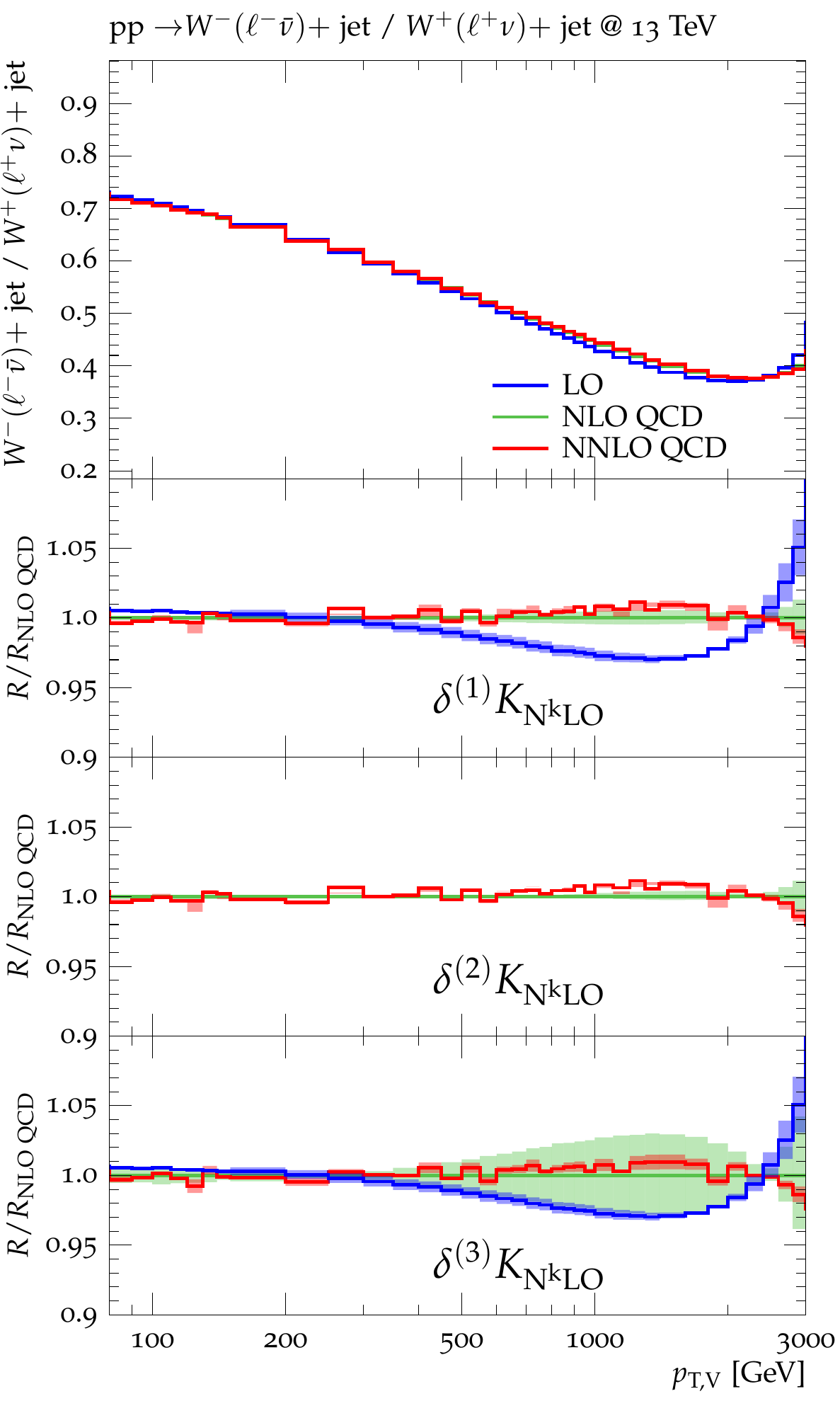}
    \includegraphics[width=\ratiotextwidthapp\textwidth]{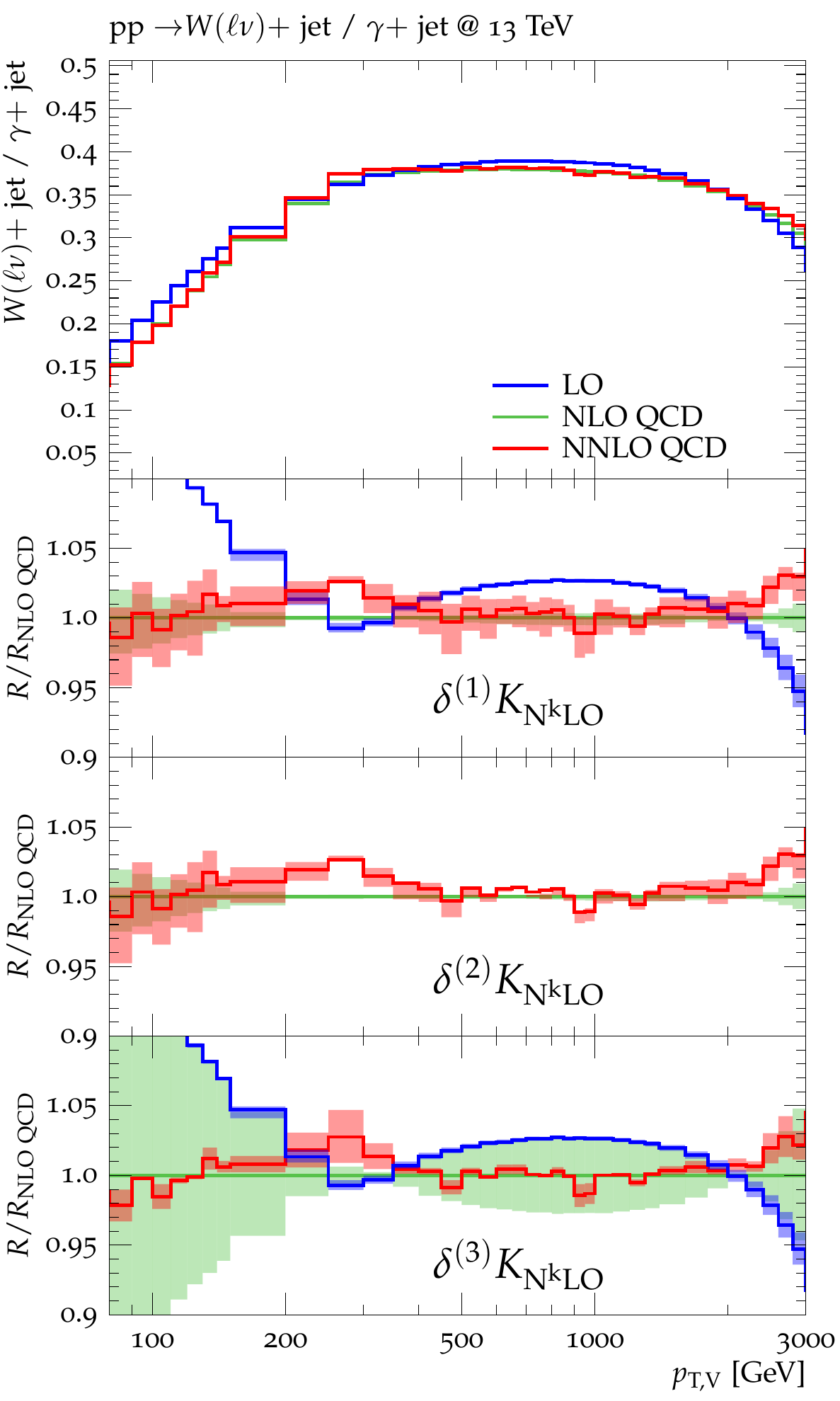}
\caption{
Ratios of $p_\rT$-distributions for various $pp\to V$+jet processes at 
LO and NLO QCD. 
The related scale uncertainties \refpar{eq:var1Kfact}, shape uncertainties 
\refpar{eq:dKqcd2} and process-correlation uncertainties \refpar{eq:dKqcd3}
are correlated amongst all
processes as discussed in \refse{se:qcd}.
}
\label{fig:app_ratios_qcd}
\end{figure*}
%%%%%%%%%%%%%%%%%%%%

%\clearpage
%%%%%%%%%%%%%%%%%%%%
\begin{figure*}[p] %[htbp!]   
\centering
  \includegraphics[width=\ratiotextwidthapp\textwidth]{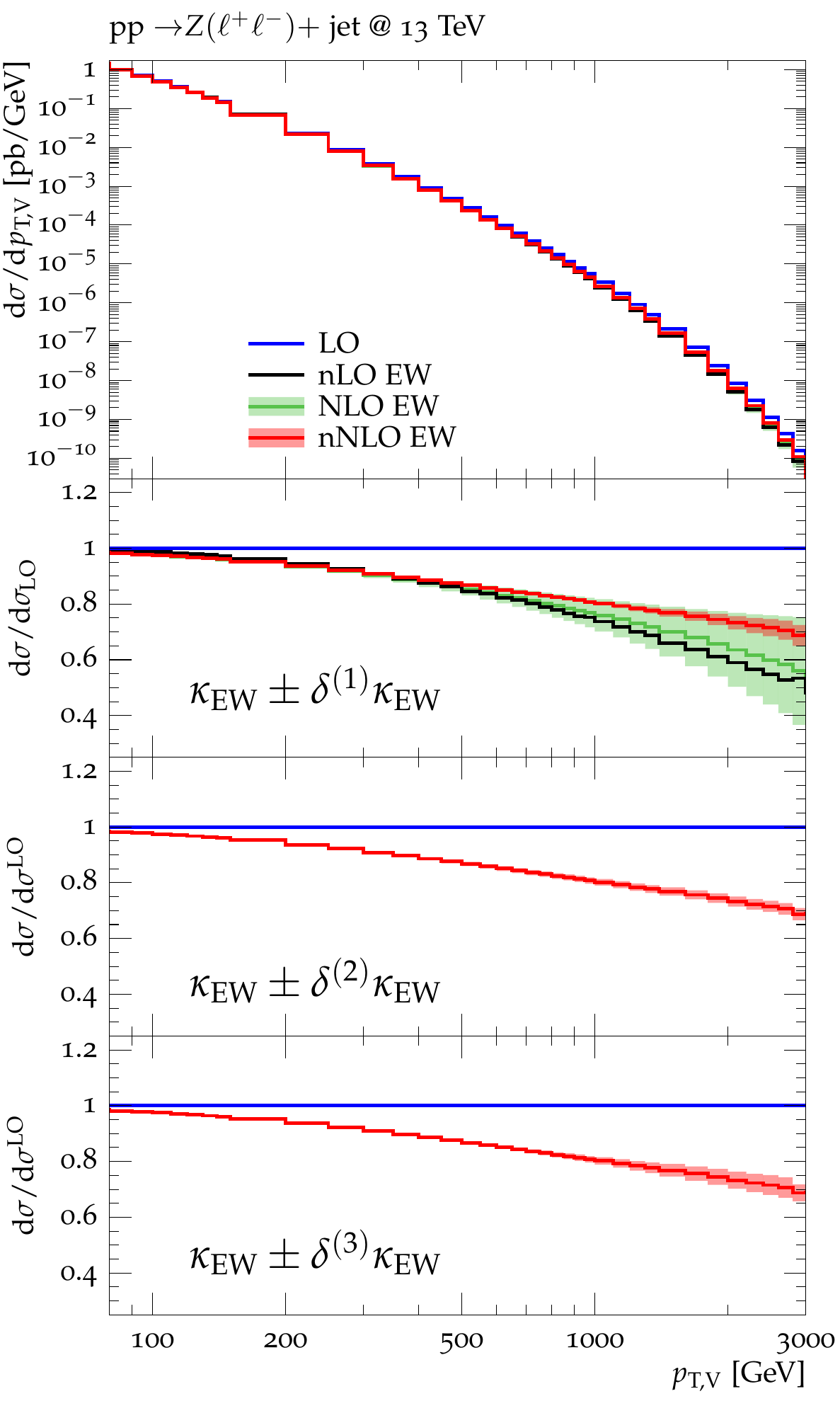}
 \includegraphics[width=\ratiotextwidthapp\textwidth]{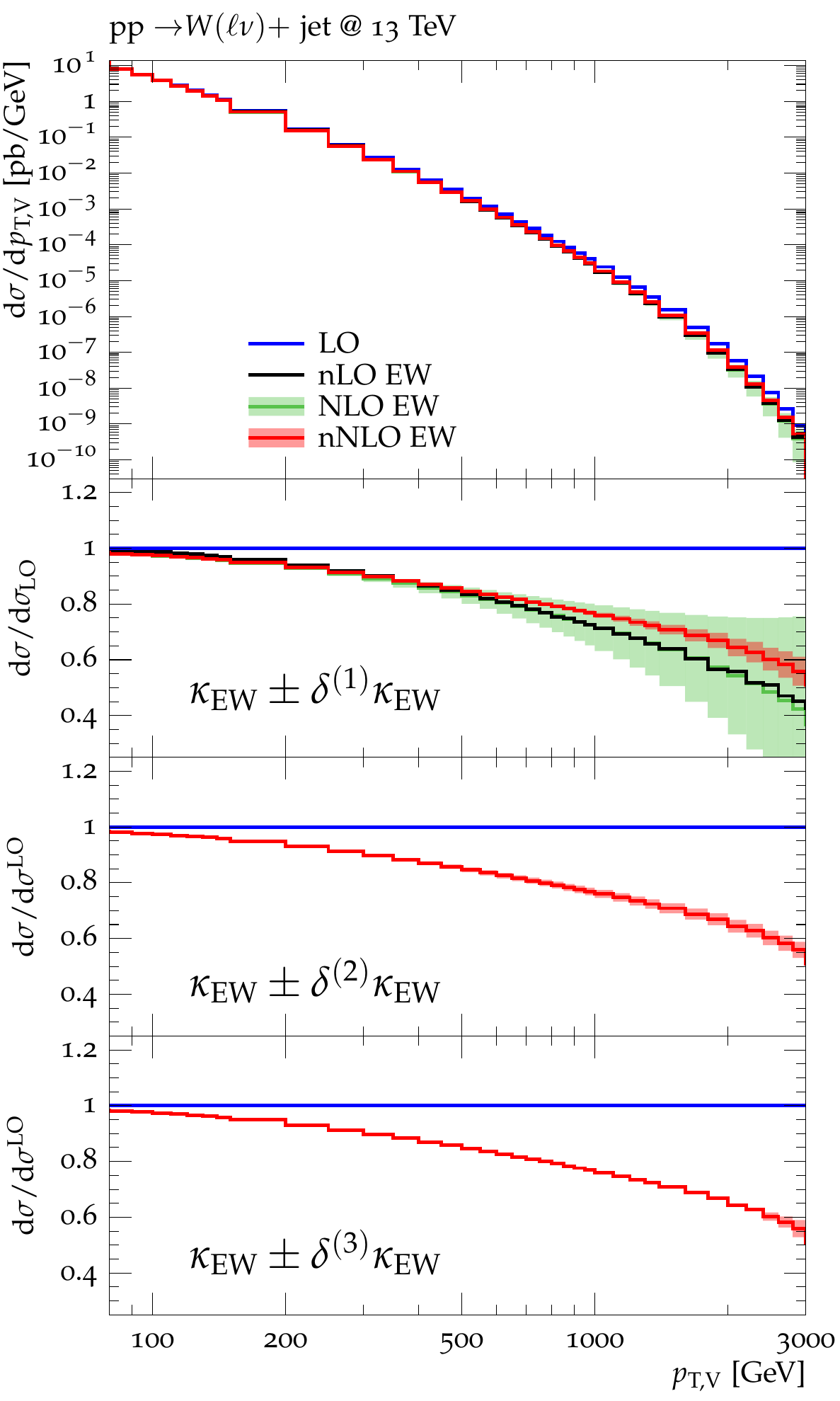}

  \includegraphics[width=\ratiotextwidthapp\textwidth]{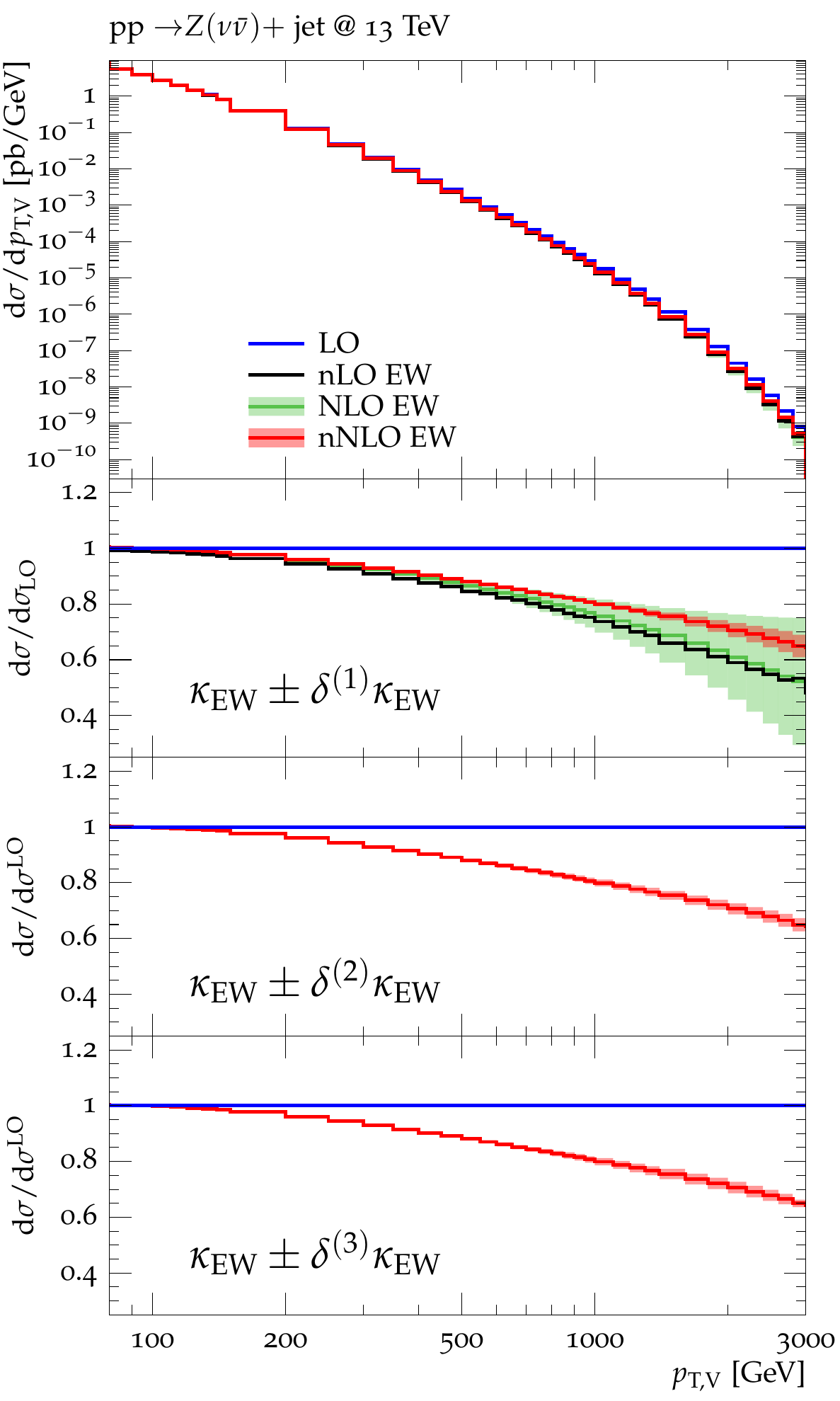}
  \includegraphics[width=\ratiotextwidthapp\textwidth]{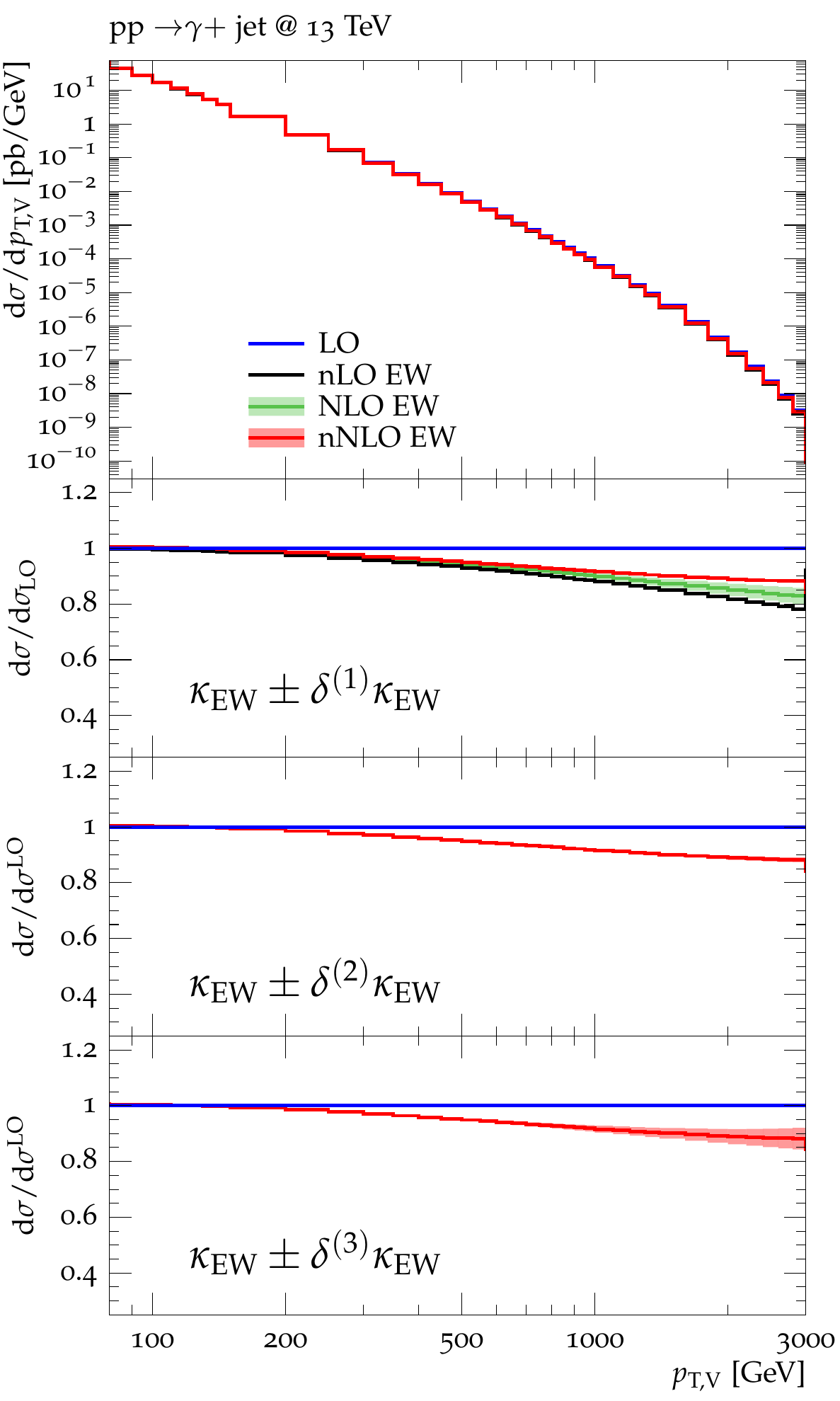}
\caption{
Higher-order EW predictions and uncertainties for different $pp\to V+$\,jet 
processes at 13\,TeV.
The main frames display absolute predictions at LO (blue), NLO EW (green) and nNLO EW (red),
as well as NLL Sudakov logarithms at NLO (black). The latter are dubbed nLO EW.
In the ratio plots all results are normalised to LO.
The bands correspond to the three types of EW uncertainties, $\delta^{(i)}\kappa^{(V)}_{\EW}$.
At nNLO EW (red bands)
they are defined
in Eqs.~\refpar{eq:dkappaEW1}, \refpar{eq:dkappaEW2} and \refpar{eq:dkappaEW3},
while at NLO EW (green band) only the uncertainty $\delta^{(1)}\kappa^{(V)}_{\NLO\,\EW}$, 
defined in \refeq{eq:EWuncertainties3}, is plotted.
}
\label{fig:app_EW_error}
\end{figure*}
%%%%%%%%%%%%%%%%%%%%

%%%%%%%%%%%%%%%%%%%%
\begin{figure*}[p]   
\centering
  \includegraphics[width=\ratiotextwidthapp\textwidth]{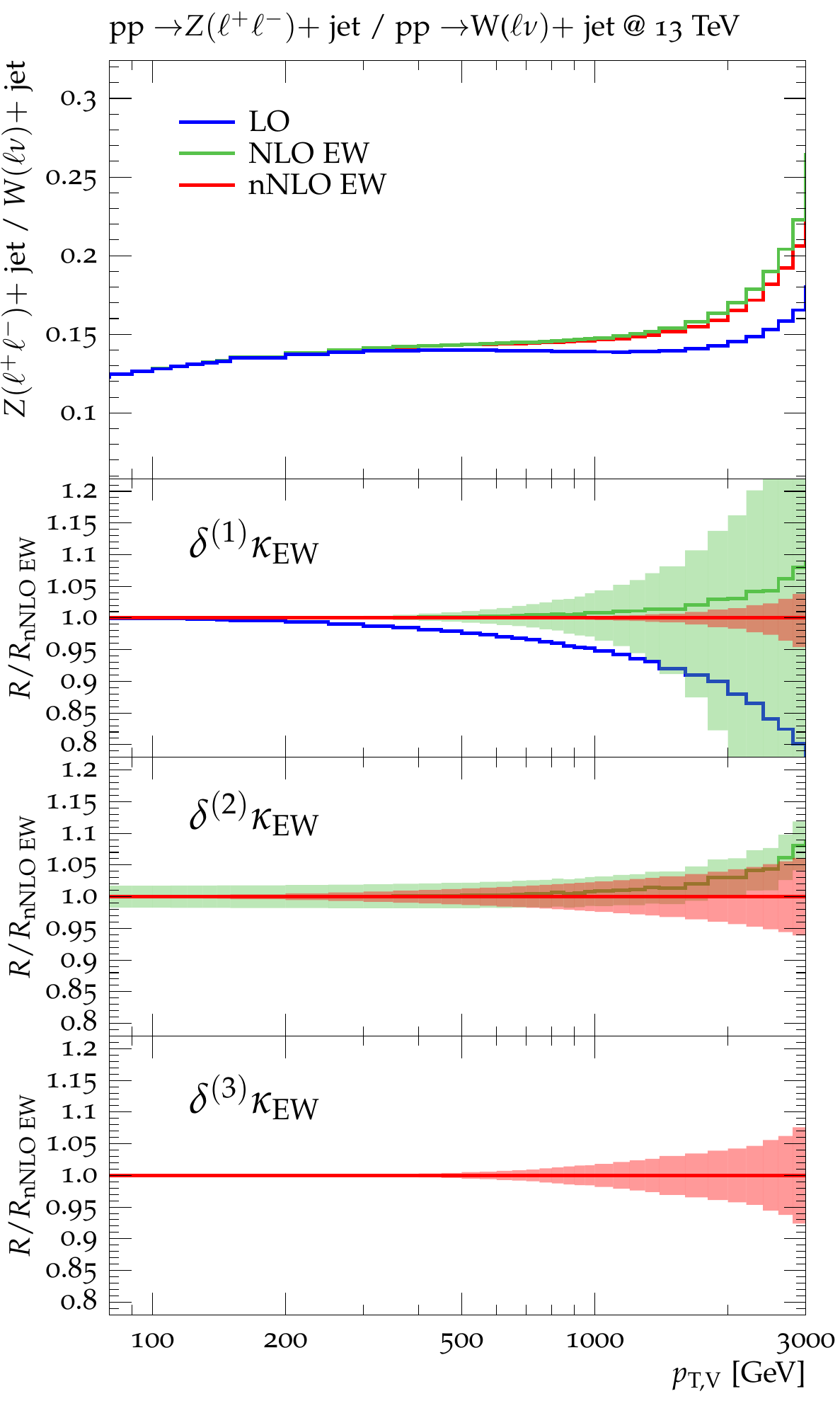}
    \includegraphics[width=\ratiotextwidthapp\textwidth]{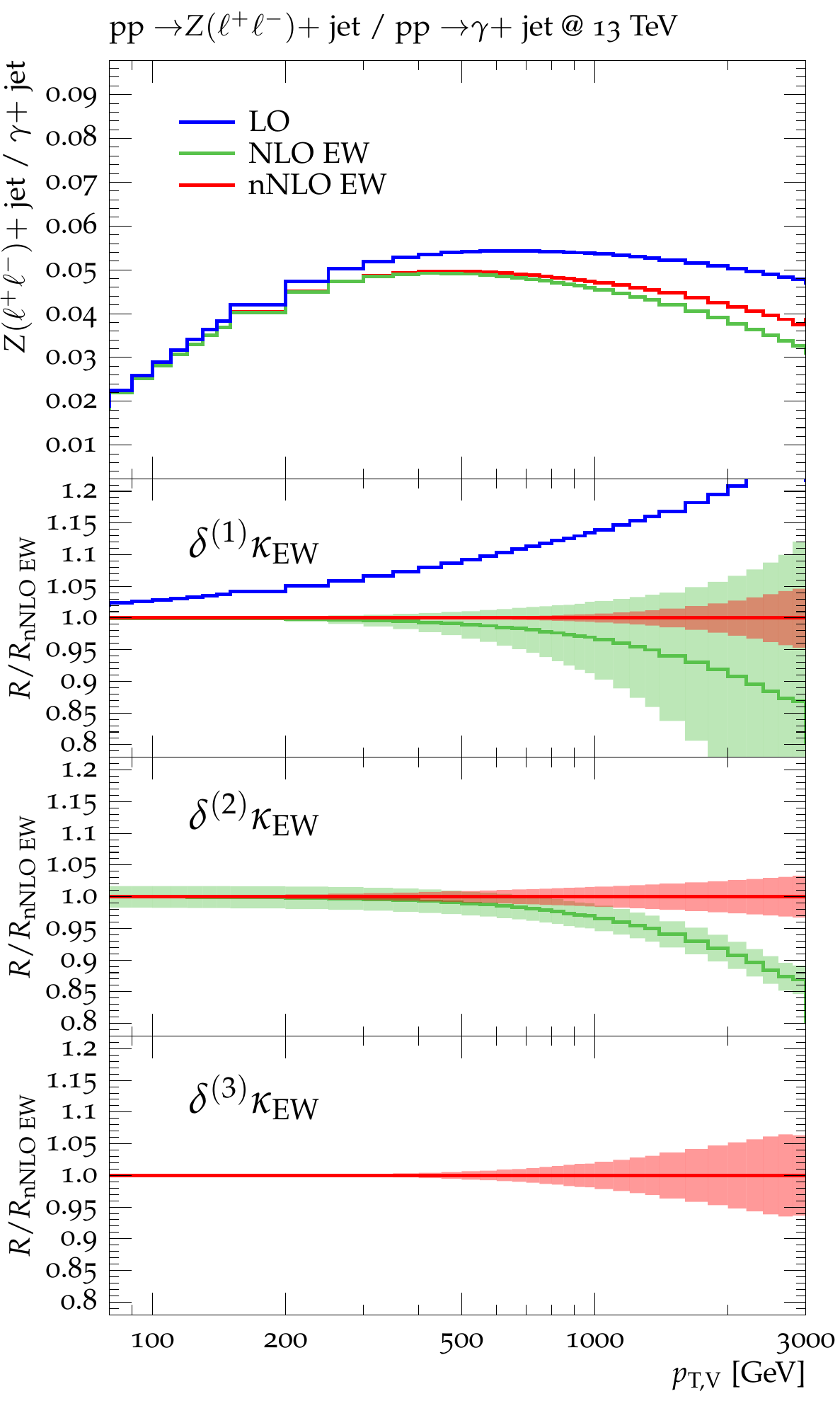}\\[6mm]
        \includegraphics[width=\ratiotextwidthapp\textwidth]{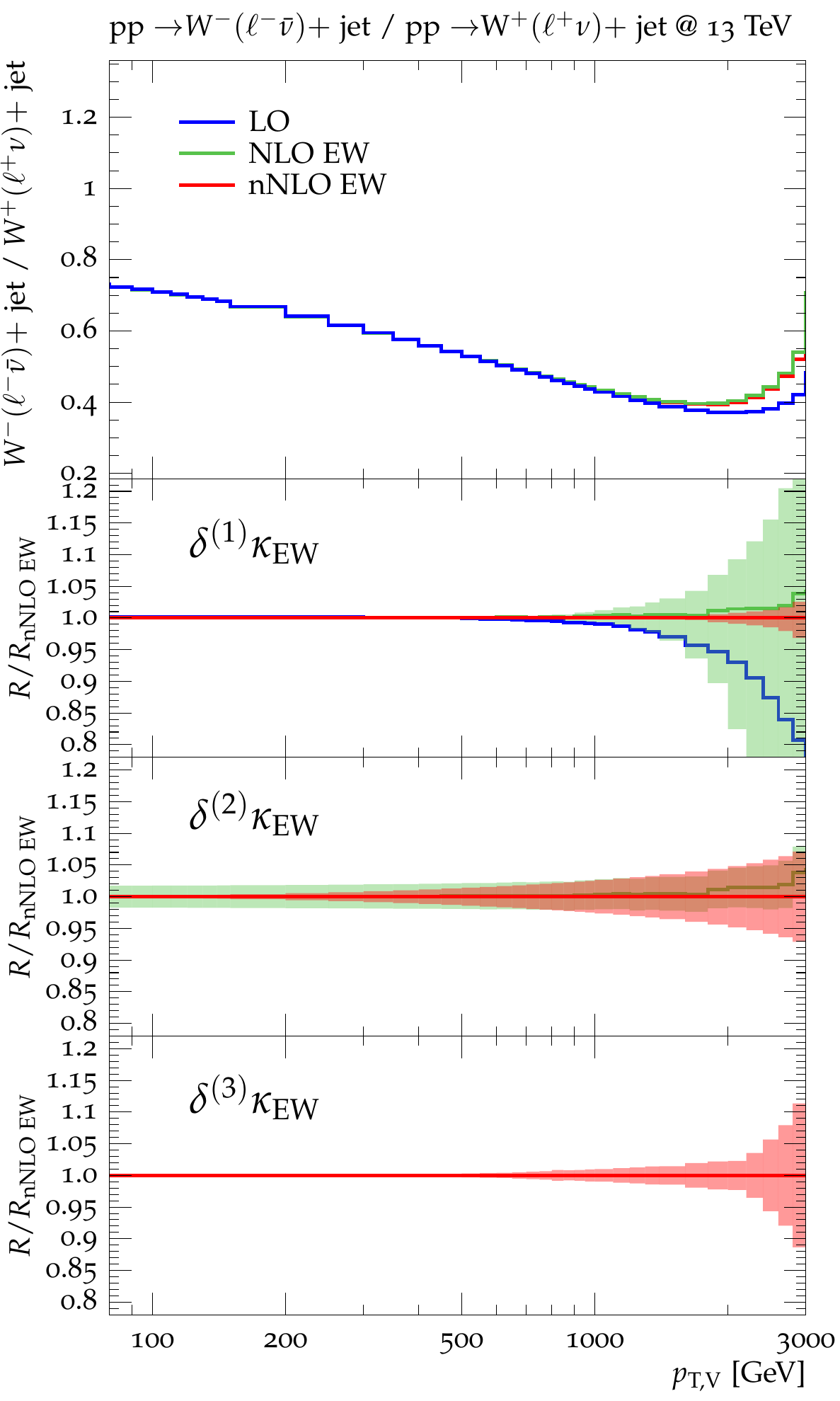}
    \includegraphics[width=\ratiotextwidthapp\textwidth]{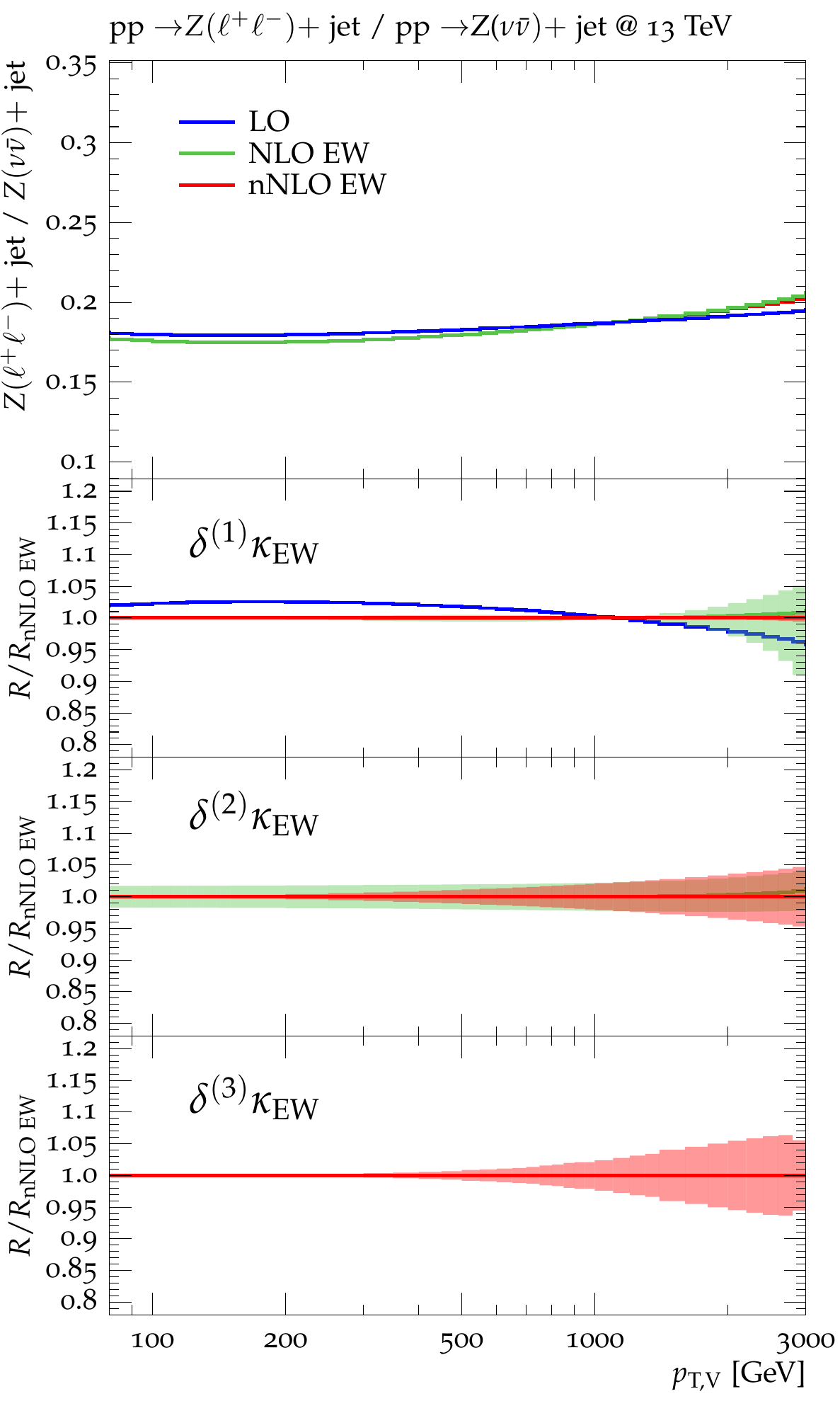}
\caption{
Ratios of $p_\rT$-distributions for various $pp\to V$+jet processes at LO, NLO EW
and nNLO EW.
The related EW uncertainties, $\delta^{(i)}\kappa^{(V)}_{\EW}$,
are defined in~Eqs.~\refpar{eq:dkappaEW1}, \refpar{eq:dkappaEW2} and \refpar{eq:dkappaEW3}
at nNLO and in \refeq{eq:EWuncertainties3} at NLO.
The uncertainty $\delta^{(1)}\kappa^{(V)}_{\EW}$
is correlated amongst processes, while $\delta^{(2)}\kappa^{(V)}_{\EW}$
and $\delta^{(3)}\kappa^{(V)}_{\EW}$ are uncorrelated.
}
\label{fig:app_ratios_ew}
\end{figure*}
%%%%%%%%%%%%%%%%%%%%

\end{appendix}

%\pagebreak
\providecommand{\href}[2]{#2}%\begingroup\raggedright\begin{thebibliography}{10}

%\clearpage
%\bibliographystyle{spphys}
\bibliographystyle{epjc}
\bibliography{dm}

%%%%%%%%%%%%%%%%%%%%%%%%%%%%%%%%%%%%%%%%%%%%%%%%%%%%%%%%%%%%%%%%%%%
\end{document}